\documentclass[mti,review,accept,pdftex,moreauthors]{Definitions/mdpi} 

\firstpage{1} 
\pubvolume{1}
\issuenum{1}
\articlenumber{0}
\pubyear{2023}
\copyrightyear{2023}
\externaleditor{Academic Editor: Firstname Lastname}
\datereceived{12 May 2023} 
\daterevised{4 June 2023} 
\dateaccepted{6 July 2023} 
\datepublished{17 July 2023} 
\hreflink{https://doi.org/} 

\usepackage{adjustbox}
\usepackage{multirow}
\usepackage{longtable}
\usepackage{arydshln}
\usepackage{comment}
\usepackage{graphicx}
\usepackage{caption}
\usepackage[labelformat=simple]{subcaption}

\DeclareCaptionLabelFormat{subcaptionlabel}{\normalfont(\textbf{#2}\normalfont)}
\Title{Encoding Variables, Evaluation Criteria, and Evaluation Methods for Data Physicalisations: A Review} 

\TitleCitation{Encoding Variables, Evaluation Criteria, and Evaluation Methods for Data Physicalisations:  {A Review} 
}

\Author{{Champika} 
 Ranasinghe $^{1,}$*\orcidA{} and Auriol Degbelo $^{2}$\orcidB{}}

\AuthorNames{Champika Ranasinghe and Auriol Degbelo}

\AuthorCitation{Ranasinghe, C.; Degbelo, A.}

\address{%
$^{1}$ \quad Data Management \& Biometrics, Faculty of Electrical Engineering, Mathematics \& Computer Science, University of Twente, 7500 AE Enschede, The Netherlands 
\\
$^{2}$ \quad Chair of 
 Geoinformatics, Faculty of Environmental Sciences, TU Dresden, 01062 Dresden,  Germany
; auriol.degbelo@tu-dresden.de}

\corres{Correspondence: c.m.eparanasinghe@utwente.nl}

\abstract{Data physicalisations, or physical visualisations, represent data physically, using variable properties of physical media.  
As an emerging area, Data physicalisation research needs conceptual foundations to support thinking about and designing new physical representations of data and 
 evaluating them. Yet, it remains unclear at the moment (i) what encoding variables are at the designer's disposal during the creation of physicalisations, (ii) what evaluation criteria could be useful, and (iii) what methods can be used to evaluate physicalisations. This article addresses these three questions through a narrative review and a systematic review. The narrative review draws on the literature from Information Visualisation, HCI and Cartography to provide a holistic view of encoding variables for data. The systematic review looks closely into the evaluation criteria and methods that can be used to evaluate data physicalisations. Both reviews offer a conceptual framework for researchers and designers interested in designing and {evaluating} data physicalisations. {The framework can be used as a common vocabulary to describe physicalisations and to identify design opportunities. We also proposed a seven-stage model for designing and evaluating physical data representations. The model can be used to guide the design of physicalisations and ideate along the stages identified. The evaluation criteria and methods extracted during the work can inform the assessment of existing and future data physicalisation artefacts.}}


\begin{document}

\section{Introduction}
\label{sec:introduction}

Data physicalisation or the physical visualisation of data focuses on representing data using geometric or material properties of physical media \cite{Jansen2015}. While data visualisations primarily focus on the sense of vision and creating data representations that can be ``seen'', data physicalisations have the potential to create data representations that can not only be seen, but also can be touched, smelled, heard, or tasted. Thus they enable new ways of interacting with data and multisensory data experiences \cite{van2022physicalizing}. They are becoming a means for narrowing the gap between people and data and have shown cognitive benefits that come along with their tangible nature (e.g., be effective for self-reflection, attention, and access to data) \cite{kyung2020dayclo_p19, ju2019bookly_p6, menheere2021laina_p58, stusak2016_p39, stusak2015_p55, dragicevic2021data}. Furthermore, data physicalisations have the potential to reach audiences (such as for example people with disabilities) that are difficult to reach with traditional data visualisations. Although physical representations of data have existed for many years, data physicalisation is still emerging as a research area \cite{Jansen2015}. In recent years, there has been a growing interest in establishing theoretical and design foundations for data physicalisation. For example, design principles and guidelines for physicalizing data have started to emerge (e.g., \citet{hogan2017towards, sauve2022physecology, bae2022making, sosa2018data, hogan2018data, willett2016embedded}). Yet, the development of design guidelines and theoretical foundations for data physicalisation research still has a long way to go. In this research, we aim to address two important dimensions of data physicalisation research: encoding variables and evaluation.

 \textit{\textbf{Why encoding variables and evaluation}}: 
Encoding variables (i.e., the properties of the material used to encode data) are a key design dimension of any data communication activity. This is especially important for designing multisensory and immersive data experiences. Design guidelines for data visualisations assume non-disabled populations; thus, the resulting visualisations inhibit people with intellectual and developmental disabilities (IDD) from accessing and effectively engaging with data \cite{wu2021understanding}. Understanding encoding variables that are perceivable through different human sensory channels is important for designing inclusive and accessible data representations for these user groups. Although visual variables have been well explored and are established in visualisation research, a shared understanding and a common vocabulary for variables of other perceptual modalities (beyond the sense of vision) still need to be established, especially to develop guidelines for designing multisensory and immersive data experiences \cite {roberts2010using, hogan2017towards, hogan2018data, jansen2016psycho_p2}. There have been some efforts in the past to provide a partial inventory of encoding variables and develop a grammar about experiential encoding variables (e.g., \cite{hogan2018toward, stusakvariables, oehlberg2018encoding, moere2008beyond, nesbitt2001modeling}). However, a compilation of encoding variables for all human sensory modalities and their usage is not available yet. Thus, finding out what multisensory encoding variables are available and which ones are practically used in data physicalisations is useful for filling this research gap. 

In addition, evaluation is important for researchers to assess the quality and impact of their data physicalisations (for example, to ensure that the users perceive the data embedded in physical representations and what short-term and long-term impact they could have on people). Methods and criteria for evaluating data visualisations, especially evaluating their ability to effectively and efficiently analyse and discover information, are well established. Physicalisations substantially differ from visualisations, for example, in their ability to engage people, spark interest, trigger interaction, and stimulate emotions. Therefore, new criteria for describing and evaluating the value of physicalisations are emerging.  For example, \citet{Wang2019} introduced a model for describing the value of a physicalisation based on its creativity and its ability to engage beyond the raw information content (engagement related to affective, physical, intellectual, and social). However, the evaluation methods and criteria that are currently used, as well as the aspects that are evaluated, are not sufficiently known. For example, there is no common knowledge about what evaluation methods are available and used for evaluating aspects related to information discovery/analysis, hedonic aspects, or more open-ended aspects (such as, for example, behavioural stimulation or initiating a social dialogue). This paper tries to fill these two research gaps by seeking answers to the following research questions (RQs):

\begin{itemize}
    \item RQ1: Which encoding variables can be used to create data physicalisations? 
    \item RQ2: Which evaluation criteria are relevant to the study of data physicalisations? 
    \item RQ3: Which evaluation methods are relevant to the study of data physicalisations? 
\end{itemize}

 \textit{\textbf{Method and contributions}}:
Methodically, these questions are examined through two complementary reviews. Encoding variables (also called `perceptual variables' \cite{roberts2010using}) are mentioned at different places in the literature, often using different terms to refer to the same notion, and the same term to refer to different notions, due to the interdisciplinary nature of data physicalisation research. For this reason, a systematic review is not appropriate to answer RQ1. Instead, a narrative review holds more potential to cover the breadth of ideas originating from the overlapping fields with data physicalisation research. A narrative review (see e.g., \cite{Snyder2019,Grant2009}) identifies potentially relevant research that has implications for a topic and synthesizes these using meta-narratives. This narrative review uses knowledge from the Visualisation, HCI and Cartography literature and yields a synthesis of the scattered literature on encoding variables into a coherent framework (Contribution 1). \textcolor{black}{The Cartography literature was included in the narrative review because the two fields of Information Visualisation and Cartography (i) share an object of study (i.e., maps), and (ii) there is evidence of their mutual interplay. Most importantly, the use of non-visual modalities to communicate geographic information has been extensively studied by cartographers (e.g., for the design of tactile \cite{wabinski2022guidelines} and sonic \cite{laakso2010sonic} maps), and some of the insights in that context can benefit data physicalisation research.} Answering RQ2 and RQ3 is done through a systematic review of papers published between 2009 and 2022. The systematic review helps to learn about evaluation criteria and evaluation methods relevant to data physicalisation research (Contribution 2).

\section{Existing Design Spaces for Physicalisations}
\label{sec:designspaces}

Previous work has suggested several design spaces/concepts that describe the dimensions that characterize data physicalisations. Thus, these design dimensions can be used to guide the design and evaluation of data physicalisations. We first analysed these design spaces to understand the extent to which they covered the two aspects of our focus: encoding variables and evaluation. Although each individual design framework uses different terms, they cover a total of 13 different design dimensions, as outlined in Table \ref{tab:existing-design-dimensions}. \textcolor{black}{The exact terms used by each design framework are summarised in Appendix \ref{sec:appendixA} and Table \ref{tab:existingDims}}. These distinct dimensions include \textit{Data} 
 that describes the nature of data represented by the physicalisation, \textit{Audience}, which refers to the type of the target audience, {\textit{Representational Intent}} that describes the  purpose of the physicalisation (e.g., the effective and efficient discovery of information, evoking specific feelings, and initiating social dialogue), \textit{Representational Material} that refers to the material used for the physicalisation, \textit{Sensory Modality} that refers to the human sensory channel used to perceive data, \textit{Encoding Variables} that describes the physical variables used to encode data, \textcolor{black}{\textit{Representational Fidelity}} that describes the metaphorical relationship between data and the materials used to encode data), \textit{Interaction} that describes the type and the nature of interactions that the physicalisation allows, \textit{Proximity to the Data Referent}, which refers to the degree of embodiment (i.e., proximity/situatedness) of the data physicalisation with respect to the data they represent (data referent), \textit{Proximity to the User} that describes the degree of embodiment (proximity/situatedness) of the data physicalisation with respect to the user/user's environment, \textit{Physical Setup} that details the distribution of components of the physicalisation (i.e., the physical setup of components), \textit{Mobility} that indicates whether the physicalisation is bound to a specific location or not, and \textit{Narrative formulation} that describes how data physicalisation facilitates the discovery of information through its external physical form and through any interactive affordances it provides 
 \cite{VandeMoere2009}. 

\begin{table}[H]
\caption{Summary of design dimensions covered by existing design spaces for data physicalisation.}
\label{tab:existing-design-dimensions}
\begin{adjustbox}{max width=\textwidth}
\begin{tabular}{p{4cm} p{1cm}p{1cm}p{1cm}p{1cm}p{1cm}p{1cm}p{1cm}p{1cm}p{1cm}p{1cm}p{1cm}p{1cm}p{0.5cm}p{0.5cm}}
\toprule
 &
  \multicolumn{13}{c}{\textbf{\textit{Design Dimensions} 
}} \\ \cline{2-15} 
\textbf{\textit{Design Space/Framework}} &
  \multicolumn{1}{l}{\rotatebox{90} {\textbf{Data}}} &
  \multicolumn{1}{l}{\rotatebox{90} {\textbf{Audience}}} &
  \multicolumn{1}{l}{\rotatebox{90} {{\color{black} \textbf{Representational Intent}}}} &
  \multicolumn{1}{l}{\rotatebox{90} {\textbf{Representational Material}}} &
  \multicolumn{1}{l}{\rotatebox{90} {\textbf{Sensory Modalities}}} &
  \multicolumn{1}{l}{\rotatebox{90} {\textbf{Encoding Variables}}} &
  \multicolumn{1}{l}{\rotatebox{90} {{\color{black}\textbf{Representational Fidelity}}}} &
  \multicolumn{1}{l}{\rotatebox{90} {\textbf{Interaction}}} &
  \multicolumn{1}{l}{\rotatebox{90} {\textbf{Proximity---Data Ref.}}} &
  \multicolumn{1}{l}{\rotatebox{90} {\textbf{Proximity---User}}} &
  \multicolumn{1}{l}{\rotatebox{90} {\textbf{Physical Setup}}} &
  \multicolumn{1}{l}{\rotatebox{90} {\textbf{Mobility}}} &
  \multicolumn{1}{l}{\rotatebox{90} {\textbf{Narrative Formulation}}} &
  \rotatebox{90} {\color{black} \textbf{Evaluation}} 
  \\ \midrule
  
\multicolumn{1}{l}{\begin{tabular}{@{}c@{}}Mutisensory Design Space \cite{nesbitt2001modeling} \end{tabular}} & 
  \multicolumn {1}{l}{} &
  \multicolumn{1}{l}{} &
  \multicolumn{1}{l}{} &
  \multicolumn{1}{l}{} &
  \multicolumn{1}{l}{\checkmark} &
  \multicolumn{1}{l}{\checkmark} &
  \multicolumn{1}{l}{} &
  \multicolumn{1}{l}{} &
  \multicolumn{1}{l}{} &
  \multicolumn{1}{l}{} &
  \multicolumn{1}{l}{} &
  \multicolumn{1}{l}{} &
  \multicolumn{1}{l}{} &\multicolumn{1}{l}{}
   \\ \midrule  
  
\multicolumn{1}{l}{\begin{tabular}{@{}c@{}}Data Sculpture Domain Model \cite{zhao2008embodiment} \end{tabular}} & 
  \multicolumn {1}{l}{} &
  \multicolumn{1}{l}{} &
  \multicolumn{1}{l}{\checkmark} &
  \multicolumn{1}{l}{} &
  \multicolumn{1}{l}{} &
  \multicolumn{1}{l}{} &
  \multicolumn{1}{l}{} &
  \multicolumn{1}{l}{} &
  \multicolumn{1}{l}{} &
  \multicolumn{1}{l}{} &
  \multicolumn{1}{l}{} &
  \multicolumn{1}{l}{} &
  \multicolumn{1}{l}{} &\multicolumn{1}{l}{}
  
   \\ \midrule
\multicolumn{1}{l}{\begin{tabular}{@{}c@{}}Embodiment Model \cite{zhao2008embodiment} \end{tabular}} & 
  \multicolumn {1}{l}{} &
  \multicolumn{1}{l}{} &
  \multicolumn{1}{l}{} &
  \multicolumn{1}{l}{} &
  \multicolumn{1}{l}{} &
  \multicolumn{1}{l}{} &
  \multicolumn{1}{l}{\checkmark} &
  \multicolumn{1}{l}{} &
  \multicolumn{1}{l}{} &
  \multicolumn{1}{l}{} &
  \multicolumn{1}{l}{} &
  \multicolumn{1}{l}{} &
  \multicolumn{1}{l}{} &\multicolumn{1}{l}{}
   \\ \midrule

\multicolumn{1}{l}{\begin{tabular}{@{}c@{}}Data Sculpture Design Taxonomy \cite{VandeMoere2009} \end{tabular}} &
  \multicolumn{1}{l}{} &
  \multicolumn{1}{l}{} &
  \multicolumn{1}{l}{} &
  \multicolumn{1}{l}{} &
  \multicolumn{1}{l}{} &
  \multicolumn{1}{l}{} &
  \multicolumn{1}{l}{\checkmark} &
  \multicolumn{1}{l}{} &
  \multicolumn{1}{l}{} &
  \multicolumn{1}{l}{} &
  \multicolumn{1}{l}{} &
  \multicolumn{1}{l}{} & \checkmark & \multicolumn{1}{l}{}
  
   \\ \midrule
\multicolumn{1}{l}{\begin{tabular}{@{}c@{}}Framework for Situated and Embedded Data Representations \cite{willett2016embedded} \end{tabular}} &
  \multicolumn{1}{l}{} &
  \multicolumn{1}{l}{} &
  \multicolumn{1}{l}{} &
  \multicolumn{1}{l}{} &
  \multicolumn{1}{l}{} &
  \multicolumn{1}{l}{} &
  \multicolumn{1}{l}{} &
  \multicolumn{1}{l}{} &
  \multicolumn{1}{l}{\checkmark} &
  \multicolumn{1}{l}{} &
  \multicolumn{1}{l}{} &
  \multicolumn{1}{l}{} &
  \multicolumn{1}{l}{} &\multicolumn{1}{l}{}
   \\ \midrule
\multicolumn{1}{l}{\begin{tabular}{@{}c@{}}Framework for Multisensory Data Representation \cite{hogan2017towards} \end{tabular}} &
  \multicolumn{1}{l}{\checkmark} &
  \multicolumn{1}{l}{} &
  \multicolumn{1}{l}{\checkmark} &
  \multicolumn{1}{l}{\checkmark} &
  \multicolumn{1}{l}{\checkmark} &
  \multicolumn{1}{l}{} &
  \multicolumn{1}{l}{} &
  \multicolumn{1}{l}{\checkmark} &
  \multicolumn{1}{l}{} &
  \multicolumn{1}{l}{} &
  \multicolumn{1}{l}{} &
  \multicolumn{1}{l}{} &
  \multicolumn{1}{l}{} &\multicolumn{1}{l}{}
   \\ \midrule
\multicolumn{1}{l}{\begin{tabular}{@{}c@{}}Framework for Multisensorial Immersive Analytics  \cite{mccormack2018multisensory}\end{tabular}} &
  \multicolumn{1}{l}{\checkmark} &
  \multicolumn{1}{l}{} &
  \multicolumn{1}{l}{\checkmark} &
  \multicolumn{1}{l}{} &
  \multicolumn{1}{l}{\checkmark} &
  \multicolumn{1}{l}{\checkmark} &
  \multicolumn{1}{l}{} &
  \multicolumn{1}{l}{} &
  \multicolumn{1}{l}{} &
  \multicolumn{1}{l}{} &
   \multicolumn{1}{l}{\checkmark} &
  \multicolumn{1}{l}{} &
  \multicolumn{1}{l}{} &\multicolumn{1}{l}{}
   \\ \midrule

\multicolumn{1}{l}{\begin{tabular}{@{}c@{}}Physecology \cite{sauve2022physecology} \end{tabular}} &
  \multicolumn{1}{l}{\checkmark} &
   \multicolumn{1}{l}{} &
  \multicolumn{1}{l}{} &
  \multicolumn{1}{l}{} &
  \multicolumn{1}{l}{} &
  \multicolumn{1}{l}{\checkmark} &
  \multicolumn{1}{l}{} &
  \multicolumn{1}{l}{\checkmark} &
  \multicolumn{1}{l}{} &
  \multicolumn{1}{l}{\checkmark} &
  \multicolumn{1}{l}{\checkmark} &
  \multicolumn{1}{l}{} &
  \multicolumn{1}{l}{} &\multicolumn{1}{l}{}
   \\ \midrule
\multicolumn{1}{l}{\begin{tabular}{@{}c@{}}Cross-Disciplinary Design Space \cite{bae2022making} \end{tabular}} &
  \multicolumn{1}{l}{\checkmark} &
  \multicolumn{1}{l}{\checkmark} &
  \multicolumn{1}{l}{\checkmark} &
  \multicolumn{1}{l}{\checkmark} &
  \multicolumn{1}{l}{\checkmark} &
  \multicolumn{1}{l}{\checkmark} &
  \multicolumn{1}{l}{} &
  \multicolumn{1}{l}{\checkmark} &
  \multicolumn{1}{l}{} &
  \multicolumn{1}{l}{\checkmark} &
  \multicolumn{1}{l}{} &
  \multicolumn{1}{l}{\checkmark} &
  \multicolumn{1}{l}{} &\multicolumn{1}{l}{}

   \\ \midrule
\multicolumn{1}{l}{\begin{tabular}{@{}c@{}}\color {black}Design Elements in Data Physicalisation \cite{dumivcic2022design} \end{tabular}} &
  \multicolumn{1}{l}{\checkmark} &
  \multicolumn{1}{l}{} &
  \multicolumn{1}{l}{\checkmark} &
  \multicolumn{1}{l}{\checkmark} &
  \multicolumn{1}{l}{\checkmark} &
  \multicolumn{1}{l}{\color{black}\checkmark} &
  \multicolumn{1}{l}{\checkmark} &
  \multicolumn{1}{l}{\checkmark} &
  \multicolumn{1}{l}{} &
  \multicolumn{1}{l}{} &
  \multicolumn{1}{l}{} &
  \multicolumn{1}{l}{} &
  \multicolumn{1}{l}{} &\multicolumn{1}{l}{\color{black}}

   \\ \midrule

\multicolumn{1}{l}{\begin{tabular}{@{}c@{}}\color {black}This Paper \end{tabular}} &
  \multicolumn{1}{l}{} &
  \multicolumn{1}{l}{} &
  \multicolumn{1}{l}{} &
  \multicolumn{1}{l}{} &
  \multicolumn{1}{l}{} &
  \multicolumn{1}{l}{\color{black}\checkmark} &
  \multicolumn{1}{l}{} &
  \multicolumn{1}{l}{} &
  \multicolumn{1}{l}{} &
  \multicolumn{1}{l}{} &
  \multicolumn{1}{l}{} &
  \multicolumn{1}{l}{} &
  \multicolumn{1}{l}{} &\multicolumn{1}{l}{\color{black}\checkmark}

   \\ \midrule
\multicolumn{1}{l}{\begin{tabular}{@{}c@{}}\textit{N 
} \end{tabular}} &
  \multicolumn{1}{l} {5} &
  \multicolumn{1}{l}{1} &
  \multicolumn{1}{l}{5} &
  \multicolumn{1}{l}{3} &
  \multicolumn{1}{l}{5} &
  \multicolumn{1}{l}{6} &
  \multicolumn{1}{l}{3} &
  \multicolumn{1}{l}{4} &
  \multicolumn{1}{l}{1} &
  \multicolumn{1}{l}{2} &
  \multicolumn{1}{l}{2} &
  \multicolumn{1}{l}{1} &
  \multicolumn{1}{l}{1} &\multicolumn{1}{l}{1}

   \\ \bottomrule

\end{tabular}
\end{adjustbox}

\end{table}

\color{black}
As Table \ref{tab:existing-design-dimensions} shows, existing design frameworks, especially the full-scale design spaces for data physicalisation that were introduced recently (e.g., \cite{ bae2022making, sauve2022physecology}) identify \textit{Encoding Variables} as a key design dimension. Nonetheless, they left it under-specified. In particular, the encoding variables that are available for each perceptual modality and how these variables have been practically used in existing data physicalisations are not fully discussed. Furthermore, none of the frameworks covers evaluation aspects (see Table \ref{tab:existing-design-dimensions}). That is, the criteria to evaluate the merits of a physicalisation or the methods that can be used to evaluate them remain largely unexplored. Previous research on data physicalisation \mbox{(e.g., \cite{hogan2018data,Jansen2015})} also recommends these two aspects as important aspects that need further exploration and detail. The two gaps can now be addressed through a narrative and a systematic review.
\color{black}

\section{Narrative Review: Encoding Variables for Physicalisations} 
\label{sec:narrativereview}
Which encoding variables can be used to create physicalisations (RQ1)?
The starting point for the review conducted to answer this question is \citet{Jansen2015}'s definition: ``A data physicalisation (or simply physicalisation) is a physical artefact whose geometry or material properties encode data''. This definition suggests one important axis for data physicalisation research, namely, that of data encoding. The data encoding axis is referred to in the literature as the \textit{representation} dimension (see \cite{Yi2007,roth2013interactive}). Representation happens through a \textit{representational medium}, i.e., an artefact that is used to encode and store information. Representational media make use of one or more \textit{representational material} and have different \textit{information channels} (i.e., ``perceptual aspect of some medium which can be used to carry information'' \cite{Bernsen1994}). Colour, shape, and orientation are examples of information channels. Information channels are manipulated through one or more \textit{variables}: visual variables (properties of visual information channels), haptic variables (properties of haptic information channels), olfactory variables (properties of olfactory information channels), and so on.  

%

The choice of a material inevitably restricts the space of possibilities regarding the encoding variables. For instance, the choice of sound as the material for a scenario precludes the use of visual variables to encode information for that scenario. The number of materials that can be used to encode data is potentially infinite. \citet{hogan2017towards} have provided 37 examples (e.g., glass, water, bread, electronic motors, infrared light, and many more) based on a review of 154 physicalisations. Once a material is chosen, several variables (physically, these can include five variable types related to sensory channels, and one variable type related to change) are at the designer's disposal. These are briefly reviewed below and summarised in Table \ref{tab:encodingvariables}. Most definitions for the variables start intentionally with `variations/changes of...' to stress the fact that a property by itself is not a variable, it is used as a variable, when changes in this property communicate information. The number of potential variables per sensory channel is put in brackets next to each encoding variable.


\textbf{Physical variables ($\infty$):} 
Physical variables are variations in material properties that are used to encode information. An exhaustive listing of these variables is still an area of ongoing research, but a few candidates were brought forth in previous work. Hence, the number of physical variables is initialized to infinity for now. \citet{Jansen2015} mentioned smoothness, hardness (called compliance in \cite{stusakvariables}) and sponginess as examples of physical variables. Additional examples include viscosity \cite{rasmussen2012shape}, permeability \cite{rasmussen2012shape}, \mbox{slipperiness \cite{stusakvariables}}, weight \cite{stusakvariables,oehlberg2018encoding}, reflectance \cite{oehlberg2018encoding}, density \cite{oehlberg2018encoding}, thermal diffusivity \cite{oehlberg2018encoding}, stiffness \cite{oehlberg2018encoding}, pyrotechnic color \cite{oehlberg2018encoding}, tensile strength \cite{oehlberg2018encoding}, electrical resistance \cite{oehlberg2018encoding}, and thermal \mbox{expansion \cite{oehlberg2018encoding}}. An important remark about physical variables is that, while the encoding activity (\mbox{i.e., what \cite{zhao2008embodiment}} calls data mapping) is done using material properties, the decoding can only be done using sensory information channels. Consider, for instance, viscosity. Though it is a property of the material, information encoded using it can be perceived through the haptic and the visual information channels.

\begin{table}[H]
\caption{A synthesis 
 of encoding variables for data physicalisations. Variables not highlighted were extracted from textbooks as well as the scientific and grey literature during the narrative review. Variables highlighted in \textbf{bold} are additional variables that were identified while annotating papers during the systematic review.}
\label{tab:encodingvariables}
\begin{adjustbox}{max width=\textwidth}
\begin{tabular}{lp{12cm}}
\toprule
\textit{\textbf{Variable Type}}              & \textit{\textbf{Options}}                                                                                                                                          \\ \midrule
 
\\
Physical variables  & density, electrical resistance, hardness/compliance, permeability,
pyrotechnic colour, reflectance, slipperiness, smoothness, sponginess, stiffness, tensile strength, thermal diffusivity, 
thermal expansion, viscosity, weight, \textbf{material}      \\ 

\\  
Visual   variables    & visual location, colour hue, colour value, colour saturation, visual size, visual shape, visual orientation, visual arrangement,  visual texture, crispness, resolution, visual numerousness                  \\
\\
Haptic   variables    & vibration amplitude, vibration frequency, pressure/force--strength, temperature, resistance, friction, kinesthetic location, tangible size, tangible elevation, tangible shape, tangible texture, tangible orientation, tangible location, \textbf{tangible arrangement}, \textbf{tangible numerousness} \\
\\
Sonic   variables     & sound source location, loudness, pitch, register, timbre, attack/decay, rhythmic patterns                                                                                                \\ 
\\
Olfactory   variables & scent type, scent direction, scent saturation, airflow rate, air quality                                                                \\ 
\\
Gustatory   variables & taste type, temperature of the taste carrier                                                                                                               \\
\\
Dynamic   variables   & perception time, temporal order, duration, temporal frequency, rate of change, synchronization, \textbf{change pattern}                                                          \\
 \bottomrule

\end{tabular}
\end{adjustbox}

\end{table}

\textbf{Visual variables (13):} Seven visual variables were originally proposed by \cite{Bertin1983}. These were extended to a list of 12 by \cite{maceachren2004maps}, and recently synthesized in \cite{Roth2017a,White2017}. The following definitions are largely taken from \cite{White2017}: \textit{visual size} (variations in the length, area, volume or repetitions of a symbol); \textit{visual shape} (variations in the appearance or form of a symbol); \textit{color hue} (variations in the dominant wavelength of visible light, e.g., red, blue, and green); \textit{color value} (light or dark variations of a single hue); \textit{color saturation} (the intensity of a single hue); \textit{visual orientation} (variations in the direction or angle of rotation of a symbol); \textit{(visual) pattern arrangement} (variations in the distribution of individual marks that make up a symbol); \textit{(visual) pattern texture} (variations in coarseness of the pattern within a symbol); \textit{transparency} (variations in the blend level of a symbol and a background layer); \textit{crispness} (variations in the sharpness of boundaries); \textit{resolution} (variations in the level of detail at which the map symbol is displayed); and \textit{visual location} (variations in the x, y position of a symbol relative to a frame of reference). 
Next to these `atomic' visual variables, previous work has also pondered the question of `composite' visual variables. 
\citet{maceachren2004maps} proposed to consider `pattern' as a higher-level visual variable consisting of units that have shape, size, orientation, texture, and arrangement. This is already reflected in the naming of the variables above (e.g., pattern arrangement, pattern texture). \citet{Caivano1990} proposed three dimensions of texture, namely, directionality (i.e., dimension that depends on the proportionality of units of texture), size (i.e., surface of the texturing element) and density (i.e., relation of the texturing elements to the background). This suggests that texture itself is a composite variable. Finally, \citet{Kraak2020} mentioned \textit{`numerousness'} (arrangement combined with size) as a composite variable used in dot density maps. Hence, numerousness was included as the 13th visual variable in Table \ref{tab:encodingvariables}.

\textbf{Haptic variables (13):} A few haptic variables were mentioned in \cite{Paneels2010}. These were the following: actuator position, force--strength, 
 vibration frequency, and surface texture. A similar list is found in \cite{mccormack2018multisensory}: force, position, vibration, texture, and temperature. An earlier, much more comprehensive suggestion of haptic variables was proposed by \cite{Griffin2001}. She proposed that haptic sensations can be decomposed into three categories of variables: those derived from touch (tactile), those derived from kinesthesia (kinesthetic), and those derived from visual analogues (i.e., variables that can be perceived by both vision and touch). Tactile sensations are perceived when the skin comes into contact with an object; kinesthetic sensations are stimulated by bodily movements and tensions. Based on these lists and the summary of \cite{White2017}, the following haptic variables can be mentioned. There are four \textit{tactile} variables: \textit{vibration amplitude} (also called force, see \cite{Novich2015}), \textit{vibration frequency} (also called flutter, \mbox{see \cite{Griffin2001,White2017}}, or speed), \textit{pressure} (changes in the perceived physical force exerted upon a surface or body), and \textit{temperature} (changes in the perceived temperature of a surface). The perceived intensity of vibration patterns is a function of both their amplitude and frequency (see \cite{Chouvardas2008}). There are three \textit{kinesthetic} variables: \textit{resistance} (felt when attempting to deform a surface, e.g. push a button), \textit{friction} (felt when the hand moves across or through a surface), and \textit{kinesthetic location} (changes in the location of the hand in relation to the body). Haptic variables derived from their \textit{visual analogues} include the following: \textit{tangible size} (changes in length, area, or volume), \textit{tangible elevation} (changes in z locations), \textit{tangible shape} (changes in form), \textit{tangible texture/grain} (changes in patterns), \textit{tangible orientation} (changes in alignment), and \textit{tangible location} (changes in x,y locations) (Tangible 
 location generalizes what \cite{Paneels2010} called the actuator position; tangible texture is synonymous with surface texture from \cite{Paneels2010}; the terms `pressure' \cite{Griffin2001} and `force-strength' \cite{Paneels2010} describe the same reality from different perspectives: the encoder uses the strength of the force to communicate data, and the decoder perceives a pressure). The adjective `tangible' is added to make clear that the information can be perceived by the haptic senses. For instance, a bar chart printed on a T-shirt \cite{perovich2021clothing_p15} can be perceived by the eyes (visual size), but not perceived through the hands or kinesthetic. Thus, in that example, visual size is used to communicate information while tangible size is not. 
%
%

\textbf{Olfactory variables (5):}
\citet{Patnaik2019} discussed how scent can be used to convey data using introduced olfactory marks and a few olfactory variables. Olfactory marks (i.e., glyph, bouquet, or burst) are analogous to visual marks (i.e., points, lines, or polygons) and refer to the most primitive blocks that can be used to encode scent. Attributes of these marks form the olfactory variables and include the following: the \textit{scent type} (i.e., the signature of the mark), the \textit{direction of the mark} (e.g., changes in the position in space where the scent originates), the \textit{saturation}, a.k.a. chemo-intensity (changes in the concentration of odour molecules in the air), the \textit{airflow rate}, a.k.a. kinetic intensity, the \textit{air quality} (e.g., humidity, temperature, and other non-olfactory properties of the air that can be used to encode information), and the temporal pattern, a.k.a. scent animation. Since dynamic variables are discussed separately as an orthogonal dimension to all other variables, the temporal pattern is not listed here as an olfactory variable. For a related discussion on the olfactory design space, see \cite{maggioni2020smell}. The four key dimensions identified in \cite{maggioni2020smell}---namely, chemical, emotional, spatial, and temporal---overlap to a great extent with the variables from \cite{Patnaik2019}. 

\textbf{Gustatory variables (2):} How can the gustatory channel be used to encode information? This is a slightly different question from ``which properties of food can be used to encode information?'' (For example, one may use the food's shape and colour to communicate information as discussed in \cite{mueller2021datadelight}. Shape and colour are \textit{visual} variables, not gustatory variables) This has been discussed, for example, in \cite{wang2016edibilization,mueller2021datadelight}. This is also different from the question ``how do people describe taste sensations?'', which was discussed in previous work (e.g., \cite{obrist2014temporal}). In that respect, two gustatory variables can be mentioned: the \textit{signature of the taste carrier} (i.e., changes in the taste type) and the \textit{temperature of the taste carrier} (e.g., hot or cold). Note that `taste type' here does not only refer to the basic taste types mentioned in the literature (sweet, salty, sour, bitter, and umami, see e.g., \cite{Kikut-Ligaj2015}), but more broadly to any taste that can be uniquely distinguished from another one. For instance, nominal data values can be mapped to different types of unique tastes, while ordinal values can be mapped to different types of unique temperatures (e.g., the hotter, the higher). In practice, taste variables are used in conjunction with other modalities (e.g., smell and sight) during food consumption, and there is documented evidence in the literature that inputs from other modalities (e.g., visual) to affect gustatory perception \cite{gal2007cross,Harrar2013,VanDoorn2014}.

\textbf{Sonic variables (6):} Sonification is the transformation of data relations into perceived relations in an acoustic signal for communication or interpretation purposes (see \cite{kramer2010sonification}). The following properties of sound mentioned in \cite{Krygier1994,White2017} can be used to this end: \textit{sound source location} (variations in the perception of the placement of the sound's source in a two/three-dimensional space (that perception depends on the physical location of the sound source, the environmental acoustics, and the shape of the ear, see \cite{Madhyastha1995})), \textit{loudness} (variations in the magnitude of the sound), \textit{pitch} (variations in the frequency of the sound, i.e., highness or lowness), \textit{register} (variations in the location of a pitch within a range of pitches), \textit{timbre} (variations in the general prevailing characteristic or quality of the sound), and \textit{attack/decay} (variations in the time needed by the sound to reach its maximum or minimum). Duration (variations in the length of time during which a sound or silence is heard), rate of change (variations in relation between the duration of sound and silence over time), and order (variations in the sequence of sounds over time) were also mentioned in \cite{Krygier1994,White2017} as sonic variables, but are not included here, because these are dynamic variables discussed below. Finally, rhythmic patterns, mentioned for example in \cite{mccormack2018multisensory}, can be used to encode information. Rhythms result from grouping separated sounds into periodic patterns, see e.g., \cite{palomaki06meaningsconveyed,Bernard2022}. In principle, they can be generated through a combination of other variables (e.g., pitch, timbre, duration, and order). Nonetheless, since they can be used on their own to communicate information (e.g., at least theoretically, nothing prevents the use of different patterns to communicate variations in the data), they are mentioned in the table. Rhythmic patterns are of a composite sonic variable. Melodic patterns, which are the basic units for musification (see \cite{visi2014unfolding}), are an example of rhythmic patterns.

%

\begin{figure}[H]
    \centering
    \includegraphics[width=\textwidth]{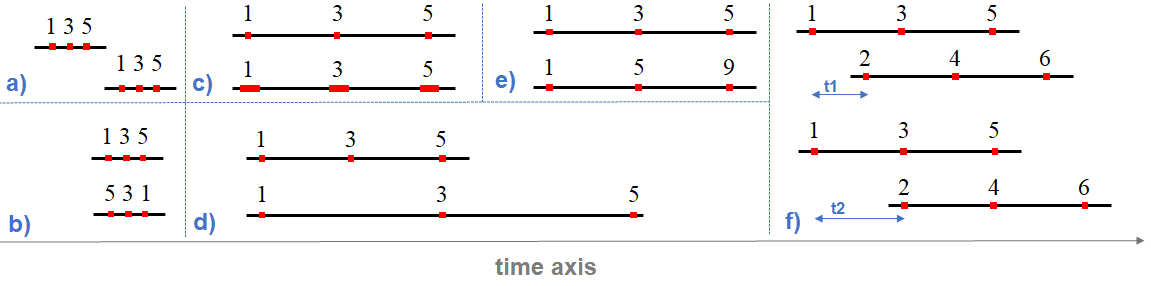}
    \caption{Dynamic 
 variables illustrated: numbers (e.g., 1, 3, and 5) stand for examples of representational states, and the space between them stands for a time interval. (\textbf{a}) Two examples of perception times; (\textbf{b}) two examples of temporal orders (chronological, reverse chronological); (\textbf{c}) two examples of duration; (\textbf{d}) two examples of temporal frequency; (\textbf{e}) two examples of rates of changes; (\textbf{f}) two examples of synchronizations (lags t1 and t2) between two time series.}
    \label{fig:dynamicvariables}
\end{figure}

\textbf{Dynamic variables (6):} Representing change over time is a recurrent need during the creation of artefacts encoding information, and dynamic variables are useful to this end. Dynamic variables are helpful when designing animations and self-reconfigurable physicalisations. Discussions on dynamic variables in the literature have focused separately on visualisations \cite{carpendale2003considering,maceachren2004maps,kobben1995evaluating,DiBiase1992,blok1998dynamic}, sound \cite{Krygier1994,White2017}, and scent \cite{Patnaik2019}. Nonetheless, given that dynamic variables are orthogonal to all other variables, an account that abstracts from the specifics of sensory modalities is needed. The review of previous descriptions led to the observation that a unifying notion across all modalities is currently missing. We propose the concept of the \textit{representational state} (or state for short) to fill this gap. A representational state refers to the particular condition of a representation (i.e., visualisation, tactile/kinesthetic sensation, scent, taste sensation, and sound) at a given point in time. Then, similarly to other variables (visual variables are properties of visual marks, sonic variables are properties of sound, and so on), dynamic variables could be conceptualised as properties of representational states. Six variables inspired from the works mentioned above and illustrated in Figure \ref{fig:dynamicvariables} are relevant: \textit{perception time} ((variations in the moments in time (a.k.a. temporal locations) the user perceives representational states); \textit{temporal order} (variations in the sequences in which the representational states are perceived); \textit{duration} (variations in the temporal life of the representational states, i.e., how long a representational state is perceived); and \textit{temporal frequency} (variations in the temporal distances between representational states, i.e., how fast/slow new representational states are communicated to users) (This variable can also be called `rate of occurrences of representational states' or `number of identifiable representational states per unit time'); \textit{rate of change} (variations in the difference in magnitude of change per unit time for a sequence of representational states); and \textit{synchronization} or phase correspondence (variations in the temporal correspondences of two or more time series). Synchronization is useful to highlight the potential relationship between two phenomena (see e.g., \cite{maceachren2004maps}). 

\begin{figure}[H]
    \centering
    \includegraphics[width=\textwidth]{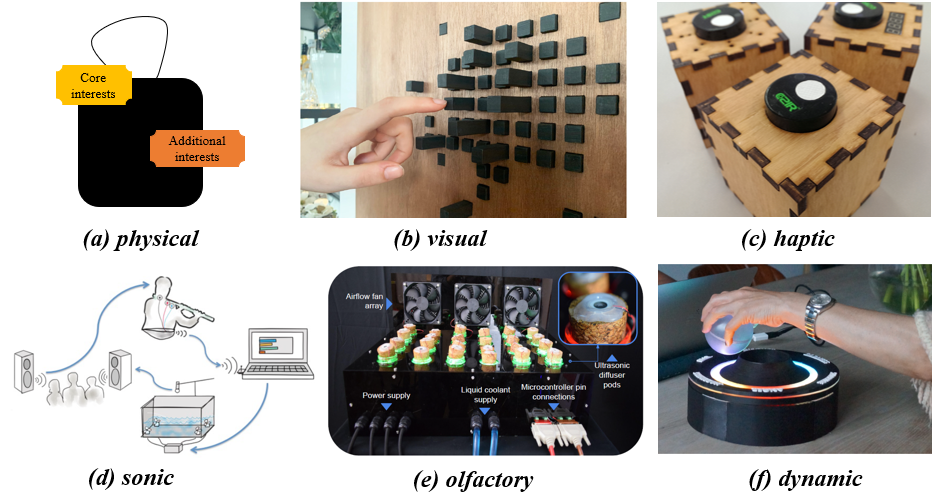}
    \caption{Examples 
 of encoding variables from papers of the systematic review. \textit{Physical}: different types of material are used to represent the users' core academic interests (Yellow stands here for `folding paper') and their additional research interests (Orange stands here for `acrylic'). For the original figure, see \cite{panagiotidou2020data_p13}. \textit{Visual}: the average effort of users during a running segment is encoded as the length of a pin on the board \cite{drogemuller2021_p59}. \textit{Haptic}: indoor air quality data is encoded as vibration in the haptic probe from \cite{hogan2017visual_p34}. \textit{Sonic}: the muscle tension of flutists is used to create live water sounds as they play their flutes \cite{pon2017_p49}. \textit{Olfactory}: the fan's speed is used to control the airflow rate \cite{batch2020scents_p29}. \textit{Dynamic}: the LED ring encircling the device fades in/out slowly or quickly to convey if the overall emotional experience of a participant is positive or negative \cite{pepping2020_p45}}. 
\label{fig:encodingvariables}
\end{figure}

Figure \ref{fig:encodingvariables} shows examples for different variable types (examples selected from or corpus).

\section{Systematic Review: Encoding Variables, Evaluation Criteria, and Methods} 
\label{sec:systematicreview}
The systematic review presented in this section is an attempt to \textcolor{black}{understand the evaluation criteria and methods used to assess data physicalisations (RQ2 and RQ3). In addition to the annotation of articles with evaluation criteria/methods, and since the encoding variables are a dimension related to representation (see Section \ref{sec:narrativereview}), we have included all dimensions from Table \ref{tab:existing-design-dimensions} that touch on an aspect of data representation (as opposed to interaction) in the annotation process. This is useful to assess how representational dimensions relate to each other on the one hand and to the evaluation dimension on the other hand. Hence, the annotation focused on the following dimensions: data type represented, representational material, representational intent, representational fidelity, and the encoding variables}. A by-product of the review is to learn about the completeness of the encoding variables derived from the current textbooks (mostly originating from work on visualisation). The remainder of this section describes the procedure used for the literature search, the screening criteria, the annotation of the papers, and the coding schemes.

\subsection{Searching and Retrieving Publications}\label{sub:searchAndretrieve}
We employed an analytical approach using a representative sample of publications on empirical work on data physicalisations and followed a systematic procedure similar to previous CHI reviews \cite{baykal2020collaborative, salminen2020literature, pettersson2018bermuda, bargas2011old, koelle2020social}. 
We used the ACM Digital Library and Scopus as the scientific repositories for our search, as many of the publications related to data physicalisation are included in these outlets. We limited our search to articles written in English and published from 2009 to 2022. The search was carried out in February 2022. We used \textit{data physicalization} and \textit{physical visualization} as keywords for the search and we searched within article title, abstract, and author keywords. The following search queries were used:    

Search query for the ACM full-text collection :

\begin{adjustwidth}{1cm}{}
		``query'': {Title:(``data physicalization''; ``physical visualization'') OR Abstract:(``data physicalization''; ``physical visualization'') OR Keyword:(``data physicalization''; ``physical visualization'')} ``filter'': {Publication Date: (01/01/2009 TO 03/31/2022)}, {ACM Content: DL}
\end{adjustwidth}

Search query for Scopus :

\begin{adjustwidth}{1cm}{}
		( TITLE-ABS-KEY (``Data Physicalization'' )  OR  TITLE-ABS-KEY (``Physical Visualization'' ) )  AND  PUBYEAR  >  2008  AND  PUBYEAR  <  2023 
\end{adjustwidth}

This initial search yielded 228 articles (\textit{n} 
 = 228): 77 from ACM and 151 from Scopus.

\subsection{Screening and Paper Selection}
\label{sub:screening}
All articles retrieved went through a screening using inclusion/exclusion criteria:

\begin{itemize}
    
    \item \textit{Criteria 1}: Articles that were not original peer-reviewed articles or that were not full papers (to ensure that the papers had a complete full scale evaluation of a data physicalisation) (e.g., late breaking works, workshops, pictorials, posters, speeches, doctoral consortium papers, etc.) were excluded. 
    \item \textit{Criteria 2}: Only the articles that discussed an artefact of data physicalisation and empirically evaluated that physicalisation were selected. Therefore, publications that introduced frameworks, theories, processes, opinions, methodologies, concepts, and reviews, as well as publications that did not empirically evaluate a physicalisation, were excluded.
    \item \textit{Criteria 3}: Articles that discussed augmented physicalisations (for example, \cite{herman2021multi}) were excluded from the analysis.
    \item \textit{Criteria 4}: Articles that discussed the same data physicalisation discussed in another article were removed, as our objective was to review different data physicalisation artefacts.
    
\end{itemize}

\begin{figure}[H]

\begin{adjustwidth}{-\extralength}{0cm}
\centering 
  \includegraphics[scale=.47]{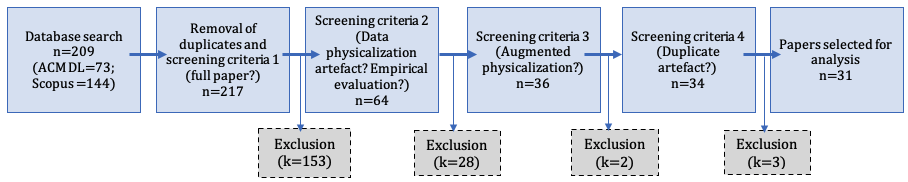}
\end{adjustwidth}
        \caption{Paper screening procedure.}
        \label{fig:screening}
\end{figure}

The removal of the duplicates and the application of inclusion/ exclusion criteria \mbox{1 resulted} in 64 articles. The subsequent screening using criteria 2, 3, and 4, which resulted in 36, 34, and \mbox{31 articles}, respectively (Figure \ref{fig:screening}). Therefore, our final set of papers selected for the analysis consisted of 31 articles (n = 31). The majority of the selected publications were from CHI (\mbox{n = 9}), followed by TEI (n = 6), IEEE TVCG (Transactions on Visualisation and Computer Graphics) (n = 4), IEEE CG\&A (Computer Graphics and Applications) (n = 3), DIS (n = 3), NordiCHI (n = 2), and AI \& Society, Elsevier C\&G (Computers and Graphics), ASSETS and SVR (n = 1 each). These 31 articles included 50 different data physicalisation artefacts (some papers contributed to more than one data physicalisation, see Table 
 \ref{tab:paperoverview}). For example, \cite{houben2016physikit_p25} contains four physicalisations that used four different physical modalities (light, vibration, movement, and air) to encode data.

\subsection{Paper Annotation} 
Every paper was coded independently by two coders, and the coding results of all papers were discussed afterwards among the two coders. When there were conflicting codes, the reasons for the individual decisions were discussed before resolving the inconsistencies. We looked not at papers, but at artefacts mentioned in these papers. One paper could, therefore, present more than one physicalisation. That is, even if an evaluation criterion was used multiple times in one paper to assess several physicalisations, we still counted that the criterion had been used multiple times. The rationale was simple: given the exploratory nature of the work, the number of artefacts over which a criterion has been used matters more at this stage than the number of different authors/research groups who used that criterion/method. 

\color{black}
\subsection{Coding Schemes}
\label{subsec:codingschemes}
We used the following coding schemes to annotate the artefacts with respect to their evaluation criteria/methods and all the dimensions of Table \ref{tab:existing-design-dimensions} that touch upon an aspect of data representation.  

\color{black}
\textit{Data scale}: This refers to the type of data that is encoded. Drawing on \citet{stevens1946theory}'s seminal taxonomy, a three-fold classification for data has become widely used in Information Visualisation and HCI research: nominal (categorical data without a natural order or rank); ordinal (ranked categories); and numerical (quantitative data).

\textit{Type of representational material}: \citet{bae2022making} draw a useful distinction between electronic and non-electronic material: an electronic material has a least one electronic component, while a non-electronic material has none. An electronic component is an entity that has the ability to control electric current (e.g., microcontrollers, sensors, computers). 

\textit{Representational fidelity}: This refers to the metaphorical distance between the physicalisation and the data it represents. \citet{VandeMoere2009} proposed three types of representational fidelity: iconic (the physicalisation bears some relationship to the data being represented through a defined metaphorical relationship); indexical (the physicalisation bears a direct relationship [either physical or causal] to the data being represented); and symbolic (the physicalisation bears no resemblance to the data being represented, and the relationship between the two must be learned using a defined convention). A detailed discussion of the concept of metaphorical distance \textcolor{black}{and examples for each type of representation fidelity are available} in \cite{zhao2008embodiment}. Though only three types of representational fidelity were proposed in \cite{VandeMoere2009}, a fourth type was identified during the annotation of the articles. The representational fidelity is dynamic if it can vary between iconic, indexical, and symbolic within the physicalisation (not necessarily automatically). For example, in PhysiAir \cite{houben2016physikit_p25}, air (coming from a fan) was used to represent ambient air (e.g., CO\textsubscript{2} level or NO\textsubscript{2} level) quality (thus, it can be considered ``iconic''). However, PhysiAir can also be configured so that air can also represent other ambient parameters such as humidity and temperature (thus, in that case the fidelity becomes ``symbolic''). Therefore, the representation fidelity in \mbox{PhysiAir \cite{houben2016physikit_p25}} can vary between iconic and symbolic depending on the configuration; thus, the fidelity is ``dynamic'' (in this case via manual reconfiguration).  

\textit{Representational intent}: This refers to the system designer's intention for encoding the data (see \cite{hogan2017towards}). Utilitarian representations were defined as those that `target a specific audience to reveal data insight related to an explicit task' \cite{hogan2017towards}; casual representations instead are `intended for a much broader audience and the exploration of data may be more open-ended and not related to a work task' \cite{hogan2017towards}. The dimension of intent was also mentioned \mbox{in \cite{dragicevic2021data,djavaherpour2021data}}. \citet{dragicevic2021data} distinguished between the motivation to discover/present and the motivation to enjoy. Nonetheless, the utilitarian/casual distinction was preferred in this work because, as \citet{dragicevic2021data} noted, it is challenging to determine in hindsight whether a physicalisation was created for the purpose of analysing data (discovery) or for the sole purpose of communicating/teaching data insights (presentation). \textcolor{black}{A similar argument applies to \citet{djavaherpour2021data}'s distinction between physicalisations with a pragmatic goal (present information in a way that allows the user to thoroughly understand the data) or artistic intent (communicate a concern, rather than show data). The two goals are not mutually exclusive, which makes it challenging to know in hindsight if the motivation of the designer was one solely or the other.}  In the work, a simple rule was used to classify a physicalisation as utilitarian or casual a posteriori. The intent is classified as utilitarian if the physicalisation is designed to support a specific task and the evaluation (e.g., efficiency, effectiveness/understanding) concerning that task has been done. Otherwise, it is classified as casual.

\textit{Evaluation criteria and methods}:
For the evaluation criteria, we used the performance-related criteria mentioned in \cite{Nielsen2012} and UX-related criteria mentioned in \cite{Saket2016} as a starting point for the annotation. As for evaluation techniques, the methods to evaluate the UX identified in \cite{pettersson2018bermuda} were used as a starting point. 

\color{black}
\textit{Encoding variables}: We used the dimensions identified and described in Section \ref{sec:narrativereview} to annotate the articles with the encoding variables. The guiding questions used to identify the presence/absence of a variable type in a physicalisation a posteriori are presented in Appendix \ref{sec:guidelines}.
\color{black}

\section{Systematic Review: Results}
\label{sec:results}
The main objective of the systematic review was to explore how the data physicalisations were evaluated (i.e. what evaluation criteria are relevant (RQ2) and what evaluation methods can be used (RQ3)). In addition, we also wanted to explore the representation dimension (i.e., what encoding variables have been used in practice, how the dimensions related to representation relate to each other and to the evaluation dimension). The following sections thus present the results of the systematic review along these lines: evaluation criteria/methods (Section \ref{subsec:criteria}),  connection between evaluation criteria and the intention of the data physicalisations from Section \ref{subsecEvalCriteriaAndIntent}, and the lessons learned about the encoding variables (\mbox{Section \ref{subsec:lessonsvariables}}) and the interrelationships observed between the dimensions related to representation and evaluation (Section \ref{subsec:interrelationships}).

\subsection{Evaluation Criteria and Methods}
\label{subsec:criteria}
The systematic analysis of the physicalisation artefacts revealed several evaluation criteria that can be used to assess the impact of data physicalisations (Table \ref{tab:evaluationcriteria}) and a wide variety of methods to collect data about these criteria (Table \ref{tab:evamethods}). This subsection summarises these evaluation criteria and the methods used to implement them. 

Table \ref{tab:evaluationcriteria} summarises the evaluation criteria used to assess the data physicalisations in our sample, ordered based on the frequency of use (most to less frequent). There is the intuition that thecriteria used in HCI/Information Visualisation can be used to evaluate data physicalisations when appropriate (see e.g., \cite{hogan2017towards}), but an open question is whether there are some criteria that could be distinctive to data physicalisation research. It can be seen that UX and performance-related criteria that are widely used in HCI/ Information Visualisation were also used to evaluate data physicalisations. We also discovered several evaluation criteria that seemed particular to data physicalisations. Thus, we grouped the evaluation criteria into those that wer enot particular to, and those that seemed particular to data physicalisation research (the dashed line in Table \ref{tab:evaluationcriteria} serves that purpose). 

\begin{table}[H]
\caption{Summary 
 of evaluation criteria (N $=$ number of physicalisations that were evaluated using the criterion). Criteria above the dashed line (except physical engagement) can be used to evaluate both visualisations and physicalisations; criteria below the dashed line seem unique to data physicalisation research.}
\label{tab:evaluationcriteria}
\newcolumntype{C}{>{\raggedright\arraybackslash}X}
\begin{tabularx}{\textwidth}{llCC}
\toprule
\textbf{Evaluation Criteria}                           & \textbf{Example Papers} & \textbf{N}  & \textbf{\%} \\ \midrule
engagement                                    &                & 16 & 32 \\
\multicolumn{1}{c}{physical   engagement}     &  \cite{perovich2021chemicals_p9,ren2021comparing_p11,hurtienne2020_p47,lee2021adio_p56,menheere2021laina_p58}                & 5  & 10  \\
\multicolumn{1}{c}{intellectual   engagement} &   \cite{ju2019bookly_p6,perovich2021chemicals_p9,kyung2020dayclo_p19,houben2016physikit_p25,sauve2020_p43,hurtienne2020_p47}               & 9  & 18 \\
\multicolumn{1}{c}{social   engagement}       &   \cite{perovich2021chemicals_p9,sauve2020_p43,hurtienne2020_p47}               & 3  & 6  \\
\multicolumn{1}{c}{affective   engagement}    &   \cite{ju2019bookly_p6,perovich2021chemicals_p9,perovich2021clothing_p15,kyung2020dayclo_p19,boem2018vitalmorph_p42,hurtienne2020_p47}               & 8  & 16 \\
\multicolumn{1}{c}{engagement   over time}    &      \cite{ju2019bookly_p6,kyung2020dayclo_p19,houben2016physikit_p25,sauve2020_p43,lee2021adio_p56,menheere2021laina_p58}          & 9  & 18 \\
user experience                             &                \cite{daniel2019cairnform_p8,veldhuis2020coda_p10,ren2021comparing_p11,ang2019physicalizing_p24,lopez-garcia2021_p28,hogan2017visual_p34,pon2017_p49,keefe2018weather_p52,menheere2021laina_p58,drogemuller2021_p59,cuya2021_p61} & 15 & 30 \\
utility                                       &      \cite{stusak2014activity_p3,ju2019bookly_p6, daniel2019cairnform_p8,ang2019physicalizing_p24,houben2016physikit_p25, taher2015exploring_p37, suzuki2017_p38,pepping2020_p45}          & 15 & 30 \\
effectiveness (question answering)          &      \cite{jansen2016psycho_p2,daniel2019cairnform_p8,ren2021comparing_p11,ang2019physicalizing_p24,batch2020scents_p29,jansen2013_p36,stusak2015_p55,drogemuller2021_p59,cuya2021_p61}          & 13 & 26 \\
efficiency (question answering)               &      \cite{jansen2016psycho_p2,ren2021comparing_p11,ang2019physicalizing_p24,batch2020scents_p29,jansen2013_p36,drogemuller2021_p59,cuya2021_p61}          & 9  & 18 \\
potential for self-reflection               &                \cite{stusak2014activity_p3, ju2019bookly_p6, kyung2020dayclo_p19,sauve2020_p43,menheere2021laina_p58} & 8  & 16 \\
understanding (qualitative)                   &               & 7  & 14 \\
\multicolumn{1}{c}{personal understanding}                  &      \cite{ang2019physicalizing_p24,lopez-garcia2021_p28,sauve2020_p43,hurtienne2020_p47}           & 6  & 12 \\
\multicolumn{1}{c}{collaborative understanding}             &  \cite{veldhuis2020coda_p10}              & 1  & 2  \\
attitude change/behavioral stimulation                      &                \cite{ju2019bookly_p6,veldhuis2020coda_p10,lopez-garcia2021_p28,hurtienne2020_p47,lee2021adio_p56,menheere2021laina_p58} & 7  & 14  \\
memorability                                  &   \cite{daniel2019cairnform_p8,ren2021comparing_p11,stusak2016_p39,hurtienne2020_p47,stusak2015_p55}             & 6  & 12 \\
enjoyment/satisfaction                        &             \cite{batch2020scents_p29,stusak2016_p39,pepping2020_p45}     & 4  & 8  \\
motivational potential                      &                \cite{stusak2014activity_p3} & 4  & 8  \\
ease of use                                 &             \cite{batch2020scents_p29,stusak2016_p39,pepping2020_p45}   & 4  & 8  \\
design parameters  &     
\cite{lopez-garcia2021_p28,daniel2019cairnform_p8}   & 3  & 6  \\
learning curve/ease of learning             &     
\cite{batch2020scents_p29,pepping2020_p45}   & 2  & 4  \\
social acceptance/ease of adoption                            &             \cite{batch2020scents_p29,boem2018vitalmorph_p42}     & 2  & 4  \\
size judgment                                    &             \cite{jansen2016psycho_p2}     & 2  & 4  \\
confidence                                    &             \cite{batch2020scents_p29}     & 1  & 2  \\
creativity                                    &              \cite{hurtienne2020_p47}  & 1  & 2  \\\hdashline
users' reactions                            &       \cite{perovich2021clothing_p15,hogan2017visual_p34,taher2015exploring_p37,boem2018vitalmorph_p42}         & 8  & 16 \\
orientation consistency                     & \cite{sauve2020change_p1}               & 6  & 12 \\
quality of the design                       &                \cite{stusak2014activity_p3,boem2018vitalmorph_p42} & 5  & 10 \\
potential for self-expression               &                \cite{stusak2014activity_p3,panagiotidou2020data_p13} & 5  & 10 \\
\multicolumn{1}{c}{representational possibilities}             &     \cite{panagiotidou2020data_p13}           & 1  & 2  \\
\multicolumn{1}{c}{representational precision}                  &     \cite{panagiotidou2020data_p13}          & 1  & 2  \\
quality of the information content          &                \cite{stusak2014activity_p3} & 4  & 8  \\
aesthetics of the physicalisation           &                \cite{stusak2014activity_p3} & 4  & 8  \\
remote awareness of physiological states    &              \cite{boem2018vitalmorph_p42,pepping2020_p45}  & 2  & 4  \\\bottomrule
\end{tabularx}

\end{table}

Table \ref{tab:evamethods} summarises the evaluation methods used to evaluate data physicalisations, which are presented in order of frequency of use (most to less frequently used). It can be seen that the methods that were widely used in HCI/ Information Visualisation were frequently used in data physicalisation research. While both lab-based and field-based experiments were used in equal frequencies in our sample, we discovered that the percentage of longitudinal/ repeated studies was significantly low compared to one-time studies.

It is useful for a data physicalisation researcher to understand the methods that have been employed to evaluate each criterion. Table \ref{tab:criteriaAndMethods} presents the evaluation methods used to assess a criterion (the evaluation criteria are presented in their order of appearance in Table \ref{tab:evaluationcriteria} (i.e., most to less frequent in the sample)). A reference next to an evaluation method stands for an example of an article implementing it. A detailed description of the meaning of the evaluation criteria is provided in Appendix \ref{sec:definition_eva_criteria_method}. A broad discussion on the takeaways and implications of the results of our review of evaluation criteria and methods is available in Section \ref{sec:discussion}.

\begin{table}[H]
      \caption{Summary 
 of evaluation methods (N $=$ number of physicalisations that were evaluated using a method).}
    \label{tab:evamethods}

\begin{adjustwidth}{-\extralength}{0cm}
\centering 
\newcolumntype{C}{>{\raggedright\arraybackslash}X}
\begin{tabularx}{\fulllength}{llCC}
\toprule
\textbf{Evaluation   Methods }           & \textbf{Example Papers} & \textbf{N}  & \textbf{\%} \\ \midrule
field-based                        &    \cite{stusak2014activity_p3,ju2019bookly_p6, daniel2019cairnform_p8,perovich2021chemicals_p9,panagiotidou2020data_p13,perovich2021clothing_p15,kyung2020dayclo_p19,houben2016physikit_p25,lopez-garcia2021_p28,hogan2017visual_p34,boem2018vitalmorph_p42,sauve2020_p43,pepping2020_p45,pon2017_p49,lee2021adio_p56,menheere2021laina_p58,drogemuller2021_p59}            & 28 & 56 \\
lab-based                          &    \cite{sauve2020change_p1,jansen2016psycho_p2,daniel2019cairnform_p8,ren2021comparing_p11,ang2019physicalizing_p24,lopez-garcia2021_p28,batch2020scents_p29,taher2015exploring_p37,suzuki2017_p38,stusak2016_p39,hurtienne2020_p47,stusak2015_p55,drogemuller2021_p59,cuya2021_p61}            & 27 & 54 \\ \hdashline
semi-structured   interviews       &   \cite{stusak2014activity_p3,ju2019bookly_p6,daniel2019cairnform_p8,perovich2021chemicals_p9,veldhuis2020coda_p10,ren2021comparing_p11,perovich2021clothing_p15,kyung2020dayclo_p19,ang2019physicalizing_p24,houben2016physikit_p25,jansen2013_p36,taher2015exploring_p37,suzuki2017_p38,stusak2016_p39,sauve2020_p43,pepping2020_p45,hurtienne2020_p47,pon2017_p49,lee2021adio_p56,menheere2021laina_p58}             & 32 & 64 \\
self-developed   questionnaires    &   \cite{jansen2016psycho_p2,stusak2014activity_p3,veldhuis2020coda_p10,panagiotidou2020data_p13,ang2019physicalizing_p24,lopez-garcia2021_p28,batch2020scents_p29,taher2015exploring_p37,stusak2016_p39,hurtienne2020_p47,pon2017_p49,stusak2015_p55,drogemuller2021_p59,cuya2021_p61}             & 21 & 42 \\
video   recording                  &   \cite{sauve2020change_p1,daniel2019cairnform_p8,veldhuis2020coda_p10,hogan2017visual_p34,jansen2013_p36, taher2015exploring_p37,suzuki2017_p38,stusak2016_p39,hurtienne2020_p47,pon2017_p49,drogemuller2021_p59}             & 19 & 38 \\
information retrieval tasks      &  \cite{sauve2020change_p1,jansen2016psycho_p2,daniel2019cairnform_p8,ren2021comparing_p11,ang2019physicalizing_p24,batch2020scents_p29,jansen2013_p36,taher2015exploring_p37,stusak2016_p39,cuya2021_p61}              & 18 & 36 \\
audio recording                  &    \cite{sauve2020change_p1,veldhuis2020coda_p10,lopez-garcia2021_p28,hogan2017visual_p34,taher2015exploring_p37,sauve2020_p43,hurtienne2020_p47,menheere2021laina_p58}            & 16 & 32 \\
live user observation            &    \cite{sauve2020change_p1,ren2021comparing_p11,perovich2021clothing_p15,stusak2016_p39,boem2018vitalmorph_p42,keefe2018weather_p52}            & 14 & 28 \\
interaction logging              &    \cite{ju2019bookly_p6,kyung2020dayclo_p19,houben2016physikit_p25,lee2021adio_p56}            & 7  & 14 \\
experience sampling/diary study  &    \cite{houben2016physikit_p25,sauve2020_p43,menheere2021laina_p58}              & 6  & 12 \\
interaction tasks                &    \cite{veldhuis2020coda_p10,taher2015exploring_p37,stusak2016_p39,drogemuller2021_p59,cuya2021_p61}            & 6  & 12 \\
standardized questionnaires      &  \cite{daniel2019cairnform_p8,lopez-garcia2021_p28,pepping2020_p45,hurtienne2020_p47}              & 5  & 10 \\
unstructured interviews                   &               \cite{boem2018vitalmorph_p42,jansen2016psycho_p2,batch2020scents_p29} & 4  & 8  \\
contextual inquiry               &    \cite{houben2016physikit_p25}            & 4  & 8  \\
focus group                      &   \cite{veldhuis2020coda_p10,hogan2017visual_p34}             & 4  & 8  \\
micro-phenomenological interview &    \cite{hogan2017visual_p34}             & 3  & 6 \\
repGrid technique                &     \cite{hogan2017visual_p34}            & 3  & 6  \\
ratio estimation                            &       \cite{jansen2016psycho_p2}         & 2  & 4  \\
constant sum                                &       \cite{jansen2016psycho_p2}         & 2  & 4  \\
sketch of participant's movements                   &               \cite{ren2021comparing_p11} & 1  & 2  \\
social interaction with a confederate of the researcher                  &    \cite{hurtienne2020_p47}            & 1  & 2  \\
think aloud              &    \cite{veldhuis2020coda_p10}            & 1  & 2  \\
post-it note feedback            &    \cite{perovich2021chemicals_p9}            & 1  & 2  \\\hdashline
one-time                           &    \cite{sauve2020change_p1,jansen2016psycho_p2,stusak2014activity_p3,perovich2021chemicals_p9, veldhuis2020coda_p10,ren2021comparing_p11, jansen2013_p36,hurtienne2020_p47,pon2017_p49,keefe2018weather_p52,drogemuller2021_p59,cuya2021_p61}                & 35 & 70 \\
longitudinal/repeated              &   \cite{ju2019bookly_p6,kyung2020dayclo_p19,houben2016physikit_p25,hogan2017visual_p34,stusak2016_p39,stusak2015_p55,lee2021adio_p56,menheere2021laina_p58}             & 15 & 30 \\ \bottomrule
\end{tabularx}%
\end{adjustwidth}
\end{table}

\label{subsubsec:criteria_and_method}


\subsection{Connecting Evaluation Criteria and Utilitarian/Casual Intents}\label{subsecEvalCriteriaAndIntent}
\begin{figure}[H]

    \includegraphics[scale=.4]{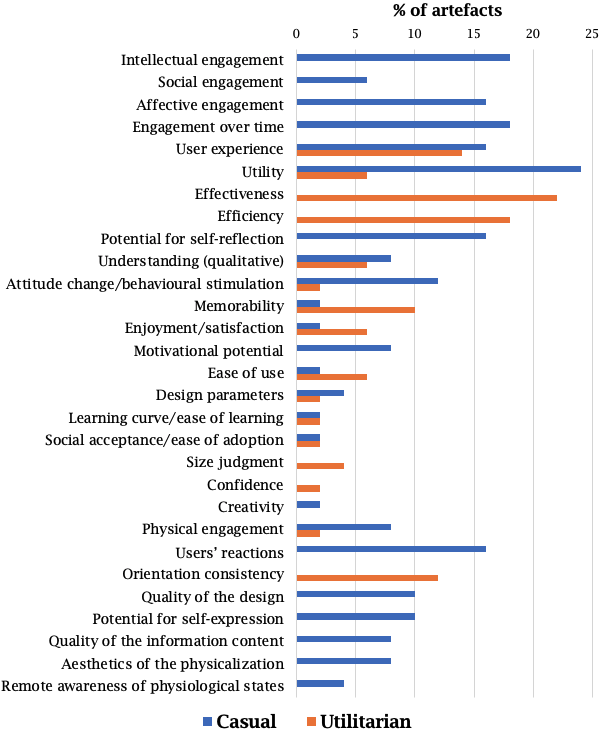}
    \caption{Evaluation criteria used to evaluate physicalisations with casual and utilitarian intents.} 
    \label{fig:evaluationCasualUtil}
\end{figure}

Previous work has identified the evaluation of physicalisations with open-ended intents as an open research challenge. For instance, \citet{hogan2017towards} commented that ``more attention is needed to evaluate representations whose purpose is more open-ended'' and \citet{Jansen2015} pointed out that finding appropriate ways of studying how people engage in data exploration when no clear task is defined is a pending evaluation-specific challenge for data physicalisation research. Thus, we also analysed how the criteria from our sample related to the utilitarian and casual representational intents (Figure \ref{fig:evaluationCasualUtil}). The findings outlined in Figure \ref{fig:evaluationCasualUtil} can inform designers about how they can evaluate their physicalisations, should these have a similar type of intent. It shows how the criteria were used so far to evaluate one type of intent (either utilitarian or casual) or both:

\begin{itemize}
    \item Criteria used for physicalisations with a casual intent: intellectual engagement, social engagement, affective engagement, the potential for self-reflection, motivational potential, creativity, user's reactions, quality of the design, potential for self-expression, quality of the information content, aesthetics, and remote awareness of physiological states.
    \item Criteria used for physicalisations with a utilitarian intent: effectiveness, efficiency, size judgement, confidence, and orientation consistency.
    \item Criteria used for both types of physicalisations: user experience, utility, understanding (qualitative), attitude change/behavioural stimulation, memorability, enjoyment/satisfaction, ease of use, design parameters, learning curve/ease of learning, social acceptance/ ease of adoption, and physical engagement. 
\end{itemize}

\begin{table}[H] \tablesize{\footnotesize}
\caption{Evaluation criteria and methods used to evaluate them.  Criteria above the dashed line (except physical engagement) can be used to evaluate both visualisations and physicalisations; criteria below the dashed line seem unique to data physicalisation research.}
\label{tab:criteriaAndMethods}

\begin{adjustwidth}{-\extralength}{0cm}
\centering 
\newcolumntype{C}{>{\raggedright\arraybackslash}X}
\begin{tabularx}{\fulllength}{CC}

\toprule \multicolumn{1}{c}{\textbf{Criteria}} & \multicolumn{1}{c}{\textbf{Evaluated Through}}  \\ \midrule

Intellectual engagement \cite{Wang2019} &
  semi-structured interviews \cite{ju2019bookly_p6,perovich2021chemicals_p9}, diary studies \cite{houben2016physikit_p25}, contextual inquiries \cite{houben2016physikit_p25}, and/or self-developed questionnaires \cite{hurtienne2020_p47} \\
Social engagement \cite{Wang2019} &
  semi-structured interviews \cite{sauve2020_p43}, self-developed questionnaires \cite{hurtienne2020_p47} and/or the use of a confederate \cite{hurtienne2020_p47}\\
Affective engagement \cite{Wang2019} &
  semi-structured interviews (e.g., \cite{ju2019bookly_p6,perovich2021chemicals_p9,panagiotidou2020data_p13}), user observations \cite{panagiotidou2020data_p13}, self-developed questionnaires \cite{hurtienne2020_p47}, and standardized questionnaires (AttrakDiff \cite{hurtienne2020_p47}, PANAS-X \cite{hurtienne2020_p47}) \\
Engagement over time &
  interaction logging \cite{ju2019bookly_p6,kyung2020dayclo_p19,lee2021adio_p56}, repeated interviews \cite{kyung2020dayclo_p19,lee2021adio_p56}, diary studies \cite{houben2016physikit_p25} and/or contextual inquiries \cite{houben2016physikit_p25} \\
User experience \cite{hassenzahl2004interplay,law2009} &
  standardized questionnaires (UEQ-S \cite{daniel2019cairnform_p8}, AttrakDiff \cite{lopez-garcia2021_p28}), self-developed questionnaires \cite{pon2017_p49}, semi-structured interviews \cite{veldhuis2020coda_p10,ren2021comparing_p11}, a RepGrid study \cite{hogan2017visual_p34}, and user observations \cite{keefe2018weather_p52} \\
Utility \cite{Nielsen2012,Roth2015} &
  semi-interviews \cite{stusak2014activity_p3,ju2019bookly_p6,houben2016physikit_p25,suzuki2017_p38}, self-developed questionnaires \cite{stusak2014activity_p3,ang2019physicalizing_p24} and a standardized questionnaire (the USE questionnaire \cite{pepping2020_p45}) \\
Effectiveness (question answering) \cite{ANSI2001} &
  the accuracy with which participants completed information retrieval tasks \cite{daniel2019cairnform_p8,stusak2016_p39} and interaction tasks \cite{taher2015exploring_p37} \\
Efficiency (question answering) \cite{Nielsen2012,ANSI2001} &
  the time taken by participants to complete information retrieval tasks \cite{ren2021comparing_p11} and/or interaction tasks \cite{taher2015exploring_p37} \\
Potential for self-reflection &
  self-developed questionnaires \cite{stusak2014activity_p3} and/or semi-structured interviews \cite{ju2019bookly_p6,kyung2020dayclo_p19} \\
Understanding (qualitative) & qualitative feedback during an interview \cite{ren2021comparing_p11} or as a rating on a self-developed questionnaire \cite{ang2019physicalizing_p24,hurtienne2020_p47} \\
Attitude change/behavioural stimulation &
   semi-structured interviews \cite{stusak2014activity_p3,ju2019bookly_p6}, think-aloud feedback \cite{veldhuis2020coda_p10}, self-developed questionnaires \cite{lopez-garcia2021_p28}, interaction logging \cite{ju2019bookly_p6,lee2021adio_p56}, user observations \cite{hurtienne2020_p47}, video recording of the interaction with the physicalisation \cite{veldhuis2020coda_p10} \\
Memorability \cite{Saket2016,stusak2015_p55} &
recall tasks \cite{daniel2019cairnform_p8,stusak2016_p39}, recall questions \cite{ren2021comparing_p11} and/or self-developed questionnaires \cite{stusak2016_p39,stusak2015_p55}  \\
Enjoyment/satisfaction \cite{Saket2016,ANSI2001} &
  self-developed questionnaires (e.g., Likert scales \cite{batch2020scents_p29,stusak2016_p39}) and a standardized questionnaire (the USE questionnaire \cite{pepping2020_p45}) \\
Motivational potential &
  self-developed questionnaires \cite{stusak2014activity_p3} \\
Ease of use &
  self-developed questionnaires (e.g., Likert scales \cite{batch2020scents_p29,stusak2016_p39}) and a standardized questionnaire (the USE questionnaire \cite{pepping2020_p45}) \\
Design parameters & systematic variation of design parameters (e.g., motion speeds \cite{daniel2019cairnform_p8} or size \cite{lopez-garcia2021_p28}) \\
Learning curve/ease of learning & self-developed questionnaires (e.g., Likert scales \cite{batch2020scents_p29,pepping2020_p45}) \\
Social acceptance/ease of adoption \cite{batch2020scents_p29} &
 self-developed questionnaires \cite{batch2020scents_p29} and unstructured interviews \cite{boem2018vitalmorph_p42}  \\
Size judgment & the accuracy of participants on information retrieval tasks \cite{jansen2016psycho_p2}  \\
Confidence \cite{batch2020scents_p29} &
 self-developed questionnaires \cite{batch2020scents_p29} \\
Creativity \cite{Wang2019} & the use of a standardized questionnaire (AttrakDiff) \cite{hurtienne2020_p47}
  \\ \hdashline
Physical engagement \cite{Wang2019} &
 the sketching of the participants' movement patterns in \cite{ren2021comparing_p11}, semi-structured interviews \cite{perovich2021chemicals_p9,lee2021adio_p56}, self-developed questionnaires \cite{hurtienne2020_p47} and a standardized questionnaire (NASA TLX \cite{hurtienne2020_p47}) \\
Users' reactions & semi-structured interviews \cite{perovich2021clothing_p15}, unstructured interviews \cite{boem2018vitalmorph_p42}, a micro-phenomenological interview \cite{hogan2017visual_p34} and user observations \cite{perovich2021clothing_p15,taher2015exploring_p37} \\
Orientation consistency \cite{sauve2020change_p1} &
   information retrieval tasks \cite{sauve2020change_p1} \\
Quality of the design & self-developed questionnaires \cite{stusak2014activity_p3}, post-it notes feedback \cite{perovich2021chemicals_p9} and unstructured interviews \cite{boem2018vitalmorph_p42} \\
Potential for self-expression & self-developed questionnaires \cite{stusak2014activity_p3,panagiotidou2020data_p13} \\
Quality of the information content &
  self-developed questionnaires \cite{stusak2014activity_p3} \\
Aesthetics of the physicalisation &
 self-developed questionnaires  \cite{stusak2014activity_p3}  \\
Remote awareness of physiological states &
 user observations and an unstructured interview in \cite{boem2018vitalmorph_p42}, and a standardized questionnaire (the emotional awareness survey) and a semi-structured interview in \cite{pepping2020_p45} \\ \bottomrule
\end{tabularx}
\end{adjustwidth}
\end{table}

\vspace{-6pt}

\subsection{Representation Dimensions}
A second objective of this review is to explore how encoding variables have been used in practice and how dimensions related to representation relate to each other. This section presents the results: lessons learned about the encoding variables (Section \ref{subsec:lessonsvariables}) and interrelationships between the dimensions (Section \ref{subsec:interrelationships}). Table \ref{tab:paperoverview} summarises the coding of the dimensions related to representation and Table \ref{tab:frequencies} shows the frequencies of the encoding variables found in the sample.

\subsubsection{Lessons Learned about the Encoding Variables}
\label{subsec:lessonsvariables} During the coding process, we discovered the following extensions to the initial list of encoding variables:

\begin{itemize}
    \item Physical variables: \textit{Material} should be added to the list in addition to the properties of the material. A nice example can be found in \cite{panagiotidou2020data_p13}, which used the tokens' material (folding paper vs acrylic) to differently encode information related to the core academic background and the additional academic interests of the users. 
    
    \item Haptic variables: The list of haptic variables that are derived from visual analogues can be extended with at least two variables: \textit{Tangible arrangement} (variations of the distribution of individual marks that make up a symbol) and \textit{tangible numerousness} (arrangement combined with size), as both can be perceived through touch. For instance, 
    the number of squares and their size were used in the `Dressed in Data' clothes to communicate data about indoor air chemicals, and this resulted in a lace pattern \cite{perovich2021clothing_p15}. That lace pattern (both the arrangement of the squares and their numbers) can be perceived by touch. This is an example of both tangible arrangement and tangible numerousness.
    
    
    \item Dynamic variables: The list of dynamic variables should be extended with \textit{change pattern} (variations in animation/movement patterns used to communicate change) as a new variable. For instance, the PhysiMove physicalisation \cite{houben2016physikit_p25} used counterclockwise movements to indicate decreases in value, clockwise movements for increases in value, and no movement for the lack of change; \citet{keefe2018weather_p52} used different animation effects to communicate the occurrence of different weather events (i.e. rain, snow, and cloud cover) and \citet{pepping2020_p45} used the slow/fast fading of LED lights to communicate whether or not an emotional experience was positive/negative.
\end{itemize}

\begin{table}[H]
\renewcommand{\arraystretch}{1.2}
\caption{An 
 overview of the 50 data physicalisations included in the sample list of 31 academic publications (dimensions related to representation).}
\label{tab:paperoverview}
\begin{adjustbox}{max width=\textwidth}
\begin{tabular}{lll
>{\columncolor[HTML]{FFF2CC}}l 
>{\columncolor[HTML]{FFF2CC}}l 
>{\columncolor[HTML]{FFF2CC}}l 
>{\columncolor[HTML]{BDD7EE}}l 
>{\columncolor[HTML]{BDD7EE}}l 
>{\columncolor[HTML]{E2EFDA}}l 
>{\columncolor[HTML]{E2EFDA}}l 
>{\columncolor[HTML]{FDB3FF}}l 
>{\columncolor[HTML]{FDB3FF}}l 
>{\columncolor[HTML]{FDB3FF}}l 
>{\columncolor[HTML]{FDB3FF}}l 
>{\columncolor[HTML]{A9D08E}}l 
>{\columncolor[HTML]{9BC2E6}}l 
>{\columncolor[HTML]{FFE699}}l 
>{\columncolor[HTML]{68EFFF}}l 
>{\columncolor[HTML]{FFE8FA}}l 
>{\columncolor[HTML]{D9D9D9}}l 
>{\columncolor[HTML]{FFF2CC}}l }
\noalign{\hrule height 1.0pt}
 & 
   &
   &
  \multicolumn{3}{c}{\cellcolor[HTML]{FFF2CC}\textbf{Data Type}} &
  \multicolumn{2}{c}{\cellcolor[HTML]{BDD7EE}\textbf{Material}} &
  \multicolumn{2}{c}{\cellcolor[HTML]{E2EFDA}\textbf{Intent}} &
  \multicolumn{4}{c}{\cellcolor[HTML]{FDB3FF}\textbf{Fidelity}} &
  \multicolumn{7}{c}{\cellcolor[HTML]{C0C0C0}\textbf{Variables}} \\
\multirow{1}{*}{\textbf{Physicalisation}} &
  \multirow{1}{*}{\textbf{Venue}} &
  \multirow{1}{*}{\textbf{Reference}} &
  \multicolumn{1}{l}{\rotatebox{90}{\cellcolor[HTML]{FFF2CC}\textbf{Categorical}}} &
  \multicolumn{1}{l}{\rotatebox{90}{\cellcolor[HTML]{FFF2CC}\textbf{Ordinal}}} &
  {\rotatebox{90}{\textbf{Numerical}}} &
 
  \multicolumn{1}{l}{\rotatebox{90}{\cellcolor[HTML]{BDD7EE}\textbf{Electronic}}} &
   {\rotatebox{90}{\textbf{Non-Electronic}}} &
  \multicolumn{1}{l} {\rotatebox{90}{\cellcolor[HTML]{E2EFDA}\textbf{Casual}}} &
   {\rotatebox{90}{\textbf{Utilitarian}}} &
  \multicolumn{1}{l} {\rotatebox{90}{\cellcolor[HTML]{FDB3FF}\textbf{Iconic}}} &
  \multicolumn{1}{l} {\rotatebox{90}{\cellcolor[HTML]{FDB3FF}\textbf{Indexical}}} &
  \multicolumn{1}{l} {\rotatebox{90}{\cellcolor[HTML]{FDB3FF}\textbf{Symbolic}}} &
   {\rotatebox{90}{\textbf{Dynamic}}} &
  \multicolumn{1}{l} {\rotatebox{90}{\cellcolor[HTML]{A9D08E}\textbf{Visual}}} &
  \multicolumn{1}{l} {\rotatebox{90}{\cellcolor[HTML]{9BC2E6}\textbf{Haptic}}} &
  \multicolumn{1}{l} {\rotatebox{90}{\cellcolor[HTML]{FFE699}\textbf{Olfactory}}} &
  \multicolumn{1}{l} {\rotatebox{90}{\cellcolor[HTML]{68EFFF}\textbf{Gustatory}}} &
  \multicolumn{1}{l} {\rotatebox{90}{\cellcolor[HTML]{FFE8FA}\textbf{Sonic}}} &
  \multicolumn{1}{l}  {\rotatebox{90}{\cellcolor[HTML]{C0C0C0}\textbf{Dynamic}}} &
  \multicolumn{1}{l} {\rotatebox{90}{\cellcolor[HTML]{E2EFDA}\textbf{Physical}}}  \\ 
   \hline
PhysiLight &
  CHI &
  \cite{houben2016physikit_p25} &
  \multicolumn{1}{l}{\cellcolor[HTML]{FFF2CC}1} &
  \multicolumn{1}{l}{\cellcolor[HTML]{FFF2CC}1} &
  1 &
  \multicolumn{1}{l}{\cellcolor[HTML]{BDD7EE}1} &
   &
  \multicolumn{1}{l}{\cellcolor[HTML]{E2EFDA}1} &
   &
  \multicolumn{1}{l}{\cellcolor[HTML]{FDB3FF}} &
  \multicolumn{1}{l}{\cellcolor[HTML]{FDB3FF}} &
  \multicolumn{1}{l}{\cellcolor[HTML]{FDB3FF}} &
  1 &
  \multicolumn{1}{l}{\cellcolor[HTML]{A9D08E}1} &
  \multicolumn{1}{l}{\cellcolor[HTML]{9BC2E6}} &
  \multicolumn{1}{l}{\cellcolor[HTML]{FFE699}} &
  \multicolumn{1}{l}{\cellcolor[HTML]{68EFFF}} &
  \multicolumn{1}{l}{\cellcolor[HTML]{FFE8FA}} & 
  {\cellcolor[HTML]{C0C0C0}}1 & {\cellcolor[HTML]{E2EFDA}}\\ \hline
PhysiBuzz &
  CHI &
  \cite{houben2016physikit_p25} &
  \multicolumn{1}{l}{\cellcolor[HTML]{FFF2CC}1} &
  \multicolumn{1}{l}{\cellcolor[HTML]{FFF2CC}1} &
  1 &
  \multicolumn{1}{l}{\cellcolor[HTML]{BDD7EE}1} &
   &
  \multicolumn{1}{l}{\cellcolor[HTML]{E2EFDA}1} &
   &
  \multicolumn{1}{l}{\cellcolor[HTML]{FDB3FF}} &
  \multicolumn{1}{l}{\cellcolor[HTML]{FDB3FF}} &
  \multicolumn{1}{l}{\cellcolor[HTML]{FDB3FF}1} &
   &
  \multicolumn{1}{l}{\cellcolor[HTML]{A9D08E}} &
  \multicolumn{1}{l}{\cellcolor[HTML]{9BC2E6}1} &
  \multicolumn{1}{l}{\cellcolor[HTML]{FFE699}} &
  \multicolumn{1}{l}{\cellcolor[HTML]{68EFFF}} &
  \multicolumn{1}{l}{\cellcolor[HTML]{FFE8FA}} &
  {\cellcolor[HTML]{C0C0C0}}& {\cellcolor[HTML]{E2EFDA}} \\ \hline
PhysiMove &
  CHI &
  \cite{houben2016physikit_p25} &
  \multicolumn{1}{l}{\cellcolor[HTML]{FFF2CC}1} &
  \multicolumn{1}{l}{\cellcolor[HTML]{FFF2CC}1} &
  1 &
  \multicolumn{1}{l}{\cellcolor[HTML]{BDD7EE}1} &
   &
  \multicolumn{1}{l}{\cellcolor[HTML]{E2EFDA}1} &
   &
  \multicolumn{1}{l}{\cellcolor[HTML]{FDB3FF}} &
  \multicolumn{1}{l}{\cellcolor[HTML]{FDB3FF}} &
  \multicolumn{1}{l}{\cellcolor[HTML]{FDB3FF}1} &
   &
  \multicolumn{1}{l}{\cellcolor[HTML]{A9D08E}1} &
  \multicolumn{1}{l}{\cellcolor[HTML]{9BC2E6}1} &
  \multicolumn{1}{l}{\cellcolor[HTML]{FFE699}} &
  \multicolumn{1}{l}{\cellcolor[HTML]{68EFFF}} &
  \multicolumn{1}{l}{\cellcolor[HTML]{FFE8FA}} &
  {\cellcolor[HTML]{C0C0C0}}1& {\cellcolor[HTML]{E2EFDA}} \\ \hline
PhysiAir &
  CHI &
 \cite{houben2016physikit_p25} &
  \multicolumn{1}{l}{\cellcolor[HTML]{FFF2CC}1} &
  \multicolumn{1}{l}{\cellcolor[HTML]{FFF2CC}1} &
  1 &
  \multicolumn{1}{l}{\cellcolor[HTML]{BDD7EE}1} &
   &
  \multicolumn{1}{l}{\cellcolor[HTML]{E2EFDA}1} &
   &
  \multicolumn{1}{l}{\cellcolor[HTML]{FDB3FF}} &
  \multicolumn{1}{l}{\cellcolor[HTML]{FDB3FF}} &
  \multicolumn{1}{l}{\cellcolor[HTML]{FDB3FF}} &
  1 &
  \multicolumn{1}{l}{\cellcolor[HTML]{A9D08E}} &
  \multicolumn{1}{l}{\cellcolor[HTML]{9BC2E6}1} &
  \multicolumn{1}{l}{\cellcolor[HTML]{FFE699}} &
  \multicolumn{1}{l}{\cellcolor[HTML]{68EFFF}} &
  \multicolumn{1}{l}{\cellcolor[HTML]{FFE8FA}} &
 {\cellcolor[HTML]{C0C0C0}}1& {\cellcolor[HTML]{E2EFDA}} \\ \hline
Spheres &
  TVCG &
  \cite{jansen2016psycho_p2} &
  \multicolumn{1}{l}{\cellcolor[HTML]{FFF2CC}} &
  \multicolumn{1}{l}{\cellcolor[HTML]{FFF2CC}} &
   &
  \multicolumn{1}{l}{\cellcolor[HTML]{BDD7EE}} &
  1 &
  \multicolumn{1}{l}{\cellcolor[HTML]{E2EFDA}} &
  1 &
  \multicolumn{1}{l}{\cellcolor[HTML]{FDB3FF}} &
  \multicolumn{1}{l}{\cellcolor[HTML]{FDB3FF}} &
  \multicolumn{1}{l}{\cellcolor[HTML]{FDB3FF}1} &
   &
  \multicolumn{1}{l}{\cellcolor[HTML]{A9D08E}1} &
  \multicolumn{1}{l}{\cellcolor[HTML]{9BC2E6}1} &
  \multicolumn{1}{l}{\cellcolor[HTML]{FFE699}} &
  \multicolumn{1}{l}{\cellcolor[HTML]{68EFFF}} &
  \multicolumn{1}{l}{\cellcolor[HTML]{FFE8FA}} &
  {\cellcolor[HTML]{C0C0C0}}& {\cellcolor[HTML]{E2EFDA}} \\ \hline
Bars &
  TVCG &
  \cite{jansen2016psycho_p2} &
  \multicolumn{1}{l}{\cellcolor[HTML]{FFF2CC}} &
  \multicolumn{1}{l}{\cellcolor[HTML]{FFF2CC}} &
   &
  \multicolumn{1}{l}{\cellcolor[HTML]{BDD7EE}} &
  1 &
  \multicolumn{1}{l}{\cellcolor[HTML]{E2EFDA}} &
  1 &
  \multicolumn{1}{l}{\cellcolor[HTML]{FDB3FF}} &
  \multicolumn{1}{l}{\cellcolor[HTML]{FDB3FF}} &
  \multicolumn{1}{l}{\cellcolor[HTML]{FDB3FF}1} &
   &
  \multicolumn{1}{l}{\cellcolor[HTML]{A9D08E}1} &
  \multicolumn{1}{l}{\cellcolor[HTML]{9BC2E6}1} &
  \multicolumn{1}{l}{\cellcolor[HTML]{FFE699}} &
  \multicolumn{1}{l}{\cellcolor[HTML]{68EFFF}} &
  \multicolumn{1}{l}{\cellcolor[HTML]{FFE8FA}} &
   {\cellcolor[HTML]{C0C0C0}}& {\cellcolor[HTML]{E2EFDA}}\\ \hline
Figure &
  TVCG &
  \cite{stusak2014activity_p3} &
  \multicolumn{1}{l}{\cellcolor[HTML]{FFF2CC}} &
  \multicolumn{1}{l}{\cellcolor[HTML]{FFF2CC}} &
  1 &
  \multicolumn{1}{l}{\cellcolor[HTML]{BDD7EE}} &
  1 &
  \multicolumn{1}{l}{\cellcolor[HTML]{E2EFDA}1} &
   &
  \multicolumn{1}{l}{\cellcolor[HTML]{FDB3FF}} &
  \multicolumn{1}{l}{\cellcolor[HTML]{FDB3FF}} &
  \multicolumn{1}{l}{\cellcolor[HTML]{FDB3FF}1} &
   &
  \multicolumn{1}{l}{\cellcolor[HTML]{A9D08E}1} &
  \multicolumn{1}{l}{\cellcolor[HTML]{9BC2E6}1} &
  \multicolumn{1}{l}{\cellcolor[HTML]{FFE699}} &
  \multicolumn{1}{l}{\cellcolor[HTML]{68EFFF}} &
  \multicolumn{1}{l}{\cellcolor[HTML]{FFE8FA}} &
  {\cellcolor[HTML]{C0C0C0}} & {\cellcolor[HTML]{E2EFDA}}\\ \hline
Necklace &
  TVCG &
  \cite{stusak2014activity_p3} &
  \multicolumn{1}{l}{\cellcolor[HTML]{FFF2CC}} &
  \multicolumn{1}{l}{\cellcolor[HTML]{FFF2CC}} &
  1 &
  \multicolumn{1}{l}{\cellcolor[HTML]{BDD7EE}} &
  1 &
  \multicolumn{1}{l}{\cellcolor[HTML]{E2EFDA}1} &
   &
  \multicolumn{1}{l}{\cellcolor[HTML]{FDB3FF}} &
  \multicolumn{1}{l}{\cellcolor[HTML]{FDB3FF}} &
  \multicolumn{1}{l}{\cellcolor[HTML]{FDB3FF}1} &
   &
  \multicolumn{1}{l}{\cellcolor[HTML]{A9D08E}1} &
  \multicolumn{1}{l}{\cellcolor[HTML]{9BC2E6}1} &
  \multicolumn{1}{l}{\cellcolor[HTML]{FFE699}} &
  \multicolumn{1}{l}{\cellcolor[HTML]{68EFFF}} &
  \multicolumn{1}{l}{\cellcolor[HTML]{FFE8FA}} &
  {\cellcolor[HTML]{C0C0C0}} & {\cellcolor[HTML]{E2EFDA}}\\ \hline
Lamp &
  TVCG &
  \cite{stusak2014activity_p3} &
  \multicolumn{1}{l}{\cellcolor[HTML]{FFF2CC}} &
  \multicolumn{1}{l}{\cellcolor[HTML]{FFF2CC}} &
  1 &
  \multicolumn{1}{l}{\cellcolor[HTML]{BDD7EE}} &
  1 &
  \multicolumn{1}{l}{\cellcolor[HTML]{E2EFDA}1} &
   &
  \multicolumn{1}{l}{\cellcolor[HTML]{FDB3FF}} &
  \multicolumn{1}{l}{\cellcolor[HTML]{FDB3FF}} &
  \multicolumn{1}{l}{\cellcolor[HTML]{FDB3FF}1} &
   &
  \multicolumn{1}{l}{\cellcolor[HTML]{A9D08E}1} &
  \multicolumn{1}{l}{\cellcolor[HTML]{9BC2E6}1} &
  \multicolumn{1}{l}{\cellcolor[HTML]{FFE699}} &
  \multicolumn{1}{l}{\cellcolor[HTML]{68EFFF}} &
  \multicolumn{1}{l}{\cellcolor[HTML]{FFE8FA}} &
  {\cellcolor[HTML]{C0C0C0}}& {\cellcolor[HTML]{E2EFDA}} \\ \hline
Jar &
  TVCG &
  \cite{stusak2014activity_p3} &
  \multicolumn{1}{l}{\cellcolor[HTML]{FFF2CC}} &
  \multicolumn{1}{l}{\cellcolor[HTML]{FFF2CC}} &
  1 &
  \multicolumn{1}{l}{\cellcolor[HTML]{BDD7EE}} &
  1 &
  \multicolumn{1}{l}{\cellcolor[HTML]{E2EFDA}1} &
   &
  \multicolumn{1}{l}{\cellcolor[HTML]{FDB3FF}} &
  \multicolumn{1}{l}{\cellcolor[HTML]{FDB3FF}} &
  \multicolumn{1}{l}{\cellcolor[HTML]{FDB3FF}1} &
   &
  \multicolumn{1}{l}{\cellcolor[HTML]{A9D08E}1} &
  \multicolumn{1}{l}{\cellcolor[HTML]{9BC2E6}1} &
  \multicolumn{1}{l}{\cellcolor[HTML]{FFE699}} &
  \multicolumn{1}{l}{\cellcolor[HTML]{68EFFF}} &
  \multicolumn{1}{l}{\cellcolor[HTML]{FFE8FA}} &
  {\cellcolor[HTML]{C0C0C0}}& {\cellcolor[HTML]{E2EFDA}} \\ \hline
Bookly &
  CHI &
  \cite{ju2019bookly_p6} &
  \multicolumn{1}{l}{\cellcolor[HTML]{FFF2CC}} &
  \multicolumn{1}{l}{\cellcolor[HTML]{FFF2CC}} &
  1 &
  \multicolumn{1}{l}{\cellcolor[HTML]{BDD7EE}1} &
   &
  \multicolumn{1}{l}{\cellcolor[HTML]{E2EFDA}1} &
   &
  \multicolumn{1}{l}{\cellcolor[HTML]{FDB3FF}} &
  \multicolumn{1}{l}{\cellcolor[HTML]{FDB3FF}} &
  \multicolumn{1}{l}{\cellcolor[HTML]{FDB3FF}1} &
   &
  \multicolumn{1}{l}{\cellcolor[HTML]{A9D08E}1} &
  \multicolumn{1}{l}{\cellcolor[HTML]{9BC2E6}1} &
  \multicolumn{1}{l}{\cellcolor[HTML]{FFE699}} &
  \multicolumn{1}{l}{\cellcolor[HTML]{68EFFF}} &
  \multicolumn{1}{l}{\cellcolor[HTML]{FFE8FA}} &
  {\cellcolor[HTML]{C0C0C0}}& {\cellcolor[HTML]{E2EFDA}} \\ \hline
CairnFORM & TEI & \cite{daniel2019cairnform_p8}&
  \multicolumn{1}{l}{\cellcolor[HTML]{FFF2CC}} &
  \multicolumn{1}{l}{\cellcolor[HTML]{FFF2CC}1} &
   &
  \multicolumn{1}{l}{\cellcolor[HTML]{BDD7EE}1} &
   &
   \multicolumn{1}{l}{\cellcolor[HTML]{E2EFDA}} &1
   &
  \multicolumn{1}{l}{\cellcolor[HTML]{FDB3FF}} &
  \multicolumn{1}{l}{\cellcolor[HTML]{FDB3FF}} &
  \multicolumn{1}{l}{\cellcolor[HTML]{FDB3FF}1} &
   &
  \multicolumn{1}{l}{\cellcolor[HTML]{A9D08E}1} &
  \multicolumn{1}{l}{\cellcolor[HTML]{9BC2E6}1} &
  \multicolumn{1}{l}{\cellcolor[HTML]{FFE699}} &
  \multicolumn{1}{l}{\cellcolor[HTML]{68EFFF}} &
  \multicolumn{1}{l}{\cellcolor[HTML]{FFE8FA}} &
 {\cellcolor[HTML]{C0C0C0}}1 & {\cellcolor[HTML]{E2EFDA}}\\ \hline
Chemicals in the Creek &
  TVCG &
  \cite{perovich2021chemicals_p9} &
  \multicolumn{1}{l}{\cellcolor[HTML]{FFF2CC}1} &
  \multicolumn{1}{l}{\cellcolor[HTML]{FFF2CC}1} &
  1 &
  \multicolumn{1}{l}{\cellcolor[HTML]{BDD7EE}1} &
   &
  \multicolumn{1}{l}{\cellcolor[HTML]{E2EFDA}1} &
   &
  \multicolumn{1}{l}{\cellcolor[HTML]{FDB3FF}} &
  \multicolumn{1}{l}{\cellcolor[HTML]{FDB3FF}} &
  \multicolumn{1}{l}{\cellcolor[HTML]{FDB3FF}1} &
   &
  \multicolumn{1}{l}{\cellcolor[HTML]{A9D08E}1} &
  \multicolumn{1}{l}{\cellcolor[HTML]{9BC2E6}} &
  \multicolumn{1}{l}{\cellcolor[HTML]{FFE699}} &
  \multicolumn{1}{l}{\cellcolor[HTML]{68EFFF}} &
  \multicolumn{1}{l}{\cellcolor[HTML]{FFE8FA}} &
  {\cellcolor[HTML]{C0C0C0}}1& {\cellcolor[HTML]{E2EFDA}} \\ \hline
CoDa &
  TEI &
  \cite{veldhuis2020coda_p10} &
  \multicolumn{1}{l}{\cellcolor[HTML]{FFF2CC}1} &
  \multicolumn{1}{l}{\cellcolor[HTML]{FFF2CC}} &
  1 &
  \multicolumn{1}{l}{\cellcolor[HTML]{BDD7EE}1} &
   &
  \multicolumn{1}{l}{\cellcolor[HTML]{E2EFDA}} &
  1 &
  \multicolumn{1}{l}{\cellcolor[HTML]{FDB3FF}} &
  \multicolumn{1}{l}{\cellcolor[HTML]{FDB3FF}} &
  \multicolumn{1}{l}{\cellcolor[HTML]{FDB3FF}1} &
   &
  \multicolumn{1}{l}{\cellcolor[HTML]{A9D08E}1} &
  \multicolumn{1}{l}{\cellcolor[HTML]{9BC2E6}1} &
  \multicolumn{1}{l}{\cellcolor[HTML]{FFE699}} &
  \multicolumn{1}{l}{\cellcolor[HTML]{68EFFF}} &
  \multicolumn{1}{l}{\cellcolor[HTML]{FFE8FA}} &
   {\cellcolor[HTML]{C0C0C0}}& {\cellcolor[HTML]{E2EFDA}}\\ \hline
Meteorite landings physicalisation &
  TEI &
  \cite{ren2021comparing_p11} &
  \multicolumn{1}{l}{\cellcolor[HTML]{FFF2CC}} &
  \multicolumn{1}{l}{\cellcolor[HTML]{FFF2CC}} &
  1 &
  \multicolumn{1}{l}{\cellcolor[HTML]{BDD7EE}} &
  1 &
  \multicolumn{1}{l}{\cellcolor[HTML]{E2EFDA}} &
  1 &
  \multicolumn{1}{l}{\cellcolor[HTML]{FDB3FF}1} &
  \multicolumn{1}{l}{\cellcolor[HTML]{FDB3FF}} &
  \multicolumn{1}{l}{\cellcolor[HTML]{FDB3FF}} &
   &
  \multicolumn{1}{l}{\cellcolor[HTML]{A9D08E}1} &
  \multicolumn{1}{l}{\cellcolor[HTML]{9BC2E6}1} &
  \multicolumn{1}{l}{\cellcolor[HTML]{FFE699}} &
  \multicolumn{1}{l}{\cellcolor[HTML]{68EFFF}} &
  \multicolumn{1}{l}{\cellcolor[HTML]{FFE8FA}} &
 {\cellcolor[HTML]{C0C0C0}} & {\cellcolor[HTML]{E2EFDA}} \\ \hline
Data Badges &
  TVCG &
  \cite{panagiotidou2020data_p13} &
  \multicolumn{1}{l}{\cellcolor[HTML]{FFF2CC}1} &
  \multicolumn{1}{l}{\cellcolor[HTML]{FFF2CC}} &
  1 &
  \multicolumn{1}{l}{\cellcolor[HTML]{BDD7EE}} &
  1 &
  \multicolumn{1}{l}{\cellcolor[HTML]{E2EFDA}1} &
   &
  \multicolumn{1}{l}{\cellcolor[HTML]{FDB3FF}} &
  \multicolumn{1}{l}{\cellcolor[HTML]{FDB3FF}} &
  \multicolumn{1}{l}{\cellcolor[HTML]{FDB3FF}1} &
   &
  \multicolumn{1}{l}{\cellcolor[HTML]{A9D08E}1} &
  \multicolumn{1}{l}{\cellcolor[HTML]{9BC2E6}1} &
  \multicolumn{1}{l}{\cellcolor[HTML]{FFE699}} &
  \multicolumn{1}{l}{\cellcolor[HTML]{68EFFF}} &
  \multicolumn{1}{l}{\cellcolor[HTML]{FFE8FA}} &
  {\cellcolor[HTML]{C0C0C0}}& {\cellcolor[HTML]{E2EFDA}}1\\ \hline
BigBarChart &
  CG\&A &
  \cite{perovich2021clothing_p15} &
  \multicolumn{1}{l}{\cellcolor[HTML]{FFF2CC}1} &
  \multicolumn{1}{l}{\cellcolor[HTML]{FFF2CC}} &
  1 &
  \multicolumn{1}{l}{\cellcolor[HTML]{BDD7EE}1} &
   &
  \multicolumn{1}{l}{\cellcolor[HTML]{E2EFDA}1} &
   &
  \multicolumn{1}{l}{\cellcolor[HTML]{FDB3FF}} &
  \multicolumn{1}{l}{\cellcolor[HTML]{FDB3FF}} &
  \multicolumn{1}{l}{\cellcolor[HTML]{FDB3FF}1} &
   &
  \multicolumn{1}{l}{\cellcolor[HTML]{A9D08E}1} &
  \multicolumn{1}{l}{\cellcolor[HTML]{9BC2E6}1} &
  \multicolumn{1}{l}{\cellcolor[HTML]{FFE699}} &
  \multicolumn{1}{l}{\cellcolor[HTML]{68EFFF}} &
  \multicolumn{1}{l}{\cellcolor[HTML]{FFE8FA}} &
  {\cellcolor[HTML]{C0C0C0}} & {\cellcolor[HTML]{E2EFDA}}\\ \hline
DressedInData &
  CG\&A &
  \cite{perovich2021clothing_p15} &
  \multicolumn{1}{l}{\cellcolor[HTML]{FFF2CC}1} &
  \multicolumn{1}{l}{\cellcolor[HTML]{FFF2CC}} &
  1 &
  \multicolumn{1}{l}{\cellcolor[HTML]{BDD7EE}} &
  1 &
  \multicolumn{1}{l}{\cellcolor[HTML]{E2EFDA}1} &
   &
  \multicolumn{1}{l}{\cellcolor[HTML]{FDB3FF}} &
  \multicolumn{1}{l}{\cellcolor[HTML]{FDB3FF}} &
  \multicolumn{1}{l}{\cellcolor[HTML]{FDB3FF}1} &
   &
  \multicolumn{1}{l}{\cellcolor[HTML]{A9D08E}1} &
  \multicolumn{1}{l}{\cellcolor[HTML]{9BC2E6}1} &
  \multicolumn{1}{l}{\cellcolor[HTML]{FFE699}} &
  \multicolumn{1}{l}{\cellcolor[HTML]{68EFFF}} &
  \multicolumn{1}{l}{\cellcolor[HTML]{FFE8FA}} &
  {\cellcolor[HTML]{C0C0C0}} & {\cellcolor[HTML]{E2EFDA}}\\ \hline
DataShirts &
  CG\&A &
  \cite{perovich2021clothing_p15} &
  \multicolumn{1}{l}{\cellcolor[HTML]{FFF2CC}1} &
  \multicolumn{1}{l}{\cellcolor[HTML]{FFF2CC}} &
  1 &
  \multicolumn{1}{l}{\cellcolor[HTML]{BDD7EE}} &
  1 &
  \multicolumn{1}{l}{\cellcolor[HTML]{E2EFDA}1} &
   &
  \multicolumn{1}{l}{\cellcolor[HTML]{FDB3FF}} &
  \multicolumn{1}{l}{\cellcolor[HTML]{FDB3FF}} &
  \multicolumn{1}{l}{\cellcolor[HTML]{FDB3FF}1} &
   &
  \multicolumn{1}{l}{\cellcolor[HTML]{A9D08E}1} &
  \multicolumn{1}{l}{\cellcolor[HTML]{9BC2E6}1} &
  \multicolumn{1}{l}{\cellcolor[HTML]{FFE699}} &
  \multicolumn{1}{l}{\cellcolor[HTML]{68EFFF}} &
  \multicolumn{1}{l}{\cellcolor[HTML]{FFE8FA}} &
  {\cellcolor[HTML]{C0C0C0}}& {\cellcolor[HTML]{E2EFDA}} \\ \hline
DayClo &
  DIS &
  \cite{kyung2020dayclo_p19} &
  \multicolumn{1}{l}{\cellcolor[HTML]{FFF2CC}1} &
  \multicolumn{1}{l}{\cellcolor[HTML]{FFF2CC}} &
  1 &
  \multicolumn{1}{l}{\cellcolor[HTML]{BDD7EE}1} &
   &
  \multicolumn{1}{l}{\cellcolor[HTML]{E2EFDA}1} &
   &
  \multicolumn{1}{l}{\cellcolor[HTML]{FDB3FF}1} &
  \multicolumn{1}{l}{\cellcolor[HTML]{FDB3FF}} &
  \multicolumn{1}{l}{\cellcolor[HTML]{FDB3FF}} &
   &
  \multicolumn{1}{l}{\cellcolor[HTML]{A9D08E}1} &
  \multicolumn{1}{l}{\cellcolor[HTML]{9BC2E6}} &
  \multicolumn{1}{l}{\cellcolor[HTML]{FFE699}} &
  \multicolumn{1}{l}{\cellcolor[HTML]{68EFFF}} &
  \multicolumn{1}{l}{\cellcolor[HTML]{FFE8FA}} &
  {\cellcolor[HTML]{C0C0C0}} & {\cellcolor[HTML]{E2EFDA}}\\ \hline
Glyph Model &
  CG &
  \cite{ang2019physicalizing_p24} &
  \multicolumn{1}{l}{\cellcolor[HTML]{FFF2CC}1} &
  \multicolumn{1}{l}{\cellcolor[HTML]{FFF2CC}} &
  1 &
  \multicolumn{1}{l}{\cellcolor[HTML]{BDD7EE}} &
  1 &
  \multicolumn{1}{l}{\cellcolor[HTML]{E2EFDA}} &
  1 &
  \multicolumn{1}{l}{\cellcolor[HTML]{FDB3FF}1} &
  \multicolumn{1}{l}{\cellcolor[HTML]{FDB3FF}} &
  \multicolumn{1}{l}{\cellcolor[HTML]{FDB3FF}} &
   &
  \multicolumn{1}{l}{\cellcolor[HTML]{A9D08E}1} &
  \multicolumn{1}{l}{\cellcolor[HTML]{9BC2E6}1} &
  \multicolumn{1}{l}{\cellcolor[HTML]{FFE699}} &
  \multicolumn{1}{l}{\cellcolor[HTML]{68EFFF}} &
  \multicolumn{1}{l}{\cellcolor[HTML]{FFE8FA}} &
  {\cellcolor[HTML]{C0C0C0}}& {\cellcolor[HTML]{E2EFDA}} \\ \hline
Streamline Model &
  CG &
  \cite{ang2019physicalizing_p24} &
  \multicolumn{1}{l}{\cellcolor[HTML]{FFF2CC}1} &
  \multicolumn{1}{l}{\cellcolor[HTML]{FFF2CC}} &
  1 &
  \multicolumn{1}{l}{\cellcolor[HTML]{BDD7EE}} &
  1 &
  \multicolumn{1}{l}{\cellcolor[HTML]{E2EFDA}} &
  1 &
  \multicolumn{1}{l}{\cellcolor[HTML]{FDB3FF}1} &
  \multicolumn{1}{l}{\cellcolor[HTML]{FDB3FF}} &
  \multicolumn{1}{l}{\cellcolor[HTML]{FDB3FF}} &
   &
  \multicolumn{1}{l}{\cellcolor[HTML]{A9D08E}1} &
  \multicolumn{1}{l}{\cellcolor[HTML]{9BC2E6}1} &
  \multicolumn{1}{l}{\cellcolor[HTML]{FFE699}} &
  \multicolumn{1}{l}{\cellcolor[HTML]{68EFFF}} &
  \multicolumn{1}{l}{\cellcolor[HTML]{FFE8FA}} &
  {\cellcolor[HTML]{C0C0C0}} & {\cellcolor[HTML]{E2EFDA}}\\ \hline
Phys1 &
  CHI &
  \cite{sauve2020change_p1} &
  \multicolumn{1}{l}{\cellcolor[HTML]{FFF2CC}} &
  \multicolumn{1}{l}{\cellcolor[HTML]{FFF2CC}} &
   &
  \multicolumn{1}{l}{\cellcolor[HTML]{BDD7EE}} &
  1 &
  \multicolumn{1}{l}{\cellcolor[HTML]{E2EFDA}} &
  1 &
  \multicolumn{1}{l}{\cellcolor[HTML]{FDB3FF}} &
  \multicolumn{1}{l}{\cellcolor[HTML]{FDB3FF}} &
  \multicolumn{1}{l}{\cellcolor[HTML]{FDB3FF}1} &
   &
  \multicolumn{1}{l}{\cellcolor[HTML]{A9D08E}1} &
  \multicolumn{1}{l}{\cellcolor[HTML]{9BC2E6}1} &
  \multicolumn{1}{l}{\cellcolor[HTML]{FFE699}} &
  \multicolumn{1}{l}{\cellcolor[HTML]{68EFFF}} &
  \multicolumn{1}{l}{\cellcolor[HTML]{FFE8FA}} &
  {\cellcolor[HTML]{C0C0C0}} & {\cellcolor[HTML]{E2EFDA}}\\ \hline
Phys2 &
  CHI &
   \cite{sauve2020change_p1} &
  \multicolumn{1}{l}{\cellcolor[HTML]{FFF2CC}} &
  \multicolumn{1}{l}{\cellcolor[HTML]{FFF2CC}} &
   &
  \multicolumn{1}{l}{\cellcolor[HTML]{BDD7EE}} &
  1 &
  \multicolumn{1}{l}{\cellcolor[HTML]{E2EFDA}} &
  1 &
  \multicolumn{1}{l}{\cellcolor[HTML]{FDB3FF}} &
  \multicolumn{1}{l}{\cellcolor[HTML]{FDB3FF}} &
  \multicolumn{1}{l}{\cellcolor[HTML]{FDB3FF}1} &
   &
  \multicolumn{1}{l}{\cellcolor[HTML]{A9D08E}1} &
  \multicolumn{1}{l}{\cellcolor[HTML]{9BC2E6}1} &
  \multicolumn{1}{l}{\cellcolor[HTML]{FFE699}} &
  \multicolumn{1}{l}{\cellcolor[HTML]{68EFFF}} &
  \multicolumn{1}{l}{\cellcolor[HTML]{FFE8FA}} &
  {\cellcolor[HTML]{C0C0C0}} & {\cellcolor[HTML]{E2EFDA}}\\ \hline
Phys3 &
  CHI &
   \cite{sauve2020change_p1} &
  \multicolumn{1}{l}{\cellcolor[HTML]{FFF2CC}} &
  \multicolumn{1}{l}{\cellcolor[HTML]{FFF2CC}} &
   &
  \multicolumn{1}{l}{\cellcolor[HTML]{BDD7EE}} &
  1 &
  \multicolumn{1}{l}{\cellcolor[HTML]{E2EFDA}} &
  1 &
  \multicolumn{1}{l}{\cellcolor[HTML]{FDB3FF}} &
  \multicolumn{1}{l}{\cellcolor[HTML]{FDB3FF}} &
  \multicolumn{1}{l}{\cellcolor[HTML]{FDB3FF}1} &
   &
  \multicolumn{1}{l}{\cellcolor[HTML]{A9D08E}1} &
  \multicolumn{1}{l}{\cellcolor[HTML]{9BC2E6}1} &
  \multicolumn{1}{l}{\cellcolor[HTML]{FFE699}} &
  \multicolumn{1}{l}{\cellcolor[HTML]{68EFFF}} &
  \multicolumn{1}{l}{\cellcolor[HTML]{FFE8FA}} &
  {\cellcolor[HTML]{C0C0C0}} & {\cellcolor[HTML]{E2EFDA}}\\ \hline
Phys4 &
  CHI &
   \cite{sauve2020change_p1} &
  \multicolumn{1}{l}{\cellcolor[HTML]{FFF2CC}} &
  \multicolumn{1}{l}{\cellcolor[HTML]{FFF2CC}} &
   &
  \multicolumn{1}{l}{\cellcolor[HTML]{BDD7EE}} &
  1 &
  \multicolumn{1}{l}{\cellcolor[HTML]{E2EFDA}} &
  1 &
  \multicolumn{1}{l}{\cellcolor[HTML]{FDB3FF}} &
  \multicolumn{1}{l}{\cellcolor[HTML]{FDB3FF}} &
  \multicolumn{1}{l}{\cellcolor[HTML]{FDB3FF}1} &
   &
  \multicolumn{1}{l}{\cellcolor[HTML]{A9D08E}1} &
  \multicolumn{1}{l}{\cellcolor[HTML]{9BC2E6}1} &
  \multicolumn{1}{l}{\cellcolor[HTML]{FFE699}} &
  \multicolumn{1}{l}{\cellcolor[HTML]{68EFFF}} &
  \multicolumn{1}{l}{\cellcolor[HTML]{FFE8FA}} &
  {\cellcolor[HTML]{C0C0C0}} & {\cellcolor[HTML]{E2EFDA}}\\ \hline
Phys5 &
  CHI &
   \cite{sauve2020change_p1} &
  \multicolumn{1}{l}{\cellcolor[HTML]{FFF2CC}} &
  \multicolumn{1}{l}{\cellcolor[HTML]{FFF2CC}} &
   &
  \multicolumn{1}{l}{\cellcolor[HTML]{BDD7EE}} &
  1 &
  \multicolumn{1}{l}{\cellcolor[HTML]{E2EFDA}} &
  1 &
  \multicolumn{1}{l}{\cellcolor[HTML]{FDB3FF}} &
  \multicolumn{1}{l}{\cellcolor[HTML]{FDB3FF}} &
  \multicolumn{1}{l}{\cellcolor[HTML]{FDB3FF}1} &
   &
  \multicolumn{1}{l}{\cellcolor[HTML]{A9D08E}1} &
  \multicolumn{1}{l}{\cellcolor[HTML]{9BC2E6}1} &
  \multicolumn{1}{l}{\cellcolor[HTML]{FFE699}} &
  \multicolumn{1}{l}{\cellcolor[HTML]{68EFFF}} &
  \multicolumn{1}{l}{\cellcolor[HTML]{FFE8FA}} &
   {\cellcolor[HTML]{C0C0C0}}& {\cellcolor[HTML]{E2EFDA}}\\ \hline
Phys6 &
  CHI &
   \cite{sauve2020change_p1} &
  \multicolumn{1}{l}{\cellcolor[HTML]{FFF2CC}} &
  \multicolumn{1}{l}{\cellcolor[HTML]{FFF2CC}} &
   &
  \multicolumn{1}{l}{\cellcolor[HTML]{BDD7EE}} &
  1 &
  \multicolumn{1}{l}{\cellcolor[HTML]{E2EFDA}} &
  1 &
  \multicolumn{1}{l}{\cellcolor[HTML]{FDB3FF}} &
  \multicolumn{1}{l}{\cellcolor[HTML]{FDB3FF}} &
  \multicolumn{1}{l}{\cellcolor[HTML]{FDB3FF}1} &
   &
  \multicolumn{1}{l}{\cellcolor[HTML]{A9D08E}1} &
  \multicolumn{1}{l}{\cellcolor[HTML]{9BC2E6}1} &
  \multicolumn{1}{l}{\cellcolor[HTML]{FFE699}} &
  \multicolumn{1}{l}{\cellcolor[HTML]{68EFFF}} &
  \multicolumn{1}{l}{\cellcolor[HTML]{FFE8FA}} &
   {\cellcolor[HTML]{C0C0C0}}& {\cellcolor[HTML]{E2EFDA}}\\ \hline
White threads &
  TEI &
  \cite{lopez-garcia2021_p28} &
  \multicolumn{1}{l}{\cellcolor[HTML]{FFF2CC}} &
  \multicolumn{1}{l}{\cellcolor[HTML]{FFF2CC}1} &
  1 &
  \multicolumn{1}{l}{\cellcolor[HTML]{BDD7EE}} &
  1 &
  \multicolumn{1}{l}{\cellcolor[HTML]{E2EFDA}1} &
   &
  \multicolumn{1}{l}{\cellcolor[HTML]{FDB3FF}} &
  \multicolumn{1}{l}{\cellcolor[HTML]{FDB3FF}} &
  \multicolumn{1}{l}{\cellcolor[HTML]{FDB3FF}1} &
   &
  \multicolumn{1}{l}{\cellcolor[HTML]{A9D08E}1} &
  \multicolumn{1}{l}{\cellcolor[HTML]{9BC2E6}1} &
  \multicolumn{1}{l}{\cellcolor[HTML]{FFE699}} &
  \multicolumn{1}{l}{\cellcolor[HTML]{68EFFF}} &
  \multicolumn{1}{l}{\cellcolor[HTML]{FFE8FA}} &
  {\cellcolor[HTML]{C0C0C0}}& {\cellcolor[HTML]{E2EFDA}} \\ \hline
Hoop &
  TEI &
  \cite{lopez-garcia2021_p28} &
  \multicolumn{1}{l}{\cellcolor[HTML]{FFF2CC}} &
  \multicolumn{1}{l}{\cellcolor[HTML]{FFF2CC}1} &
  1 &
  \multicolumn{1}{l}{\cellcolor[HTML]{BDD7EE}} &
  1 &
  \multicolumn{1}{l}{\cellcolor[HTML]{E2EFDA}1} &
   &
  \multicolumn{1}{l}{\cellcolor[HTML]{FDB3FF}} &
  \multicolumn{1}{l}{\cellcolor[HTML]{FDB3FF}} &
  \multicolumn{1}{l}{\cellcolor[HTML]{FDB3FF}1} &
   &
  \multicolumn{1}{l}{\cellcolor[HTML]{A9D08E}1} &
  \multicolumn{1}{l}{\cellcolor[HTML]{9BC2E6}1} &
  \multicolumn{1}{l}{\cellcolor[HTML]{FFE699}} &
  \multicolumn{1}{l}{\cellcolor[HTML]{68EFFF}} &
  \multicolumn{1}{l}{\cellcolor[HTML]{FFE8FA}} &
  {\cellcolor[HTML]{C0C0C0}}& {\cellcolor[HTML]{E2EFDA}} \\ \hline
ViScent 2.0 &
  CHI &
  \cite{batch2020scents_p29} &
  \multicolumn{1}{l}{\cellcolor[HTML]{FFF2CC}1} &
  \multicolumn{1}{l}{\cellcolor[HTML]{FFF2CC}1} &
  1 &
  \multicolumn{1}{l}{\cellcolor[HTML]{BDD7EE}1} &
   &
  \multicolumn{1}{l}{\cellcolor[HTML]{E2EFDA}} &
  1 &
  \multicolumn{1}{l}{\cellcolor[HTML]{FDB3FF}} &
  \multicolumn{1}{l}{\cellcolor[HTML]{FDB3FF}} &
  \multicolumn{1}{l}{\cellcolor[HTML]{FDB3FF}} &
  1 &
  \multicolumn{1}{l}{\cellcolor[HTML]{A9D08E}} &
  \multicolumn{1}{l}{\cellcolor[HTML]{9BC2E6}1} &
  \multicolumn{1}{l}{\cellcolor[HTML]{FFE699}1} &
  \multicolumn{1}{l}{\cellcolor[HTML]{68EFFF}} &
  \multicolumn{1}{l}{\cellcolor[HTML]{FFE8FA}} &
  {\cellcolor[HTML]{C0C0C0}}1 & {\cellcolor[HTML]{E2EFDA}}\\ \hline
Auditory Probe &
  DIS &
  \cite{hogan2017visual_p34} &
  \multicolumn{1}{l}{\cellcolor[HTML]{FFF2CC}} &
  \multicolumn{1}{l}{\cellcolor[HTML]{FFF2CC}} &
  1 &
  \multicolumn{1}{l}{\cellcolor[HTML]{BDD7EE}1} &
   &
  \multicolumn{1}{l}{\cellcolor[HTML]{E2EFDA}1} &
   &
  \multicolumn{1}{l}{\cellcolor[HTML]{FDB3FF}} &
  \multicolumn{1}{l}{\cellcolor[HTML]{FDB3FF}} &
  \multicolumn{1}{l}{\cellcolor[HTML]{FDB3FF}1} &
   &
  \multicolumn{1}{l}{\cellcolor[HTML]{A9D08E}} &
  \multicolumn{1}{l}{\cellcolor[HTML]{9BC2E6}} &
  \multicolumn{1}{l}{\cellcolor[HTML]{FFE699}} &
  \multicolumn{1}{l}{\cellcolor[HTML]{68EFFF}} &
  \multicolumn{1}{l}{\cellcolor[HTML]{FFE8FA}1} &
   {\cellcolor[HTML]{C0C0C0}}& {\cellcolor[HTML]{E2EFDA}}\\ \hline
Haptic Probe &
  DIS &
  \cite{hogan2017visual_p34} &
  \multicolumn{1}{l}{\cellcolor[HTML]{FFF2CC}} &
  \multicolumn{1}{l}{\cellcolor[HTML]{FFF2CC}} &
  1 &
  \multicolumn{1}{l}{\cellcolor[HTML]{BDD7EE}1} &
   &
  \multicolumn{1}{l}{\cellcolor[HTML]{E2EFDA}1} &
   &
  \multicolumn{1}{l}{\cellcolor[HTML]{FDB3FF}} &
  \multicolumn{1}{l}{\cellcolor[HTML]{FDB3FF}} &
  \multicolumn{1}{l}{\cellcolor[HTML]{FDB3FF}1} &
   &
  \multicolumn{1}{l}{\cellcolor[HTML]{A9D08E}} &
  \multicolumn{1}{l}{\cellcolor[HTML]{9BC2E6}1} &
  \multicolumn{1}{l}{\cellcolor[HTML]{FFE699}} &
  \multicolumn{1}{l}{\cellcolor[HTML]{68EFFF}} &
  \multicolumn{1}{l}{\cellcolor[HTML]{FFE8FA}} &
   {\cellcolor[HTML]{C0C0C0}}& {\cellcolor[HTML]{E2EFDA}}\\ \hline
Visual Probe &
  DIS &
  \cite{hogan2017visual_p34} &
  \multicolumn{1}{l}{\cellcolor[HTML]{FFF2CC}} &
  \multicolumn{1}{l}{\cellcolor[HTML]{FFF2CC}} &
  1 &
  \multicolumn{1}{l}{\cellcolor[HTML]{BDD7EE}1} &
   &
  \multicolumn{1}{l}{\cellcolor[HTML]{E2EFDA}1} &
   &
  \multicolumn{1}{l}{\cellcolor[HTML]{FDB3FF}} &
  \multicolumn{1}{l}{\cellcolor[HTML]{FDB3FF}} &
  \multicolumn{1}{l}{\cellcolor[HTML]{FDB3FF}1} &
   &
  \multicolumn{1}{l}{\cellcolor[HTML]{A9D08E}1} &
  \multicolumn{1}{l}{\cellcolor[HTML]{9BC2E6}} &
  \multicolumn{1}{l}{\cellcolor[HTML]{FFE699}} &
  \multicolumn{1}{l}{\cellcolor[HTML]{68EFFF}} &
  \multicolumn{1}{l}{\cellcolor[HTML]{FFE8FA}} &
  {\cellcolor[HTML]{C0C0C0}} & {\cellcolor[HTML]{E2EFDA}}\\ \hline
Physical 3D Bar chart &
  CHI &
  \cite{jansen2013_p36} &
  \multicolumn{1}{l}{\cellcolor[HTML]{FFF2CC}1} &
  \multicolumn{1}{l}{\cellcolor[HTML]{FFF2CC}} &
  1 &
  \multicolumn{1}{l}{\cellcolor[HTML]{BDD7EE}} &
  1 &
  \multicolumn{1}{l}{\cellcolor[HTML]{E2EFDA}} &
  1 &
  \multicolumn{1}{l}{\cellcolor[HTML]{FDB3FF}} &
  \multicolumn{1}{l}{\cellcolor[HTML]{FDB3FF}} &
  \multicolumn{1}{l}{\cellcolor[HTML]{FDB3FF}1} &
   &
  \multicolumn{1}{l}{\cellcolor[HTML]{A9D08E}1} &
  \multicolumn{1}{l}{\cellcolor[HTML]{9BC2E6}1} &
  \multicolumn{1}{l}{\cellcolor[HTML]{FFE699}} &
  \multicolumn{1}{l}{\cellcolor[HTML]{68EFFF}} &
  \multicolumn{1}{l}{\cellcolor[HTML]{FFE8FA}} &
   {\cellcolor[HTML]{C0C0C0}}& {\cellcolor[HTML]{E2EFDA}}\\ \hline
EMERGE &
  CHI & \cite{taher2015exploring_p37} &
  \multicolumn{1}{l}{\cellcolor[HTML]{FFF2CC}1} &
  \multicolumn{1}{l}{\cellcolor[HTML]{FFF2CC}} &
  1 &
  \multicolumn{1}{l}{\cellcolor[HTML]{BDD7EE}1} &
   &
  \multicolumn{1}{l}{\cellcolor[HTML]{E2EFDA}1} &
   &
  \multicolumn{1}{l}{\cellcolor[HTML]{FDB3FF}} &
  \multicolumn{1}{l}{\cellcolor[HTML]{FDB3FF}} &
  \multicolumn{1}{l}{\cellcolor[HTML]{FDB3FF}1} &
   &
  \multicolumn{1}{l}{\cellcolor[HTML]{A9D08E}1} &
  \multicolumn{1}{l}{\cellcolor[HTML]{9BC2E6}1} &
  \multicolumn{1}{l}{\cellcolor[HTML]{FFE699}} &
  \multicolumn{1}{l}{\cellcolor[HTML]{68EFFF}} &
  \multicolumn{1}{l}{\cellcolor[HTML]{FFE8FA}} &
  {\cellcolor[HTML]{C0C0C0}} & {\cellcolor[HTML]{E2EFDA}}\\ \hline
FluxMarker & ASSETS & \cite{suzuki2017_p38} &
  \multicolumn{1}{l}{\cellcolor[HTML]{FFF2CC}} &
  \multicolumn{1}{l}{\cellcolor[HTML]{FFF2CC}} &
  1 &
  \multicolumn{1}{l}{\cellcolor[HTML]{BDD7EE}1} &
   &
  \multicolumn{1}{l}{\cellcolor[HTML]{E2EFDA}1} &
   &
  \multicolumn{1}{l}{\cellcolor[HTML]{FDB3FF}} &
  \multicolumn{1}{l}{\cellcolor[HTML]{FDB3FF}} &
  \multicolumn{1}{l}{\cellcolor[HTML]{FDB3FF}1} &
   &
  \multicolumn{1}{l}{\cellcolor[HTML]{A9D08E}1} &
  \multicolumn{1}{l}{\cellcolor[HTML]{9BC2E6}1} &
  \multicolumn{1}{l}{\cellcolor[HTML]{FFE699}} &
  \multicolumn{1}{l}{\cellcolor[HTML]{68EFFF}} &
  \multicolumn{1}{l}{\cellcolor[HTML]{FFE8FA}} &
  {\cellcolor[HTML]{C0C0C0}} & {\cellcolor[HTML]{E2EFDA}}\\ \hline
2D Bar Chart &
  TEI &
  \cite{stusak2016_p39} &
  \multicolumn{1}{l}{\cellcolor[HTML]{FFF2CC}1} &
  \multicolumn{1}{l}{\cellcolor[HTML]{FFF2CC}} &
  1 &
  \multicolumn{1}{l}{\cellcolor[HTML]{BDD7EE}} &
  1 &
  \multicolumn{1}{l}{\cellcolor[HTML]{E2EFDA}} &
  1 &
  \multicolumn{1}{l}{\cellcolor[HTML]{FDB3FF}} &
  \multicolumn{1}{l}{\cellcolor[HTML]{FDB3FF}} &
  \multicolumn{1}{l}{\cellcolor[HTML]{FDB3FF}1} &
   &
  \multicolumn{1}{l}{\cellcolor[HTML]{A9D08E}1} &
  \multicolumn{1}{l}{\cellcolor[HTML]{9BC2E6}1} &
  \multicolumn{1}{l}{\cellcolor[HTML]{FFE699}} &
  \multicolumn{1}{l}{\cellcolor[HTML]{68EFFF}} &
  \multicolumn{1}{l}{\cellcolor[HTML]{FFE8FA}} &
  {\cellcolor[HTML]{C0C0C0}} & {\cellcolor[HTML]{E2EFDA}}\\ \hline
3D Bar Chart &
  TEI &
  \cite{stusak2016_p39} &
  \multicolumn{1}{l}{\cellcolor[HTML]{FFF2CC}1} &
  \multicolumn{1}{l}{\cellcolor[HTML]{FFF2CC}} &
  1 &
  \multicolumn{1}{l}{\cellcolor[HTML]{BDD7EE}} &
  1 &
  \multicolumn{1}{l}{\cellcolor[HTML]{E2EFDA}} &
  1 &
  \multicolumn{1}{l}{\cellcolor[HTML]{FDB3FF}} &
  \multicolumn{1}{l}{\cellcolor[HTML]{FDB3FF}} &
  \multicolumn{1}{l}{\cellcolor[HTML]{FDB3FF}1} &
   &
  \multicolumn{1}{l}{\cellcolor[HTML]{A9D08E}1} &
  \multicolumn{1}{l}{\cellcolor[HTML]{9BC2E6}1} &
  \multicolumn{1}{l}{\cellcolor[HTML]{FFE699}} &
  \multicolumn{1}{l}{\cellcolor[HTML]{68EFFF}} &
  \multicolumn{1}{l}{\cellcolor[HTML]{FFE8FA}} &
  {\cellcolor[HTML]{C0C0C0}}& {\cellcolor[HTML]{E2EFDA}}1 \\ \hline
Vital + Morph &
  AI \& Soc &
  \cite{boem2018vitalmorph_p42} &
  \multicolumn{1}{l}{\cellcolor[HTML]{FFF2CC}} &
  \multicolumn{1}{l}{\cellcolor[HTML]{FFF2CC}} &
  1 &
  \multicolumn{1}{l}{\cellcolor[HTML]{BDD7EE}1} &
   &
  \multicolumn{1}{l}{\cellcolor[HTML]{E2EFDA}1} &
   &
  \multicolumn{1}{l}{\cellcolor[HTML]{FDB3FF}} &
  \multicolumn{1}{l}{\cellcolor[HTML]{FDB3FF}} &
  \multicolumn{1}{l}{\cellcolor[HTML]{FDB3FF}1} &
   &
  \multicolumn{1}{l}{\cellcolor[HTML]{A9D08E}1} &
  \multicolumn{1}{l}{\cellcolor[HTML]{9BC2E6}1} &
  \multicolumn{1}{l}{\cellcolor[HTML]{FFE699}} &
  \multicolumn{1}{l}{\cellcolor[HTML]{68EFFF}} &
  \multicolumn{1}{l}{\cellcolor[HTML]{FFE8FA}} &
  {\cellcolor[HTML]{C0C0C0}} & {\cellcolor[HTML]{E2EFDA}}\\ \hline
Loop &
  NordiCHI &
  \cite{sauve2020_p43} &
  \multicolumn{1}{l}{\cellcolor[HTML]{FFF2CC}1} &
  \multicolumn{1}{l}{\cellcolor[HTML]{FFF2CC}} &
  1 &
  \multicolumn{1}{l}{\cellcolor[HTML]{BDD7EE}1} &
   &
  \multicolumn{1}{l}{\cellcolor[HTML]{E2EFDA}1} &
   &
  \multicolumn{1}{l}{\cellcolor[HTML]{FDB3FF}} &
  \multicolumn{1}{l}{\cellcolor[HTML]{FDB3FF}} &
  \multicolumn{1}{l}{\cellcolor[HTML]{FDB3FF}1} &
   &
  \multicolumn{1}{l}{\cellcolor[HTML]{A9D08E}1} &
  \multicolumn{1}{l}{\cellcolor[HTML]{9BC2E6}1} &
  \multicolumn{1}{l}{\cellcolor[HTML]{FFE699}} &
  \multicolumn{1}{l}{\cellcolor[HTML]{68EFFF}} &
  \multicolumn{1}{l}{\cellcolor[HTML]{FFE8FA}} &
  {\cellcolor[HTML]{C0C0C0}} & {\cellcolor[HTML]{E2EFDA}}\\ \hline
Motiis &
  NordiCHI &
  \cite{pepping2020_p45} &
  \multicolumn{1}{l}{\cellcolor[HTML]{FFF2CC}1} &
  \multicolumn{1}{l}{\cellcolor[HTML]{FFF2CC}} &
  1 &
  \multicolumn{1}{l}{\cellcolor[HTML]{BDD7EE}1} &
   &
  \multicolumn{1}{l}{\cellcolor[HTML]{E2EFDA}1} &
   &
  \multicolumn{1}{l}{\cellcolor[HTML]{FDB3FF}} &
  \multicolumn{1}{l}{\cellcolor[HTML]{FDB3FF}} &
  \multicolumn{1}{l}{\cellcolor[HTML]{FDB3FF}1} &
   &
  \multicolumn{1}{l}{\cellcolor[HTML]{A9D08E}1} &
  \multicolumn{1}{l}{\cellcolor[HTML]{9BC2E6}1} &
  \multicolumn{1}{l}{\cellcolor[HTML]{FFE699}} &
  \multicolumn{1}{l}{\cellcolor[HTML]{68EFFF}} &
  \multicolumn{1}{l}{\cellcolor[HTML]{FFE8FA}} &
  {\cellcolor[HTML]{C0C0C0}}1 & {\cellcolor[HTML]{E2EFDA}}\\ \hline
Move\&Find &
  CG\&A &
  \cite{hurtienne2020_p47} &
  \multicolumn{1}{l}{\cellcolor[HTML]{FFF2CC}} &
  \multicolumn{1}{l}{\cellcolor[HTML]{FFF2CC}} &
  1 &
  \multicolumn{1}{l}{\cellcolor[HTML]{BDD7EE}1} &
   &
  \multicolumn{1}{l}{\cellcolor[HTML]{E2EFDA}1} &
   &
  \multicolumn{1}{l}{\cellcolor[HTML]{FDB3FF}1} &
  \multicolumn{1}{l}{\cellcolor[HTML]{FDB3FF}} &
  \multicolumn{1}{l}{\cellcolor[HTML]{FDB3FF}} &
   &
  \multicolumn{1}{l}{\cellcolor[HTML]{A9D08E}} &
  \multicolumn{1}{l}{\cellcolor[HTML]{9BC2E6}1} &
  \multicolumn{1}{l}{\cellcolor[HTML]{FFE699}} &
  \multicolumn{1}{l}{\cellcolor[HTML]{68EFFF}} &
  \multicolumn{1}{l}{\cellcolor[HTML]{FFE8FA}} &
  {\cellcolor[HTML]{C0C0C0}} & {\cellcolor[HTML]{E2EFDA}}\\ \hline
Torrent &
  TEI &
  \cite{pon2017_p49} &
  \multicolumn{1}{l}{\cellcolor[HTML]{FFF2CC}} &
  \multicolumn{1}{l}{\cellcolor[HTML]{FFF2CC}} &
  1 &
  \multicolumn{1}{l}{\cellcolor[HTML]{BDD7EE}1} &
   &
  \multicolumn{1}{l}{\cellcolor[HTML]{E2EFDA}1} &
   &
  \multicolumn{1}{l}{\cellcolor[HTML]{FDB3FF}} &
  \multicolumn{1}{l}{\cellcolor[HTML]{FDB3FF}} &
  \multicolumn{1}{l}{\cellcolor[HTML]{FDB3FF}1} &
   &
  \multicolumn{1}{l}{\cellcolor[HTML]{A9D08E}1} &
  \multicolumn{1}{l}{\cellcolor[HTML]{9BC2E6}} &
  \multicolumn{1}{l}{\cellcolor[HTML]{FFE699}} &
  \multicolumn{1}{l}{\cellcolor[HTML]{68EFFF}} &
  \multicolumn{1}{l}{\cellcolor[HTML]{FFE8FA}1} &
  {\cellcolor[HTML]{C0C0C0}}1 & {\cellcolor[HTML]{E2EFDA}}\\ \hline
Weather Report &
  CG\&A &
  \cite{keefe2018weather_p52} &
  \multicolumn{1}{l}{\cellcolor[HTML]{FFF2CC}1} &
  \multicolumn{1}{l}{\cellcolor[HTML]{FFF2CC}1} &
  1 &
  \multicolumn{1}{l}{\cellcolor[HTML]{BDD7EE}1} &
   &
  \multicolumn{1}{l}{\cellcolor[HTML]{E2EFDA}1} &
   &
  \multicolumn{1}{l}{\cellcolor[HTML]{FDB3FF}1} &
  \multicolumn{1}{l}{\cellcolor[HTML]{FDB3FF}} &
  \multicolumn{1}{l}{\cellcolor[HTML]{FDB3FF}} &
   &
  \multicolumn{1}{l}{\cellcolor[HTML]{A9D08E}1} &
  \multicolumn{1}{l}{\cellcolor[HTML]{9BC2E6}1} &
  \multicolumn{1}{l}{\cellcolor[HTML]{FFE699}} &
  \multicolumn{1}{l}{\cellcolor[HTML]{68EFFF}} &
  \multicolumn{1}{l}{\cellcolor[HTML]{FFE8FA}} &
  {\cellcolor[HTML]{C0C0C0}}1& {\cellcolor[HTML]{E2EFDA}} \\ \hline
Physical bar chart &
  CHI &
  \cite{stusak2015_p55} &
  \multicolumn{1}{l}{\cellcolor[HTML]{FFF2CC}1} &
  \multicolumn{1}{l}{\cellcolor[HTML]{FFF2CC}} &
  1 &
  \multicolumn{1}{l}{\cellcolor[HTML]{BDD7EE}} &
  1 &
  \multicolumn{1}{l}{\cellcolor[HTML]{E2EFDA}} &
  1 &
  \multicolumn{1}{l}{\cellcolor[HTML]{FDB3FF}} &
  \multicolumn{1}{l}{\cellcolor[HTML]{FDB3FF}} &
  \multicolumn{1}{l}{\cellcolor[HTML]{FDB3FF}1} &
   &
  \multicolumn{1}{l}{\cellcolor[HTML]{A9D08E}1} &
  \multicolumn{1}{l}{\cellcolor[HTML]{9BC2E6}1} &
  \multicolumn{1}{l}{\cellcolor[HTML]{FFE699}} &
  \multicolumn{1}{l}{\cellcolor[HTML]{68EFFF}} &
  \multicolumn{1}{l}{\cellcolor[HTML]{FFE8FA}} &
  {\cellcolor[HTML]{C0C0C0}} & {\cellcolor[HTML]{E2EFDA}}\\ \hline
ADIO &
  CHI &
  \cite{lee2021adio_p56} &
  \multicolumn{1}{l}{\cellcolor[HTML]{FFF2CC}1} &
  \multicolumn{1}{l}{\cellcolor[HTML]{FFF2CC}} &
  1 &
  \multicolumn{1}{l}{\cellcolor[HTML]{BDD7EE}1} &
   &
  \multicolumn{1}{l}{\cellcolor[HTML]{E2EFDA}1} &
   &
  \multicolumn{1}{l}{\cellcolor[HTML]{FDB3FF}} &
  \multicolumn{1}{l}{\cellcolor[HTML]{FDB3FF}} &
  \multicolumn{1}{l}{\cellcolor[HTML]{FDB3FF}1} &
   &
  \multicolumn{1}{l}{\cellcolor[HTML]{A9D08E}1} &
  \multicolumn{1}{l}{\cellcolor[HTML]{9BC2E6}1} &
  \multicolumn{1}{l}{\cellcolor[HTML]{FFE699}} &
  \multicolumn{1}{l}{\cellcolor[HTML]{68EFFF}} &
  \multicolumn{1}{l}{\cellcolor[HTML]{FFE8FA}} &
  {\cellcolor[HTML]{C0C0C0}}& {\cellcolor[HTML]{E2EFDA}} \\ \hline
Laina &
  DIS &
  \cite{menheere2021laina_p58} &
  \multicolumn{1}{l}{\cellcolor[HTML]{FFF2CC}} &
  \multicolumn{1}{l}{\cellcolor[HTML]{FFF2CC}} &
  1 &
  \multicolumn{1}{l}{\cellcolor[HTML]{BDD7EE}1} &
   &
  \multicolumn{1}{l}{\cellcolor[HTML]{E2EFDA}1} &
   &
  \multicolumn{1}{l}{\cellcolor[HTML]{FDB3FF}1} &
  \multicolumn{1}{l}{\cellcolor[HTML]{FDB3FF}} &
  \multicolumn{1}{l}{\cellcolor[HTML]{FDB3FF}} &
   &
  \multicolumn{1}{l}{\cellcolor[HTML]{A9D08E}1} &
  \multicolumn{1}{l}{\cellcolor[HTML]{9BC2E6}1} &
  \multicolumn{1}{l}{\cellcolor[HTML]{FFE699}} &
  \multicolumn{1}{l}{\cellcolor[HTML]{68EFFF}} &
  \multicolumn{1}{l}{\cellcolor[HTML]{FFE8FA}} &
  {\cellcolor[HTML]{C0C0C0}} & {\cellcolor[HTML]{E2EFDA}}\\ \hline
Physical graph &
  CHI &
  \cite{drogemuller2021_p59} &
  \multicolumn{1}{l}{\cellcolor[HTML]{FFF2CC}} &
  \multicolumn{1}{l}{\cellcolor[HTML]{FFF2CC}} &
   &
  \multicolumn{1}{l}{\cellcolor[HTML]{BDD7EE}} &
  1 &
  \multicolumn{1}{l}{\cellcolor[HTML]{E2EFDA}} &
  1 &
  \multicolumn{1}{l}{\cellcolor[HTML]{FDB3FF}} &
  \multicolumn{1}{l}{\cellcolor[HTML]{FDB3FF}1} &
  \multicolumn{1}{l}{\cellcolor[HTML]{FDB3FF}} &
   &
  \multicolumn{1}{l}{\cellcolor[HTML]{A9D08E}1} &
  \multicolumn{1}{l}{\cellcolor[HTML]{9BC2E6}1} &
  \multicolumn{1}{l}{\cellcolor[HTML]{FFE699}} &
  \multicolumn{1}{l}{\cellcolor[HTML]{68EFFF}} &
  \multicolumn{1}{l}{\cellcolor[HTML]{FFE8FA}} &
  {\cellcolor[HTML]{C0C0C0}} & {\cellcolor[HTML]{E2EFDA}}\\ \hline
Visuo-haptic interface &
  SVR &
  \cite{cuya2021_p61} &
  \multicolumn{1}{l}{\cellcolor[HTML]{FFF2CC}} &
  \multicolumn{1}{l}{\cellcolor[HTML]{FFF2CC}} &
  1 &
  \multicolumn{1}{l}{\cellcolor[HTML]{BDD7EE}1} &
   &
  \multicolumn{1}{l}{\cellcolor[HTML]{E2EFDA}} &
  1 &
  \multicolumn{1}{l}{\cellcolor[HTML]{FDB3FF}} &
  \multicolumn{1}{l}{\cellcolor[HTML]{FDB3FF}} &
  \multicolumn{1}{l}{\cellcolor[HTML]{FDB3FF}1} &
   &
  \multicolumn{1}{l}{\cellcolor[HTML]{A9D08E}} &
  \multicolumn{1}{l}{\cellcolor[HTML]{9BC2E6}1} &
  \multicolumn{1}{l}{\cellcolor[HTML]{FFE699}} &
  \multicolumn{1}{l}{\cellcolor[HTML]{68EFFF}} &
  \multicolumn{1}{l}{\cellcolor[HTML]{FFE8FA}} &
  {\cellcolor[HTML]{C0C0C0}}& {\cellcolor[HTML]{E2EFDA}} \\ 
\noalign{\hrule height 1.0pt}
\end{tabular}
\end{adjustbox}

\end{table}

\begin{table}[H]
\renewcommand{\arraystretch}{1.2}
\caption{Number 
 of times {(N)} the variables were found in the sample {used}.}
\label{tab:frequencies}
\setlength{\tabcolsep}{2.4mm}
\begin{tabular}{ll
>{\columncolor[HTML]{9BC2E6}}l 
>{\columncolor[HTML]{9BC2E6}}l ll}
\noalign{\hrule height 1.0pt}
\cellcolor[HTML]{A9D08E}\textit{\textbf{visual}} & \cellcolor[HTML]{A9D08E}\textit{N}   & \textit{\textbf{haptic}} & \textit{N}   & \cellcolor[HTML]{FFE8FA}\textit{\textbf{sonic}}     & \cellcolor[HTML]{FFE8FA}\textit{N}   \\ \hline
\cellcolor[HTML]{A9D08E}visual location          & \cellcolor[HTML]{A9D08E}29 & tangible location        & 29 & \cellcolor[HTML]{FFE8FA}pitch                       & \cellcolor[HTML]{FFE8FA}1  \\ \hline
\cellcolor[HTML]{A9D08E}visual size              & \cellcolor[HTML]{A9D08E}28 & tangible size            & 26 & \cellcolor[HTML]{FFE8FA}timbre                      & \cellcolor[HTML]{FFE8FA}1  \\ \hline
\cellcolor[HTML]{A9D08E}colour hue               & \cellcolor[HTML]{A9D08E}17 & tangible arrangement     & 7 & \cellcolor[HTML]{FFE8FA}rythmic patterns            & \cellcolor[HTML]{FFE8FA}1  \\ \hline
\cellcolor[HTML]{A9D08E}visual arrangement       & \cellcolor[HTML]{A9D08E}9 & tangible numerousness    & 7 & \cellcolor[HTML]{FFE699}\textit{\textbf{olfactory}} & \cellcolor[HTML]{FFE699}\textit{N}   \\ \hline
\cellcolor[HTML]{A9D08E}visual numerousness             & \cellcolor[HTML]{A9D08E}8 & tangible shape           & 4  & \cellcolor[HTML]{FFE699}air quality                 & \cellcolor[HTML]{FFE699}1  \\ \hline
\cellcolor[HTML]{A9D08E}visual shape             & \cellcolor[HTML]{A9D08E}7 & tangible orientation     & 4  & \cellcolor[HTML]{FFE699}scent saturation            & \cellcolor[HTML]{FFE699}1  \\ \hline
\cellcolor[HTML]{A9D08E}colour value             & \cellcolor[HTML]{A9D08E}4  & vibration amplitude      & 3  & \cellcolor[HTML]{FFE699}airflow rate                & \cellcolor[HTML]{FFE699}1  \\ \hline
\cellcolor[HTML]{A9D08E}visual orientation       & \cellcolor[HTML]{A9D08E}4  & vibration frequency      & 2  & \cellcolor[HTML]{FFE699}scent type                  & \cellcolor[HTML]{FFE699}1  \\ \hline
\cellcolor[HTML]{A9D08E}visual texture           & \cellcolor[HTML]{A9D08E}1  & force-strength           & 3  & \cellcolor[HTML]{D9D9D9}\textit{\textbf{dynamic}}   & \cellcolor[HTML]{D9D9D9}\textit{N} \\ \hline
\cellcolor[HTML]{E2EFDA}\textit{\textbf{physical}}                       & {\cellcolor[HTML]{E2EFDA}}\textit{N}                         & tangible texture         & 2  & \cellcolor[HTML]{D9D9D9}perception time             & \cellcolor[HTML]{D9D9D9}9 \\ \hline
\cellcolor[HTML]{E2EFDA}material type                                    & \cellcolor[HTML]{E2EFDA}1                          & temperature              & 1  & \cellcolor[HTML]{D9D9D9}change pattern              & \cellcolor[HTML]{D9D9D9}6 \\ \hline
\cellcolor[HTML]{E2EFDA}weight                                           & {\cellcolor[HTML]{E2EFDA}}1                          & resistance               & 1  & \cellcolor[HTML]{D9D9D9}temporal frequency          & \cellcolor[HTML]{D9D9D9}2  \\ \noalign{\hrule height 1.0pt}

\end{tabular}

\end{table}


In addition, though previous work (e.g., \cite{jansen2016psycho_p2}) mentioned the ambiguity surrounding the use of `size' as an encoding variable (as size can have different aspects), a systematic account of possible usage is still lacking. Our annotations led to the following dimensions of “size” for data physicalisation research: length \cite{panagiotidou2020data_p13,sauve2020change_p1}, height \cite{perovich2021clothing_p15,jansen2016psycho_p2,stusak2014activity_p3,ren2021comparing_p11,sauve2020change_p1, jansen2013_p36,taher2015exploring_p37,stusak2015_p55}, diameter \cite{jansen2016psycho_p2,stusak2014activity_p3,daniel2019cairnform_p8,ren2021comparing_p11,sauve2020_p43}, area \cite{perovich2021clothing_p15}, surface area \cite{panagiotidou2020data_p13}, and volume \cite{ju2019bookly_p6,stusak2016_p39,ang2019physicalizing_p24}. Size was also used, not as an encoding variable, but to denote the overall size of the physicalisation. This use of size in the sense of a design parameter that influences the user experience was investigated in \cite{lopez-garcia2021_p28} (the authors used `scale' as a synonym for size in their work). The multiplicity of interpretations for `size' suggests a necessary precisification by the authors investigating it: either as a variable or as a design parameter. 

\subsubsection{Interrelationships between Dimensions}
\label{subsec:interrelationships}
Several pipelines describing the process of creating physicalisations were proposed in previous work, which include, for example, the extended version of the infovis pipeline to accommodate the physical rendering of data \cite{jansen2013interaction}, the data sensification workflow that focuses on encoding data in the experience people have with representations \cite{hogan2018data}, and the pipeline for the digital fabrication of physicalisations from \cite{djavaherpour2021data}. While these pipelines are valuable, none has explicity linked the dimensions we have examined in our systematic review. Hence, we looked into possible connections between the dimensions examined as a first step towards a theory of \textbf{\textit{representation}} in data physicalisation research. Such a theory would inform researchers and designers about the consequences of their choices during the process of building and evaluating physicalisations (e.g., how the choices made at early stages impact the options available at later stages). There are four important elements of theory development according to \cite{Whetten1989}: (1) extract key concepts, (2) identify the (causal) relationships between these concepts, (3) elaborate on the rationales for these relationships, and (4) clarify the range of application of the theory. We address the four elements in turn. 

\textit{Key concepts}: These are the dimensions considered during the annotation of the articles: all dimensions from existing design spaces touching on data representation, plus the evaluation dimension (see Section \ref{sec:systematicreview}). {The interaction concept is key to the design of data physicalisations (see Table \ref{tab:existing-design-dimensions}) and is hence included in the model, even if it was not explicitly examined during the work.}

\begin{figure}[H]

    \includegraphics[width=\textwidth]{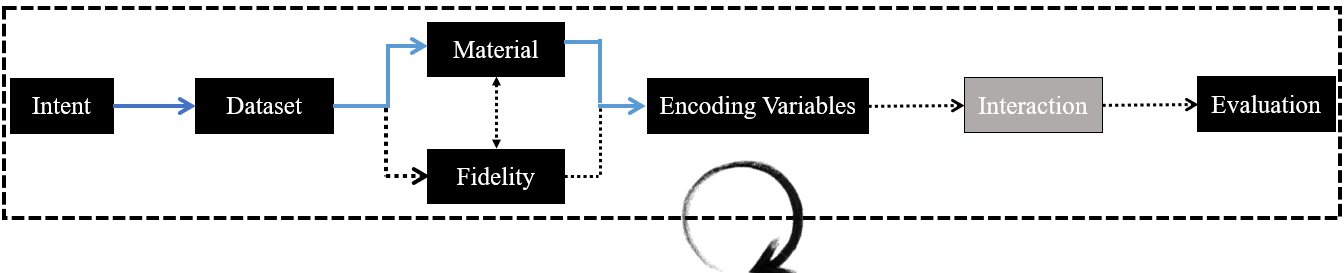}
    \caption{A model connecting the dimensions investigated during the systematic review. Blue arrows indicate a statistically significant association between two dimensions. The interaction dimension is coloured grey because we did not study this dimension in our systematic review. The process is iterative, but arrows describing iterations are omitted in the figure to ease readability.} 
    \label{fig:researchmodel}
\end{figure}

\textit{Relationships}: We proposed to link the dimensions considered sequentially into a seven-stage model for designing and evaluating data physicalisations, as shown in Figure~\ref{fig:researchmodel}.
\color{black}

The steps in this seven-stage model are the following:
\begin{enumerate}
\item Form the {intention (goal/purpose)} of the representation (casual, utilitarian, see \mbox{Section \ref{subsec:codingschemes});}
\item Select a dataset (categorical, ordinal, numerical ,or a mix of these);
\item Choose the representational fidelity (iconic, indexical, symbolic, or dynamic, see Section \ref{subsec:codingschemes});
\item Choose a material (examples in Section \ref{sec:narrativereview});
\item Pick the encoding variables (examples in Section \ref{sec:narrativereview}). {The choice of the encoding variables entails the choice of sensory modalities.}
\item Design the interaction (not discussed in this article, but useful references can be found in \cite{rasmussen2012shape,sauve2022physecology});
\item Evaluate the artefact (examples in Section \ref{subsec:criteria}).
\end{enumerate}

In a nutshell, the researcher interested in studying data physicalisations starts with a purpose and selects a dataset in line with that purpose. Afterwards, they choose a representational material, which is a choice that is strongly tied to the choice of the representational metaphor. Since both representational material and fidelity strongly determine each other, they are given an equal footing on the diagram. The choice of the encoding variables follows that of the material. In practice, the design of the interaction happens concurrently with the design of other aspects of the representation, but since the choice of the encoding variables (e.g., visual vs sonic) constrains the interaction possibilities, they are shown sequentially. The systematic evaluation of the artefact happens last (and is the step that distinguishes the researcher from the designer in this model). The meaning of the arrow \textbf{$\longrightarrow$} is `precedes and constrains'. The whole process is \textbf{iterative}, which means that designers can come back to any stage from any stage, but, for simplicity, arrows representing iterations are not shown in Figure \ref{fig:researchmodel}.

\textit{Range of application}: The relationships proposed above and tested below are based on the operationalizations of the concepts related to data physicalisation described in \mbox{Section \ref{subsec:codingschemes}.} They may not be valid for other operationalisations (e.g., other taxonomies for data type \cite{djavaherpour2021data} or representational intent  \cite{dragicevic2021data,djavaherpour2021data}).

\textit{Quantitative analysis}:
We computed Fisher's exact test \cite{weisstein2022fisher} (and, when appropriate, used Pearson's chi-squared test instead) and the Cramér's V for all pairs of dimensions (Table \ref{tab:interrelationships}). A significant value for the Fisher's exact test or chi-squared test indicates a nonrandom association between two categorical variables, while Cramér's V indicates the strength of the association (0 $=$ no association; 1 $=$ complete association). We now report on the findings for all pairs of consecutive dimensions of the model:

\begin{itemize}
    \item Intent---dataset (\emph{p}-value $<$ 0.001; Cramér's V $=$ 0.67): There were differences in proportions for nearly all types of datasets. Most notably, physicalisations with a casual intent used the combination of categorical and ordinal and numerical datasets more often than those with a utilitarian intent; they also used numerical data much more often than those with a utilitarian intent. The nonrandom association observed here could be due to some bias in the sample: all physicalisations where the type of dataset was `not documented' were those having a utilitarian intent (these physicalisations were used to investigate the theoretical properties of physicalisation in \cite{sauve2020change_p1,jansen2016psycho_p2,drogemuller2021_p59}: orientation consistency, size judgment, and graph physicalisation).  
    \item Dataset---material type (\emph{p}-value $<$ 0.001; Cramér's V $=$ 0.65): Physicalisations encoding three types of datasets (categorical and ordinal and numerical) all used electronic material. The nonrandom association observed here could also be due to some bias in the sample: all physicalisations where the type of dataset was `not documented' were those using non-electronic material (investigation of theoretical properties).
    \item Dataset---representational fidelity: The Fisher's exact test between the data type and the representational fidelity was not significant. Nonetheless, the association between the number of datasets and the fidelity was significant (\emph{p}-value $=$ 0.01; Cramér's V $=$ 0.46). In particular, there was no physicalisation with two/three datasets that had an indexical fidelity (i.e., the physicalisation
bore a direct relationship [physical or causal] to the data being represented) in our sample. 
    \item  Material type---encoding variables (\emph{p}-value $<$ 0.001; Cramér's V $=$ 0.69): Physicalisations combining variables beyond the visual and haptic dimensions (e.g., visual and sonic and haptic and olfactory) all used electronic material.
    \item \textls[-15]{Representational fidelity---encoding variables: The Fisher's exact test was not significant.}
    \item Encoding variables---evaluation criteria: We grouped the evaluation criteria into three categories: traditional, novel and traditional and novel. `Traditional' refers to the criteria above the dashed line (except physical engagement), whereas `novel' refers to physical engagement and criteria below the dashed line. The Fisher's exact test was not significant. 
\end{itemize}

\begin{table}[H]
\renewcommand{\arraystretch}{1.2}

       \caption{Relationships between the different dimensions. A number within a cell is the Cramér’s V (strength of the association) for a statistically significant association (i.e., a possible non-random association between two dimensions). A `-' indicates statistically non-significant associations. To improve readability, some names were abbreviated in the table: evaluation (evaluation criteria), n\_modalities (number of modalities), n\_datasets (number of datasets), material (material type).}
    \label{tab:interrelationships}
\begin{adjustbox}{max width=\textwidth}
\begin{tabular}{llllllllll}
\toprule
                       & \textbf{Intent}                   & \textbf{Fidelity}                 & \textbf{Evaluation}     & \textbf{Variables}                & \textbf{n\_Modalities}   & \textbf{Dynamicity}                                      & \textbf{Data\_Type}               & \textbf{n\_Datasets }    & \textbf{Material}           \\ \hline
intent                 & \cellcolor[HTML]{9B9B9B} & -                        & 0.57                     & -                        & 0.36                     & -                                               & 0.67                     & -                        & 0.49                     \\ \hline
fidelity               & -                        & \cellcolor[HTML]{9B9B9B} & -                        & -                        & -                        & -                                               & -                        & 0.46                     & -                        \\ \hline
evaluation   & 0.57                     & -                        & \cellcolor[HTML]{9B9B9B} & -                        & -                        & -                                               & 0.62                     & -                        & 0.40                     \\ \hline
variables              & -                        & -                        & -                        & \cellcolor[HTML]{9B9B9B} & 1                        & 1                                               & 0.55                     & 0.72                     & 0.69                     \\ \hline
n\_modalities & 0.36                     & -                        & -                        & 1                        & \cellcolor[HTML]{9B9B9B} & -                                               & -                        & -                        & 0.49                     \\ \hline
dynamicity             & -                        & -                        & -                        & 1                        & -                        & \cellcolor[HTML]{9B9B9B}{\color[HTML]{656565} } & 0.79                     & 0.71                     & 0.47                     \\ \hline
data\_type             & 0.67                     & -                        & 0.62                     & 0.55                     & -                        & 0.79                                            & \cellcolor[HTML]{9B9B9B} & 1                        & 0.65                     \\ \hline
n\_datasets   & -                        & 0.46                     & -                        & 0.72                     & -                        & 0.71                                            & 1                        & \cellcolor[HTML]{9B9B9B} & 0.41                     \\ \hline
material         & 0.49                     & -                        & 0.40                     & 0.69                     & 0.49                     & 0.47                                            & 0.65                     & 0.41                     & \cellcolor[HTML]{9B9B9B} \\ \noalign{\hrule height 1.0pt}

\end{tabular}
\end{adjustbox}

\end{table}

\textls[-15]{Table \ref{tab:interrelationships} summarises the results from the analysis. The key observations from Table \ref{tab:interrelationships} are the following:  }

\begin{itemize}
\item The Cramer's V between  n\_modalities/variables, n\_datasets/data\_type, and dynamicity/variables was 1 because the dimensions were derived from one another. In particular, n\_modalities counted the number of encoding variables used,  n\_datasets counted the number of data types used, and `dynamicity' documented whether (or not) dynamic variables were part of the encoding variables. 
\item Only non-random associations between two consecutive dimensions are highlighted in Figure \ref{fig:researchmodel}. Nonetheless, the data suggests that there were more non-random associations (e.g., intent/evaluation, intent/data, and data/material). Overall, the material dimension exhibited significant correlations with other non-derived dimensions most often (4/5: intent, evaluation, variables, and data type), followed by the data type dimension (4/5: intent, evaluation, variables, and material) and the intent dimension (3/5: evaluation, data type, and material). The fidelity dimension correlated with other non-derived dimensions the least often. 

\end{itemize}

Overall, the fact that the material/data type/intent dimensions exhibied nonrandom associations with other dimensions is in line with intuition. The number of datasets to encode and the combination of encoding variables emerged as determinants to watch, but since this study is the first to assess the interrelationships between these different representational dimensions and given the size of the sample, more work is needed to unveil the exact nature of the influences between dimensions. Thus, the observations above should be taken as working hypotheses \cite{chamberlin1890method} about the relationships between the representation dimensions in physicalisation research. 

\color{black}

\section{Discussion} 
\label{sec:discussion}


\color{black}
So far, this work has provided a synthesis of encoding variables for physicalisations (Section \ref{sec:narrativereview} and Table \ref{tab:encodingvariables}), a snapshot of evaluation criteria, as well as examples of methods to apply these criteria (Section \ref{subsec:criteria}) and working hypotheses about the relationships between different representational aspects of data physicalisation (representational intent, material type, representational fidelity, data type, and encoding variables, see Section \ref{subsec:interrelationships}). We now discuss general observations made about encoding variables and the evaluation criteria/methods, as well as their implications.


\color{black}
\subsection{Encoding Variables}
\label{subsec:takeaways-variables}
\textit{Takeaways}: 
One takeaway from the narrative review is that data encoding as an object of study is a fertile ground for interdisciplinary research. Indeed, several variables synthesized in Table \ref{tab:encodingvariables} were mentioned separately (and sometimes under slightly different names) in the literature on Information Visualisation, Cartography, Human--Computer Interaction, Sonification, Immersive Analytics, and Neuroscience. 
A case in point is dynamicity (the representation of change), for which the variables were `rediscovered' separately for the visual, sonic, and olfactory modalities.  
As for the systematic review, one takeaway is that inclusiveness is always realized to a certain extent, namely to the extent to which a given sensory modality is supported. In that sense, none of the physicalisations in the sample was fully inclusive (Table \ref{tab:paperoverview}). 
Finally, we have observed that only a few physicalisations actually used physical variables (i.e., changes in material properties) to convey messages about phenomena. A similar observation was made in \cite{sauve2022physecology}, who reported that information communicated through (a change in) physical or material form has been so far rare in practice. 

\textit{Implications}: Looking forward, researchers can use the framework from Table \ref{tab:encodingvariables} as a vocabulary to describe experiments assessing the effectiveness of variables (across disciplines). That is, the variables can serve as `boundary objects' \cite{star1989structure} between data physicalisation researchers and researchers from the fields mentioned just above (Information Visualisation, Cartography, Sonification, and so on). Boundary objects are concepts \textit{shared} by different communities, which can be viewed or used differently by each. For instance, a subject that can benefit from a plurality of perspectives is the study of users' perception of time-varying representations across different modalities (visual, haptic, aural, etc.). Here, the dynamic variables from Table \ref{tab:encodingvariables} can serve as boundary object between the different communities investigating the user experience of time-varying representations. Another subject that can benefit from a plurality of perspectives is the notion of variable syntactics (Interpretive flexibility is only one distinguishing characteristic of boundary objects. Another important characteristic is the arrangement of how to operate and collaborate \cite{starThisNotBoundary2010}. \mbox{As \citet{vuillemotBoundaryObjectsDesign2021}} put it, ``Groups can work on common objects locally, making them more tailored to their local use and needs, i.e. something that is not interdisciplinary, and then share it back in a way that works across the various groups''. It is challenging to exactly predict how arrangement will look like, as several disciplines use the variables. Hence, arrangement is not further specified in this article). As indicated in \cite{Roth2017a}, variable syntactics prescribe the use of a variable given a type of dataset (e.g.,  nominal, ordinal, or numerical). That is, variable syntactics tell how effective/ineffective a given variable is with respect to encoding a given data type. While variable syntactics have been suggested (mostly for the visual \cite{maceachren2004maps}, aural \cite{Krygier1994}, and dynamic variables \cite{maceachren2004maps}), researchers have so far been using different schemes while relating variables to data types. For example, `unacceptable/acceptable' was used for Bertin's original visual variable syntactics \cite{maceachren2004maps}, `not effective/effective' was used for auditory variables in \cite{Krygier1994}, and `poor/marginally effective/good' was used for visual and dynamic variables in \cite{maceachren2004maps}. Data physicalisation research will benefit from the harmonization of these schemes so that they abstract from the specifics of sensory modalities in a similar way to that done for dynamic variables in Section \ref{sec:narrativereview}. Next to researchers, designers of physicalisations can use the framework to identify design opportunities (e.g., through unexplored variables or an untried mix of variables). In that sense, the framework can be useful to support their goal, discussed in \cite{stolterman2008nature}, of creating the not-yet-existing. 

\subsection{Evaluation Criteria and Methods}
\textit{Takeaways}: \citet{Jansen2015} identified several key challenges for evaluating data physicalisations: (i) finding appropriate ways of studying how people engage in data exploration when no clear task is defined, (ii) assessing the merits of data representations that go beyond pure time and error metrics, (iii) exploring methodologies to understand how people reason, collaborate, and communicate with physicalisations, and (iv) finding fair alternative representations to use as a baseline for comparison. Our sample and analysis suggest that research that provides answers to challenges (i) and (ii) is ongoing. Notably, several criteria for evaluating  aspects that go beyond the traditional time and error metrics have emerged (see Section \ref{subsec:criteria}). However, we found fewer answers relevant to challenges (iii) and (iv) in our sample. For example, many of the papers in our corpus used on-screen visualisations (e.g., \cite{jansen2013_p36}), paper representations (e.g., \cite{stusak2016_p39}) and VR representations (e.g., \cite{ren2021comparing_p11}) as a baseline for comparing the effects of data physicalisations. The extent to which these baselines provide a fair ground for comparison still needs to be systematically assessed and discussed. 

Furthermore, the majority of the studies in our corpus (70\%, Table \ref{tab:evamethods}) were one-time studies. The evaluation of variables such as behavioural change and the impact on learning/ skills development requires long-term studies to understand the long-term effects. This long-term assessment would be relevant, for instance, to work using personal data physicalisations for teaching (e.g., \cite{perin2021whatstudentslearn}), or personal data physicalisations in real-world contexts (e.g., \cite{thudt2018selfreflection}). Besides, the evaluation of some aspects of physicalisations has not appeared in our sample, and this suggests that they could be under-explored or not explored at all. Since the systematic review has focused on representational aspects, we mention here a few, related to representation primarily. These include the following: strategies to communicate uncertainties in the underlying dataset; the impact of the material and representation fidelity on meaning-making and memorability; the connection between material properties and data types (e.g., would viscosity be a good material to represent numerical/ordinal/categorical data?); evaluating aspects related to affordances (cognitive affordances, physical affordances, sensory affordances, and functional affordances) in relation to data physicalisation; evaluating the multisensory perception of data; evaluating the interplay between representation and situatedness (e.g., physicalisations that are situated in close spatial proximity to their data referents \cite{willett2016embedded} compared to non-situated data physicalisations); and evaluating representation strategies and their adequacy for diverse user groups (e.g., children or elderly).

\textit{Implications}: We have already mentioned above that, despite the progress, there are still many unanswered questions. In particular, challenges (iii) and (iv) mentioned above deserve more attention. To these, the gap related to the long-term assessment of physicalisations' impacts and the need for more systematic accounts of the impact of representational features of physicalisations on users can be added.

\subsection{Relationships between Representational Dimensions}
\label{subsec:takeaways-interrelationships}
\textit{Takeaways}: Though the precedence links connecting the dimensions remain conjectural at this point, our analysis has highlighted that there is a non-random association between several dimensions touching on representational aspects for physicalisations. The results of the quantitative analysis suggests plausible relationships between the most important dimensions related to representation (i.e., material, data type, and intent). We were, however, surprised to see that the fidelity dimension did not seem to strongly connect with other dimensions related to representation. This may be due to the fact that the majority of the physicalisations in our sample (78\%) had a symbolic intent, and, hence, bore no resemblance to the data represented.

\textit{Implications}: Looking forward, the observations of the non-random associations encourage further research towards structural equation models for data physicalisation research. The exact nature of these non-random associations will be uncovered with more examples (and the working hypotheses mentioned in Section \ref{subsec:interrelationships} about the associations that can be used as a starting point). Models that describe the expected consequences of design choices during the process of building and evaluating physicalisations will benefit researchers and designers alike. These models will need to provide an account of the indirect relationships between dimensions. For instance, there was no significant association between encoding variables and evaluation, but there still was a non-random association between intent and evaluation (Table \ref{tab:interrelationships}). The documentation of researchers' work, using a consistent vocabulary to facilitate cross-comparison (e.g., the coding scheme from Section \ref{subsec:codingschemes}), will be needed to catalyze progress along these lines.

\color{black}

\subsection{Reflections on the Methodical Approach}
As \citet{roberts2010using} pointed out, we need a body of research that helps researchers tackle questions such as `what are the perceptual variables that are available?', `what are their limitations?', and `what guidelines are there for each variable?'. While it is clear that these questions still deserve attention, what is less clear are the methodical steps to arrive at general answers. This article has addressed the first question through a combination of a narrative and a systematic review. The narrative review has given the flexibility to draw ideas from different disciplines and reconcile differences in terminologies where appropriate. The systematic review has highlighted how work done in Information Visualisation and Human--Computer Interaction has been implementing the framework from the narrative review (Table \ref{tab:encodingvariables}). We anticipate that the framework from the narrative review can serve as a starting point for answering the other questions above. We also anticipate that replicating the study using articles from other communities will help to progressively extend that framework. For example, given the current corpus with predominant papers from the ACM Digital Library as input for the analysis, physicalisations from the Geography and Cartography communities (e.g., \cite{ballatore2019sonifying,moorman2020geospatial}), the variables that they use, and the lessons learned about them were not taken into account. Furthermore, since we wanted to learn about evaluation criteria and methods, we restricted ourselves to articles that evaluated their physicalisations in some way. Hence, some articles that could have possibly been useful to extend the list of variables (e.g., \cite{friske2020entangling}) were excluded. Replicating the study by removing this constraint would also be useful as we expand our understanding of perceptual variables for physicalisation research. In summary, though not without flaws, the combination of narrative and systematic review seems promising as we seek answers to the questions mentioned above.


\subsection{Limitations}
Our search criteria might have excluded some relevant papers, such as those that did not contain the search keywords that we used \textcolor{black}{(e.g., data sculpture, composite physicalisation, constructive visualisation)} or if they did not contain an empirical evaluation of a physicalisation. Also, we did not search for particular forms of physicalisations such as, for example, `sonification', `haptification' or `olfaction'. Consequently, our findings are dependent on the sample of papers we selected. 
\color{black}
Hence, the work does not claim to be exhaustive with respect to the evaluation criteria/methods collected. For instance, one may draw a distinction between physicalisations as designed artefacts, and physicalisations as printed artefacts. The former are physicalisations that were outcomes of a design process (some components of these physicalisations may be 3D printed and some not), while the latter denote physicalisations that were created entirely through printing (i.e., data is rendered as a physically fabricated object, see \cite{djavaherpour2021data}). Our sample is biased towards the former. Thus, we likely missed criteria relevant to the evaluation of physicalisations as printed artefacts (e.g., the accuracy of the printed artefact was proposed in \cite{allahverdi2018landscaper} to document the errors introduced by the printing process and does not appear in Table \ref{tab:evaluationcriteria}).  

\color{black}

\section{Conclusions and Future Work} 
\label{sec:conclusion}
This research provides two contributions to data physicalisation research: (i) a synthesis of the scattered literature on perceptual variables into a coherent framework, and (ii) a snapshot of evaluation criteria and methods relevant to the study of physicalisations. These two contributions can serve as a starting point for further work on the theories and guidelines for data physicalisations such as , notably, the empirical effectiveness of encoding variables for physicalisations and the applicability of perceptual variables to data communication/analysis scenarios.

A question that could guide follow-up reviews to this article is the following: `what do we know to be true of all perceptual variables, empirically?' In addition to including more examples of physicalisations, follow-up reviews could also cover more dimensions (e.g., reconfigurability discussed in \cite{Jansen2015}, the interaction discussed in \cite{Jansen2015,hogan2017towards}, and the audience mentioned in \cite{sauve2022physecology}) and the relationships, if any, with encoding variables, as well as evaluation criteria/methods.

Regarding encoding variables, there is a need for a more systematic investigation of which of these are atomic and which are composite. For instance, air quality is currently listed as an olfactory variable, but is in fact an umbrella term for many variables (e.g., air temperature and air humidity). Another direction for future research is the investigation of the effect of redundant sensorization (i.e., the combined use of several modalities) on user experience. There are works in the literature documenting the positive effects of redundant symbolization for the visual channel (e.g., \cite{gorte2022choriented,dobson1983visual}), as well as the visual and haptic channels used in combination (e.g., \cite{drogemuller2021_p59}), and more work along these lines is needed to increase our understanding of the use of redundancy during data encoding more broadly.

Finally, some criteria will benefit from a breaking down of factors that constitute them. This is the case, for example, for ``users' reactions'', ``design quality'', and the ``aesthetics of the physicalisation''. Developing standardized questionnaires that support the evaluation of these criteria, and more generally of criteria unique to data physicalisation research, is also an interesting direction for future work.

\vspace{6pt} 


\authorcontributions{Conceptualization, C.R. 
and A.D.; methodology, C.R. and A.D.; formal analysis, C.R. and A.D.; writing---original draft preparation, C.R. and A.D.; writing---review and editing, C.R. and A.D. All authors have read and agreed to the published version of the manuscript.}

\funding{Auriol Degbelo received funding from the German Research Foundation through the project NFDI4Earth (DFG project no. 460036893, \url{https://www.nfdi4earth.de/}  (accessed on 1 March 2022)
) within the German National Research Data Infrastructure (NFDI, \url{https://www.nfdi.de/}(accessed on 1 March 2022)).}

\institutionalreview{Not applicable.}

\informedconsent{Not applicable.}

\dataavailability{The data presented in this study are available in the article.}

\acknowledgments{We thank several anonymous reviewers for their helpful comments on earlier versions of this article.}


\conflictsofinterest{The authors declare no conflict of interest.} 



%
%

\appendixtitles{yes} 
\appendixstart
\appendix
  \section{Dimensions of Existing Design Spaces}
  \label{sec:appendixA}
The dimensions of existing design spaces along with their original names are summarised in Table \ref{tab:existingDims}.   
  
\begin{table}[H]
\caption{Dimensions of existing design spaces} 
\label{tab:existingDims}
\begin{adjustbox}{max width=\textwidth}
\begin{tabular}{lp{12cm}}
\toprule
\textbf{Design Space/ Framework} & \textbf{Dimensions} \\ \midrule
Multi-Sensory design space \cite{nesbitt2001modeling}                      & Sensory modalities, Encoding Variables         \\ \midrule
Data sculpture domain model \cite{zhao2008embodiment}   & Focus, Manifestation \\ \midrule
 Embodiment model \cite{zhao2008embodiment}  &   Metaphorical distance from data, Metaphorical data from reality         \\ \midrule
  Data Sculpture Design Taxonomy \cite{VandeMoere2009} & Representational fidelity, Narrative formulation fidelity          \\ \midrule
  Framework for multi-sensory data representation \cite{hogan2017towards}   &   Use of modalities (material, sensory modality), Representational intent (utilitarian, casual), Human data relations (interaction mode, type of data)         \\ \midrule
  Framework for multi-sensorial Immersive Analytics  \cite{mccormack2018multisensory}  &  Data (type of data, analytics possible), Sensory Mapping (encoding variables),   Devices, Human (human sensory channel)        \\ \midrule
  
 Physecology \cite{sauve2022physecology}  & Data type, Information communication, Interaction mechanisms, Spatial coupling, Physical setup, Audience. \\ \midrule
  Cross-disciplinary Design Space \cite{bae2022making} & Context (task, audience, location, data source), Structure (embodiment, material, encoding channel, mobility, data scalability, data duration), Interactions (interaction mediator, sense modality, data interactions)           \\ \midrule
  Design Elements in Data Physicalisation \cite{dumivcic2022design}  &  \textls[-15]{Design objective (Data form and property, Data theme and topic, design purpose, researched impact of physicalisations), Aesthetics (Design metaphor), Appearance (geometry, material), User experience (interaction, use of technology)}          \\ \bottomrule
\end{tabular}
\end{adjustbox}

\end{table}

\section{Guideline: Identifying When a Variable Type Has Been Used}
\label{sec:guidelines}
How to recognize the presence of a variable a posteriori: a variable is used if the sensory modality can be used \textit{independently} to perceive differences in data. 

\noindent Guiding questions: 
\begin{itemize}
\item Imagine I were blind; would I still perceive differences in the data?
\item Imagine I could not touch; would I still perceive differences in the data?
\item Imagine I could not smell; would I still perceive differences in the data?
\item Imagine I could not hear; would I still perceive differences in the data?
\item Imagine I could not taste; would I still perceive differences in the data?
\end{itemize}

If any of these questions is answered by ``No'', then it is evidence that \textit{only} the variable type corresponding to the sensory encoding channel mentioned in the question (visual, haptic, olfactory, sonic, and gustatory) has been used. If the question is answered by ``Yes'', it is evidence that a different sensory encoding channel from the one mentioned in the question has been used to encode data. Finding out whether or not dynamic variables are used can be done by asking the question: is animation or self-reconfiguration implemented?

\section{Definitions of Evaluation Criteria}
\label{sec:definition_eva_criteria_method}

This supplementary material provides additional details (e.g., definitions) about the evaluation criteria mentioned in the article. Some evaluation criteria were defined explicitly in the literature, and, for these, we add the references of the original articles next to their names. The remaining criteria were either (i) mentioned without explicit definition in the articles annotated, or (ii) needed relabelling to reflect the deeper notion they point at. For these criteria, we provide a tentative definition congruent with the article(s) annotated. The criteria are mentioned in their order of appearance in the article (most frequent to least frequent).
\begin{itemize}
\item[] \textbf{Criteria not particular to data physicalisation research}. 

\item Intellectual engagement \cite{Wang2019}: Refers to the ability to engage the user in intellectual activities such as recognition, analysis, and contemplation.

\item Social engagement \cite{Wang2019}: This is present when observers talk with companions, but also when laughing, gesturing, and mimicking the body postures of others. 
It was assessed, for instance, in \cite{hurtienne2020_p47} through the use of a confederate. 

\begin{itemize}
\item \textit{Confederates} are individuals recruited by lead experimenters to play the role of a bystander, participant, or teammate (see e.g., \cite{Leis2015}).
\end{itemize}

\item Affective engagement \cite{Wang2019}: Refers to the emotional experience of users. The arousing of feelings such as awe, respect, wonder, concern, fear, disgust, anger, or intimidation are indicators of an affective engagement.

\item Engagement over time: The evolution of engagement over a given time period.

\item User experience \cite{hassenzahl2004interplay,law2009}: The review of definitions by \citet{law2009} pointed out that the ISO definition of UX,``A person's perceptions and responses that result from the use or anticipated use of a product, system or service'', is in line with what most UX researchers associate to the concept. In essence, UX refers to all aspects of the users' interaction with a product. It has pragmatic attributes and hedonic attributes \cite{hassenzahl2004interplay}. 

\item Utility \cite{Nielsen2012,Roth2015}: It is the usefulness of an interface for completing the user’s desired set of objectives \cite{Roth2015}.

\item Effectiveness (question answering) \cite{ANSI2001}: In the sample analyzed, effectiveness was measured through the accuracy with which participants completed information retrieval tasks \cite{daniel2019cairnform_p8,stusak2016_p39} and interaction tasks \cite{taher2015exploring_p37}. 

\begin{itemize}
    \item Information retrieval tasks are specifically directed at retrieving information (e.g., cluster, maxima, or minima of a dataset), whereas interaction tasks are more open-ended (e.g., data analysis tasks such as annotation, filtering or navigation). Hence, not every interaction task is an information retrieval task.
\end{itemize} 

\item Efficiency (question answering) \cite{Nielsen2012,ANSI2001}: This is the time taken by participants to complete information retrieval tasks or interaction tasks.

\item Potential for self-reflection: This is the ability of the physicalisations to prompt users to think about themselves. \citet{thudt2018selfreflection} identified four types of personal reflection in the context of data physicalisation: reflection on (their) data, reflection on (their) context, reflection on (their) action, and reflection on (their) values. 

\item Understanding (qualitative): This refers to the assessment of the understanding of datasets through qualitative feedback during an interview \cite{ren2021comparing_p11} or as a rating on a self-developed questionnaire \cite{ang2019physicalizing_p24,hurtienne2020_p47}. 
\begin{itemize}
    \item This assessment may touch upon the understanding by an individual (in that case we talk about personal understanding, see \cite{ang2019physicalizing_p24,sauve2020_p43}), or a group of people (in that case, we talk about collaborative understanding, see \cite{veldhuis2020coda_p10}).
\end{itemize}

\item Attitude change/behavioural stimulation: This refers to the extent to which a physicalisation can change the attitudes of users (e.g., do they care more about a given subject?) or inspire them to take some action \cite{hurtienne2020_p47}. 

\item Memorability \cite{Saket2016,stusak2015_p55}: Memorability has different facets, for instance, recognition or recall (see \cite{Brown1977}), explicit or implicit memorability (see \cite{stusak2015_p55}), and the storage of information in short-term memory or long-term memory (see \cite{Camina2017}). It is the capability of maintaining and retrieving information \cite{Saket2016}.

\item Enjoyment/satisfaction \cite{Saket2016,ANSI2001}: Enjoyment is a feeling that causes a person to experience pleasure \cite{Saket2016}. Satisfaction denotes the freedom from discomfort and positive attitudes towards the use of the product \cite{ANSI2001}.

\item Motivational potential: The ability of the physicalisation to promote gradual changes in individuals’ behaviour or sustain the changes over time. It was evaluated through self-developed questionnaires \cite{stusak2014activity_p3}.

\item Ease of use: The perceived ease of use.

\item Design parameters: sSme studies intended to find optimal design parameters and conducted a systematic evaluation of these parameters to that end. For instance, \citet{daniel2019cairnform_p8} systematically varied motion speeds to find out the best speed to animate the CairnFORM physicalisation. \citet{lopez-garcia2021_p28} systematically varied the size of two physicalisations and assessed the impact of these changes on ease of viewing and understanding.

\item Learning curve/ease of learning: This refers to the perceived learning curve.

\item Social acceptance/ease of adoption \cite{batch2020scents_p29}: this refers to participants' opinions about the possible introduction of the physicalisation in their lives or sentiments regarding the ease of adoption of the physicalisation. 

\item Size judgment: Although this was assessed primarily through the accuracy of participants on information retrieval tasks in \cite{jansen2016psycho_p2} (and, hence, could have been said to belong to the assessment of the effectiveness of the physicalisation), we still kept this criterion as separate, because it is important for the development of theories of perceptual effectiveness of variables. Ratio estimation \cite{Cleveland1986} and constant sum \cite{Comrey1950,Spence1990} are two methods to collect data about participants' judgments.

\item Confidence \cite{batch2020scents_p29}: This refers to the self-reported confidence levels of users.

\item Creativity \cite{Wang2019}: The ability of the physicalisation to support the introduction of new and original ideas.
\end{itemize}

\begin{itemize}
\item[] \textbf{Criteria that seem particular to data physicalisation research.}

\item Physical engagement \cite{Wang2019}: It invites people to spend time touching and interacting with the data
(even if just in imagination), moving around it to
take different perspectives, bending down to read a label, and employing senses including smell and hearing.

\item Users' reactions: Some articles used the term `user reaction' \cite{taher2015exploring_p37,boem2018vitalmorph_p42} or `ad-hoc impression' \cite{hogan2017visual_p34} to refer to how the users react to a physicalisation. While there are overlaps with engagement (e.g., the user reactions mentioned in \cite{taher2015exploring_p37} could be classified as an assessment of physical engagement, and part of the reactions documented in \cite{boem2018vitalmorph_p42} could be classified as an assessment of affective engagement), we still keep this evaluation criterion as distinct, because it could be useful for exploratory studies. 

\item Orientation consistency \cite{sauve2020change_p1}: The consistency of user responses to information retrieval tasks across different orientations.


\item Quality of the design: This touches upon participants' general feedback about design decisions and material choices. It was evaluated, for instance, through self-developed questionnaires \cite{stusak2014activity_p3}, post-it note feedback, \cite{perovich2021chemicals_p9} and unstructured interviews \cite{boem2018vitalmorph_p42}.

\item Potential for self-expression: The extent to which the physicalisation can help users express some personal characteristics (e.g., academic profile or running performance). It has at least two components mentioned in \cite{panagiotidou2020data_p13}: representational possibilities (what the user can say through the physicalisation) and representational precision (how accurately they can say what they intend to say). 

\item Quality of the information content: Evaluated in \cite{stusak2014activity_p3} through self-developed questionnaires.

\item Aesthetics of the physicalisation: This touches upon the appearance of the physicalisation. It was evaluated using self-developed questionnaires in \cite{stusak2014activity_p3}. 

\item Remote awareness of physiological states: Some studies \cite{boem2018vitalmorph_p42,pepping2020_p45} explored the use of physicalisations as a means for remote monitoring. That is, a user uses a physicalisation to infer the physiological state (e.g., emotional state \cite{pepping2020_p45} or arterial blood \mbox{pressure \cite{boem2018vitalmorph_p42}}) of another distant user. 

\end{itemize}

\begin{adjustwidth}{-\extralength}{0cm}

\reftitle{References}

\PublishersNote{}
\end{adjustwidth}

\begin{thebibliography}{999}

\bibitem[Jansen et~al.(2015)Jansen, Dragicevic, Isenberg, Alexander, Karnik,
  Kildal, Subramanian, and Hornb{\ae}k]{Jansen2015}
Jansen, Y.; Dragicevic, P.; Isenberg, P.; Alexander, J.; Karnik, A.; Kildal,
  J.; Subramanian, S.; Hornb{\ae}k, K.
\newblock {Opportunities and challenges for data physicalization}.
\newblock In Proceedings of the 33rd Annual ACM Conference
  on Human Factors in Computing Systems---CHI '15,  Seoul, Republic of Korea, 23--28 April 2015;  
 Begole, B., Kim, J., Inkpen,
  K., Woo, W., Eds.;  
 pp. 3227--3236.
\newblock {\url{https://doi.org/10.1145/2702123.2702180}}.

\bibitem[van Loenhout et~al.(2022)van Loenhout, Ranasinghe, Degbelo, and
  Bouali]{van2022physicalizing}
van Loenhout, R.; Ranasinghe, C.; Degbelo, A.; Bouali, N.
\newblock Physicalizing Sustainable Development Goals Data: An Example with SDG
  7 (Affordable and Clean Energy).
\newblock In Proceedings of the CHI Conference on Human Factors in Computing
  Systems Extended Abstracts,  New Orleans, LA, USA, 30 April--6 May 2022;  
 pp. 1--7.

\bibitem[Lee et~al.(2020)Lee, Ju, Dzhoroev, Goh, Lee, and
  Park]{kyung2020dayclo_p19}
Lee, K.R.; Ju, S.; Dzhoroev, T.; Goh, G.i.; Lee, M.H.; Park, Y.W.
\newblock {DayClo: An everyday table clock providing interaction with personal
  schedule data for self-reflection}.
\newblock In Proceedings of the DIS'20: Designing Interactive Systems
  Conference 2020, Eindhoven, The Netherlands,  6--20 July 2020; Wakkary, R., Andersen, K., Odom, W., Desjardins, A.,
  Petersen, M.G., Eds.; 2020; pp. 1793--1806.
\newblock {\url{https://doi.org/10.1145/3357236.3395439}}.

\bibitem[Ju et~al.(2019)Ju, Lee, Kim, and Park]{ju2019bookly_p6}
Ju, S.; Lee, K.R.; Kim, S.; Park, Y.W.
\newblock {Bookly: An interactive everyday artifact showing the time of
  physically accumulated reading activity}.
\newblock In Proceedings of the 2019 CHI Conference on Human
  Factors in Computing Systems, Glasgow, UK,  4--9 May  2019; Brewster, S.A., Fitzpatrick, G., Cox, A.L.,
  Kostakos, V., Eds.; pp. 1--8.
\newblock {\url{https://doi.org/10.1145/3290605.3300614}}.

\bibitem[Menheere et~al.(2021)Menheere, Van~Hartingsveldt, Birkeb{\ae}k, Vos,
  and Lallemand]{menheere2021laina_p58}
Menheere, D.; Van~Hartingsveldt, E.; Birkeb{\ae}k, M.; Vos, S.; Lallemand, C.
\newblock Laina: Dynamic data physicalization for slow exercising feedback.
\newblock In Proceedings of the Designing Interactive Systems Conference 2021, virtual event,
  28 June--2 July 2021; pp. 1015--1030.
\bibitem[Stusak et~al.(2016)Stusak, Hobe, and Butz]{stusak2016_p39}
Stusak, S.; Hobe, M.; Butz, A.
\newblock {If your mind can grasp it, your hands will help}.
\newblock In Proceedings of the TEI'16: Tenth International
  Conference on Tangible, Embedded, and Embodied Interaction, Eindhoven, The Netherlands,  14--17 February 2016; Bakker, S.,
  Hummels, C., Ullmer, B., Geurts, L., Hengeveld, B., Saakes, D., Broekhuijsen,
  M., Eds.;  pp. 92--99.

\bibitem[Stusak et~al.(2015)Stusak, Schwarz, and Butz]{stusak2015_p55}
Stusak, S.; Schwarz, J.; Butz, A.
\newblock {Evaluating the memorability of physical Visualizations}.
\newblock In Proceedings of the 33rd Annual ACM Conference
  on Human Factors in Computing Systems (CHI 2015), Seoul, Republic of Korea, 23--28 April 2015; Begole, B., Kim, J.,
  Inkpen, K., Woo, W., Eds.;  pp. 3247--3250.
\newblock {\url{https://doi.org/10.1145/2702123.2702248}}.

\bibitem[Dragicevic et~al.(2021)Dragicevic, Jansen, and
  Moere]{dragicevic2021data}
Dragicevic, P.; Jansen, Y.; Moere, A.V.
\newblock {Data physicalization}. In {\em Springer Handbook of Human Computer
  Interaction}; Springer:  Berlin/Heidelberg, Germany, 
  2021.

\bibitem[Hogan and Hornecker(2017)]{hogan2017towards}
Hogan, T.; Hornecker, E.
\newblock Towards a design space for multisensory data representation.
\newblock {\em Interact. Comput.} {\bf 2017}, {\em 29},~147--167.
\newblock {\url{https://doi.org/10.1093/iwc/iww015}}.

\bibitem[Sauv{\'e} et~al.(2022)Sauv{\'e}, Sturdee, and
  Houben]{sauve2022physecology}
Sauv{\'e}, K.; Sturdee, M.; Houben, S.
\newblock Physecology: A Conceptual Framework to Describe Data Physicalizations
  in their Real-World Context.
\newblock {\em ACM Trans. -Comput.-Hum. Interact.} {\bf 2022}, {\em
  29},~1--33.

\bibitem[Bae et~al.(2022)Bae, Zheng, West, Do, Huron, and
  Szafir]{bae2022making}
Bae, S.S.; Zheng, C.; West, M.E.; Do, E.Y.L.; Huron, S.; Szafir, D.A.
\newblock Making Data Tangible: A Cross-disciplinary Design Space for Data
  Physicalization.
\newblock In Proceedings of the CHI Conference on Human Factors in Computing
  Systems, New Orleans LA USA, 30 April--6  May 2022; pp. 1--18.

\bibitem[Sosa et~al.(2018)Sosa, Gerrard, Esparza, Torres, Napper,
  et~al.]{sosa2018data}
Sosa, R.; Gerrard, V.; Esparza, A.; Torres, R.; Napper, R.;  et~al.
\newblock Data objects: Design principles for data physicalisation.
\newblock In Proceedings of the DS 92: Proceedings of the DESIGN 2018 15th
  International Design Conference,  Dubrovnik, Croatia, 24 May 2018; pp. 1685--1696.

\bibitem[Hogan(2018)]{hogan2018data}
Hogan, T.
\newblock {Data sensification: Beyond representation modality, toward encoding
  data in experience}.
\newblock In Proceedings of the Design as a Catalyst for Change---DRS
  International Conference 2018, Limerick, Ireland,  25--28 June 2018; Storni, C., Leahy, K., McMahon, M., Lloyd, P.,
  Bohemia, E., Eds. 
\newblock {\url{https://doi.org/10.21606/drs.2018.238}}.

\bibitem[Willett et~al.(2016)Willett, Jansen, and
  Dragicevic]{willett2016embedded}
Willett, W.; Jansen, Y.; Dragicevic, P.
\newblock Embedded data representations.
\newblock {\em IEEE Trans. Vis. Comput. Graph.} {\bf
  2016}, {\em 23},~461--470.

\bibitem[Wu et~al.(2021)Wu, Petersen, Ahmad, Burlinson, Tanis, and
  Szafir]{wu2021understanding}
Wu, K.; Petersen, E.; Ahmad, T.; Burlinson, D.; Tanis, S.; Szafir, D.A.
\newblock Understanding data accessibility for people with intellectual and
  developmental disabilities.
\newblock In Proceedings of the 2021 CHI Conference on Human
  Factors in Computing Systems,  Yokohama, Japan, 8--13  May  2021; pp. 1--16.

\bibitem[Roberts and Walker(2010)]{roberts2010using}
Roberts, J.C.; Walker, R.
\newblock {Using all our senses: the need for a unified theoretical approach to
  multi-sensory information visualization}.
\newblock In Proceedings of the IEEE VisWeek 2010 Workshop on The Role of
  Theory in Information Visualization, Salt Lake City, UT, USA,   24--29 October  2010.

\bibitem[Jansen and Hornbaek(2016)]{jansen2016psycho_p2}
Jansen, Y.; Hornbaek, K.
\newblock {A psychophysical investigation of size as a physical variable}.
\newblock {\em IEEE Trans. Vis. Comput. Graph.} {\bf
  2016}, {\em 22},~479--488.
\newblock {\url{https://doi.org/10.1109/TVCG.2015.2467951}}.

\bibitem[Hogan et~al.(2018)Hogan, Hinrichs, Alexander, Huron, Carpendale, and
  Hornecker]{hogan2018toward}
Hogan, T.; Hinrichs, U.; Alexander, J.; Huron, S.; Carpendale, S.; Hornecker,
  E.
\newblock Toward a design language for data physicalization.
\newblock In Proceedings of the IEEE VIS Workshop,   Estrel Hotel Berlin, Germany, 21--26 October 2018.

\bibitem[Stusak et~al.(2018)Stusak, Butz, and Tabard]{stusakvariables}
Stusak, S.; Butz, A.; Tabard, A.
\newblock Variables for Data Physicalization Units.
\newblock In Proceedings of the Toward a Design Language for Data
  Physicalization: Workshop at IEEE VIS,  Estrel Hotel Berlin, Germany, 21--26 October 2018.

\bibitem[Oehlberg and Willett(2018)]{oehlberg2018encoding}
Oehlberg, L.; Willett, W.
\newblock Encoding data through experiential material properties.
\newblock In Proceedings of the Toward a Design Language for Data
  Physicalization: Workshop at IEEE VIS,  Estrel Hotel Berlin, Germany, 21--26 October 2018.

\bibitem[Moere(2008)]{moere2008beyond}
Moere, A.V.
\newblock Beyond the tyranny of the pixel: Exploring the physicality of
  information visualization.
\newblock In Proceedings of the 2008 12th International Conference Information
  Visualisation, London, UK, 9--11 July 2008;  pp. 469--474.

\bibitem[Nesbitt(2001)]{nesbitt2001modeling}
Nesbitt, K.V.
\newblock Modeling the multi-sensory design space.
\newblock In Proceedings of the  APVis '01: Proceedings of the 2001 Asia-Pacific symposium on Information Visualisation, 
  Sydney, Australia,  1  December 2001; Volume~9, pp. 27--36.

\bibitem[Wang et~al.(2019)Wang, Segal, Klatzky, Keefe, Isenberg, Hurtienne,
  Hornecker, Dwyer, Barrass, and Rhyne]{Wang2019}
Wang, Y.; Segal, A.; Klatzky, R.; Keefe, D.F.; Isenberg, P.; Hurtienne, J.;
  Hornecker, E.; Dwyer, T.; Barrass, S.; Rhyne, T.M.
\newblock {An emotional response to the value of visualization}.
\newblock {\em IEEE Comput. Graph. Appl.} {\bf 2019}, {\em
  39},~8--17.
\newblock {\url{https://doi.org/10.1109/MCG.2019.2923483}}.

\bibitem[Snyder(2019)]{Snyder2019}
Snyder, H.
\newblock {Literature review as a research methodology: An overview and
  guidelines}.
\newblock {\em J. Bus. Res.} {\bf 2019}, {\em 104},~333--339.
\newblock {\url{https://doi.org/10.1016/j.jbusres.2019.07.039}}.

\bibitem[Grant and Booth(2009)]{Grant2009}
Grant, M.J.; Booth, A.
\newblock {A typology of reviews: an analysis of 14 review types and associated
  methodologies}.
\newblock {\em Health Inf. Libr. J.} {\bf 2009}, {\em
  26},~91--108.
\newblock {\url{https://doi.org/10.1111/j.1471-1842.2009.00848.x}}.

\bibitem[Wabiński et~al.(2022)Wabiński, Mościcka, and
  Touya]{wabinski2022guidelines}
Wabiński, J.; Mościcka, A.; Touya, G.
\newblock Guidelines for standardizing the design of tactile maps: a review of
  research and best practice.
\newblock {\em  Cartogr. J.} {\bf 2022}, {\em 59},~239--258.
\newblock {\url{https://doi.org/10.1080/00087041.2022.2097760}}.

\bibitem[Laakso and Tiina~Sarjakoski(2010)]{laakso2010sonic}
Laakso, M.; Tiina~Sarjakoski, L.
\newblock Sonic maps for hiking - use of sound in enhancing the map use
  experience.
\newblock {\em  Cartogr. J.} {\bf 2010}, {\em 47},~300--307.
\newblock {\url{https://doi.org/10.1179/000870410X12911298276237}}.

\bibitem[{Vande Moere} and Patel(2009)]{VandeMoere2009}
{Vande Moere}, A.; Patel, S.
\newblock {The physical visualization of information: designing data sculptures
  in an educational context}. In {\em Visual Information Communication}; Huang,
  M.L., Nguyen, Q.V., Zhang, K., Eds.; Springer: Boston, MA, USA, 2009; pp.
  1--23.
\newblock {\url{https://doi.org/10.1007/978-1-4419-0312-9_1}}.

\bibitem[Zhao and Moere(2008)]{zhao2008embodiment}
Zhao, J.; Moere, A.V.
\newblock Embodiment in data sculpture: A model of the physical visualization
  of information.
\newblock In Proceedings of the 3rd International Conference
  on Digital Interactive Media in Entertainment and Arts,  Athens, Greece, 10--12 September 2008; pp. 343--350.



\bibitem[McCormack et~al.(2018)McCormack, Roberts, Bach, Freitas, Itoh, Hurter,
  and Marriott]{mccormack2018multisensory}
McCormack, J.; Roberts, J.C.; Bach, B.; Freitas, C.D.S.; Itoh, T.; Hurter, C.;
  Marriott, K.
\newblock Multisensory immersive analytics. In {\em Immersive Analytics};
  Springer Nature Switzerland AG 2018;  Springer:  Berlin/Heidelberg, Germany,   
  2018; pp. 57--94.

\bibitem[Dumičić et~al.(2022)Dumičić, Thoring, Klöckner, and
  Joost]{dumivcic2022design}
Dumičić, Z.; Thoring, K.; Klöckner, H.W.; Joost, G.
\newblock \emph{Design Elements in Data Physicalization: A Systematic Literature
  Review};  Lockton, D., Lenzi, S., Hekkert, P., Oak, A., Sadaba, J., Lloyd, P., Eds.;  Design Research Society: Bilbao, Spain, 2022. 
\newblock {\url{https://doi.org/10.21606/drs.2022.660}}.

\bibitem[Yi et~al.(2007)Yi, ah~Kang, Stasko, and Jacko]{Yi2007}
Yi, J.S.; ah~Kang, Y.; Stasko, J.; Jacko, J.
\newblock {Toward a deeper understanding of the role of interaction in
  information visualization}.
\newblock {\em IEEE Trans. Vis. Comput. Graph.} {\bf
  2007}, {\em 13},~1224--1231.
\newblock {\url{https://doi.org/10.1109/TVCG.2007.70515}}.

\bibitem[Roth(2013)]{roth2013interactive}
Roth, R.E.
\newblock {Interactive maps: What we know and what we need to know}.
\newblock {\em J. Spat. Inf. Sci.} {\bf 2013}, {\em
  6},~59--115.
\newblock {\url{https://doi.org/10.5311/JOSIS.2013.6.105}}.

\bibitem[Bernsen(1994)]{Bernsen1994}
Bernsen, N.O.
\newblock {Foundations of multimodal representations: a taxonomy of
  representational modalities}.
\newblock {\em Interact. Comput.} {\bf 1994}, {\em 6},~347--371.
\newblock {\url{https://doi.org/10.1016/0953-5438(94)90008-6}}.

\bibitem[Rasmussen et~al.(2012)Rasmussen, Pedersen, Petersen, and
  Hornb{\ae}k]{rasmussen2012shape}
Rasmussen, M.K.; Pedersen, E.W.; Petersen, M.G.; Hornb{\ae}k, K.
\newblock {Shape-changing interfaces: a review of the design space and open
  research questions}.
\newblock In Proceedings of the CHI'12---Conference on Human Factors in
  Computing Systems,  Austin, TX, USA,  5--10 May 2012; Konstan, J.A., Chi, E.H., H{\"{o}}{\"{o}}k, K., Eds.;  
  pp. 735--744.
\newblock {\url{https://doi.org/10.1145/2207676.2207781}}.

\bibitem[Bertin(1983)]{Bertin1983}
Bertin, J.
\newblock {\em {Semiology of Graphics: Diagrams, Networks, Maps}};  Translated by
  William J. Berg; The University of Wisconsin Press: Madison, WI, USA, 1983.

\bibitem[MacEachren(1995)]{maceachren2004maps}
MacEachren, A.M.
\newblock {\em {How Maps Work: Representation, Visualization, and Design}};
  Guilford Press:  New York, NY, USA, 1995.

\bibitem[Roth(2017)]{Roth2017a}
Roth, R.E.
\newblock {Visual variables}. In {\em International Encyclopedia of Geography:
  People, the Earth, Environment and Technology}; Richardson, D., Castree, N.,
  Goodchild, M.F., Kobayashi, A., Liu, W., Marston, R.A., Eds.; John Wiley \&
  Sons, Ltd.: Oxford, UK,  2017; pp. 1--11.
\newblock {\url{https://doi.org/10.1002/9781118786352.wbieg0761}}.

\bibitem[White(2017)]{White2017}
White, T.
\newblock {Symbolization and the visual variables}. In {\em Geographic
  Information Science \& Technology Body of Knowledge}; Wilson, J.P., Ed.;
  Number~Q2; University Consortium for Geographic Information Science (UCGIS): Chesapeake, VA, USA,  2017. 
\newblock {\url{https://doi.org/10.22224/gistbok/2017.2.3}}.

\bibitem[Caivano(1990)]{Caivano1990}
Caivano, J.L.
\newblock Visual texture as a semiotic system.
\newblock {\em Semiotica} {\bf 1990}, {\em 80}, 239--252.
\newblock {\url{https://doi.org/10.1515/semi.1990.80.3-4.239}}.

\bibitem[Kraak et~al.(2020)Kraak, Roth, Ricker, Kagawa, and Sourd]{Kraak2020}
Kraak, M.J.; Roth, R.E.; Ricker, B.; Kagawa, A.; Sourd, G.L.
\newblock {\em {Mapping for a Sustainable World}}; The United Nations: New
  York, NY, USA,  2020; p.~79.

\bibitem[Paneels and Roberts(2010)]{Paneels2010}
Paneels, S.; Roberts, J.C.
\newblock {Review of designs for haptic data visualization}.
\newblock {\em IEEE Trans. Haptics} {\bf 2010}, {\em 3},~119--137.
\newblock {\url{https://doi.org/10.1109/TOH.2009.44}}.

\bibitem[Griffin(2001)]{Griffin2001}
Griffin, A.L.
\newblock {Feeling it out: The use of haptic visualization for exploratory
  geographic analysis}.
\newblock {\em Cartogr. Perspect.} {\bf 2001}, 12--29.
\newblock {\url{https://doi.org/10.14714/CP39.636}}.

\bibitem[Novich and Eagleman(2015)]{Novich2015}
Novich, S.D.; Eagleman, D.M.
\newblock {Using space and time to encode vibrotactile information: Toward an
  estimate of the skin's achievable throughput}.
\newblock {\em Exp. Brain Res.} {\bf 2015}, {\em 233},~2777--2788.
\newblock {\url{https://doi.org/10.1007/s00221-015-4346-1}}.

\bibitem[Chouvardas et~al.(2008)Chouvardas, Miliou, and
  Hatalis]{Chouvardas2008}
Chouvardas, V.; Miliou, A.; Hatalis, M.
\newblock {Tactile displays: Overview and recent advances}.
\newblock {\em Displays} {\bf 2008}, {\em 29},~185--194.
\newblock {\url{https://doi.org/10.1016/j.displa.2007.07.003}}.

\bibitem[Perovich et~al.(2021)Perovich, Cai, Guo, Zimmerman, Paseman, {Espinoza
  Silva}, and Brody]{perovich2021clothing_p15}
Perovich, L.J.; Cai, P.; Guo, A.; Zimmerman, K.; Paseman, K.; {Espinoza Silva},
  D.; Brody, J.G.
\newblock {Data clothing and BigBarChart: designing physical data reports on
  indoor pollutants for individuals and communities}.
\newblock {\em IEEE Comput. Graph. Appl.} {\bf 2021}, {\em
  41},~87--98.
\newblock {\url{https://doi.org/10.1109/MCG.2020.3025322}}.

\bibitem[Panagiotidou et~al.(2020)Panagiotidou, Gorucu, and {Vande
  Moere}]{panagiotidou2020data_p13}
Panagiotidou, G.; Gorucu, S.; {Vande Moere}, A.
\newblock {Data badges: making an academic profile through a DIY wearable
  physicalization}.
\newblock {\em IEEE Comput. Graph. Appl.} {\bf 2020}, {\em
  40},~51--60.
\newblock {\url{https://doi.org/10.1109/MCG.2020.3025504}}.

\bibitem[Drogemuller et~al.(2021)Drogemuller, Cunningham, Walsh, Baumeister,
  Smith, and Thomas]{drogemuller2021_p59}
Drogemuller, A.; Cunningham, A.; Walsh, J.A.; Baumeister, J.; Smith, R.T.;
  Thomas, B.H.
\newblock {Haptic and visual comprehension of a 2D graph layout through
  physicalisation}.
\newblock In Proceedings of the CHI'21: CHI Conference on Human Factors in
  Computing Systems, Virtual Event,  Yokohama, Japan, 8--13  May  2021; Kitamura, Y., Quigley, A., Isbister, K., Igarashi, T.,
  Bj{\o}rn, P., Drucker, S.M., Eds.;  pp. 463:1--463:16.
\newblock {\url{https://doi.org/10.1145/3411764.3445704}}.


\bibitem[Hogan et~al.(2017)Hogan, Hinrichs, and Hornecker]{hogan2017visual_p34}
Hogan, T.; Hinrichs, U.; Hornecker, E.
\newblock {The Visual and beyond: characterizing experiences with auditory,
  haptic and visual data representations}.
\newblock In Proceedings of the 2017 Conference on Designing
  Interactive Systems, (DIS'17), Edinburgh, UK, 10--14 June  2017; Mival, O.H., Smyth, M., Dalsgaard, P., Eds.;      pp. 797--809.
\newblock {\url{https://doi.org/10.1145/3064663.3064702}}.

\bibitem[Pon et~al.(2017)Pon, Pattison, Fyfe, Radford, and
  Carpendale]{pon2017_p49}
Pon, A.; Pattison, E.; Fyfe, L.; Radford, L.; Carpendale, S.
\newblock {\emph{Torrent}: Integrating embodiment, physicalization and
  musification in music-making}.
\newblock In Proceedings of the Tenth International
  Conference on Tangible, Embedded, and Embodied Interaction (TEI 2017), Yokohama, Japan,  20-23 March 2017;
  Inakage, M., Ishii, H., Do, E.Y.L., Steimle, J., Shaer, O., Kunze, K.,
  Peiris, R.L., Eds.;     Association for Computing Machinery, New York NY, United States; pp. 209--216.
\newblock {\url{https://doi.org/10.1145/3024969.3024974}}.

\bibitem[Batch et~al.(2020)Batch, Patnaik, Akazue, and
  Elmqvist]{batch2020scents_p29}
Batch, A.; Patnaik, B.; Akazue, M.; Elmqvist, N.
\newblock {Scents and sensibility: Evaluating information olfactation}.
\newblock In Proceedings of the CHI'20: CHI Conference on Human Factors in
  Computing Systems, Honolulu, HI, USA,  April 25 - 30, 2020; Bernhaupt, R., Mueller, F.F., Verweij, D., Andres, J.,
  McGrenere, J., Cockburn, A., Avellino, I., Goguey, A., Bj{\o}n, P.; Zhao, S.;
   et~al., Eds.;     Association for Computing MachineryNew YorkNYUnited States; pp. 1--14.
\newblock {\url{https://doi.org/10.1145/3313831.3376733}}.

\bibitem[Pepping et~al.(2020)Pepping, Scholte, van Wijland, de~Meij, Wallner,
  and Bernhaupt]{pepping2020_p45}
Pepping, J.; Scholte, S.; van Wijland, M.; de~Meij, M.; Wallner, G.; Bernhaupt,
  R.
\newblock {Motiis: fostering parents' awareness of their adolescents emotional
  experiences during gaming}.
\newblock In Proceedings of the NordiCHI '20: Shaping Experiences, Shaping
  Society, Proceedings of the 11th Nordic Conference on Human-Computer
  Interaction, Tallinn, Estonia,  October 25 - 29, 2020; Lamas, D., Sarapuu, H., L{\'{a}}rusd{\'{o}}ttir, M., Stage, J.,
  Ardito, C., Eds.;     Association for Computing Machinery New York, NY, United States; pp. 58:1--58:11.
\newblock {\url{https://doi.org/10.1145/3419249.3420173}}.

\bibitem[Patnaik et~al.(2019)Patnaik, Batch, and Elmqvist]{Patnaik2019}
Patnaik, B.; Batch, A.; Elmqvist, N.
\newblock {Information olfactation: harnessing scent to convey data}.
\newblock {\em IEEE Trans. Vis. Comput. Graph.} {\bf
  2019}, {\em 25},~726--736.
\newblock {\url{https://doi.org/10.1109/TVCG.2018.2865237}}.

\bibitem[Maggioni et~al.(2020)Maggioni, Cobden, Dmitrenko, Hornb{\ae}k, and
  Obrist]{maggioni2020smell}
Maggioni, E.; Cobden, R.; Dmitrenko, D.; Hornb{\ae}k, K.; Obrist, M.
\newblock SMELL SPACE: mapping out the olfactory design space for novel
  interactions.
\newblock {\em ACM Trans. -Comput.-Hum. Interact. (TOCHI)} {\bf
  2020}, {\em 27},~1--26.

\bibitem[Mueller et~al.(2021)Mueller, Dwyer, Goodwin, Marriott, Deng, Phan,
  Lin, Chen, Wang, and Khot]{mueller2021datadelight}
Mueller, F.F.; Dwyer, T.; Goodwin, S.; Marriott, K.; Deng, J.; Phan, H.D.; Lin,
  J.; Chen, K.T.; Wang, Y.; Khot, R.A.
\newblock {Data as delight: eating data}.
\newblock In Proceedings of the CHI'21: CHI Conference on Human Factors in
  Computing Systems, Yokohama, Japan,  May 8-13, 2021; Kitamura, Y., Quigley, A., Isbister, K., Igarashi, T.,
  Bj{\o}rn, P., Drucker, S.M., Eds.; \newblock {Association for Computing Machinery New York, NY, United States; pp.
  621:1--621:14}.
\newblock {\url{https://doi.org/10.1145/3411764.3445218}}.

\bibitem[Wang et~al.(2016)Wang, Ma, Luo, and Qu]{wang2016edibilization}
Wang, Y.; Ma, X.; Luo, Q.; Qu, H.
\newblock {Data edibilization: representing data with food}.
\newblock In Proceedings of the Extended Abstracts of the 2016 CHI Conference
  on Human Factors in Computing Systems (CHIEA'16), San Jose, CA, USA, May 7 – 12 ; Kaye, J., Druin, A., Lampe,
  C., Morris, D., Hourcade, J.P., Eds.;  Association for Computing Machinery New York, NY, United States; pp.
  409--422.
\newblock {\url{https://doi.org/10.1145/2851581.2892570}}.

\bibitem[Obrist et~al.(2014)Obrist, Comber, Subramanian, Piqueras-Fiszman,
  Velasco, and Spence]{obrist2014temporal}
Obrist, M.; Comber, R.; Subramanian, S.; Piqueras-Fiszman, B.; Velasco, C.;
  Spence, C.
\newblock {Temporal, affective, and embodied characteristics of taste
  experiences: a framework for design}.
\newblock In Proceedings of the CHI Conference on Human Factors in Computing
  Systems (CHI'14),  Toronto, ON, Canada,  26 April 2014- 1 May 2014; Jones, M., Palanque, P.A., Schmidt, A., Grossman, T., Eds.; Association for Computing Machinery New York, NY, United States; pp. 2853--2862.
\newblock {\url{https://doi.org/10.1145/2556288.2557007}}.


\bibitem[Kikut-Ligaj and Trzcieli{\'{n}}ska-Lorych(2015)]{Kikut-Ligaj2015}
Kikut-Ligaj, D.; Trzcieli{\'{n}}ska-Lorych, J.
\newblock {How taste works: cells, receptors and gustatory perception}.
\newblock {\em Cell. Mol. Biol. Lett.} {\bf 2015}, {\em 20}.
\newblock {\url{https://doi.org/10.1515/cmble-2015-0042}}.

\bibitem[Gal et~al.(2007)Gal, Wheeler, and Shiv]{gal2007cross}
Gal, D.; Wheeler, S.C.; Shiv, B.
\newblock {Cross-modal influences on gustatory perception 
 {\bf 2007}.
 \newblock{Social Science Research Network (“SSRN”),}
 \newblock{\url{https://papers.ssrn.com/sol3/papers.cfm?abstract_id=1030197}}, last accessed on 30th May 2022.}

\bibitem[Harrar and Spence(2013)]{Harrar2013}
Harrar, V.; Spence, C.
\newblock {The taste of cutlery: how the taste of food is affected by the
  weight, size, shape, and colour of the cutlery used to eat it}.
\newblock {\em Flavour} {\bf 2013}, {\em 2},~21.
\newblock {\url{https://doi.org/10.1186/2044-7248-2-21}}.

\bibitem[{Van Doorn} et~al.(2014){Van Doorn}, Wuillemin, and
  Spence]{VanDoorn2014}
{Van Doorn}, G.H.; Wuillemin, D.; Spence, C.
\newblock {Does the colour of the mug influence the taste of the coffee?}
\newblock {\em Flavour} {\bf 2014}, {\em 3},~10.
\newblock {\url{https://doi.org/10.1186/2044-7248-3-10}}.


\bibitem[Kramer et~al.(2010)Kramer, Walker, Bonebright, Cook, Flowers, Miner,
  and Neuhoff]{kramer2010sonification}
Kramer, G.; Walker, B.; Bonebright, T.; Cook, P.; Flowers, J.H.; Miner, N.;
  Neuhoff, J.
\newblock {\emph{Sonification Report: Status of the Field and Research Agenda}}; Faculty Publications, Department of Psychology, DigitalCommons@University of Nebraska - Lincoln, US. 
: {
  2010}.
\newblock 

\bibitem[Krygier(1994)]{Krygier1994}
Krygier, J.B.
\newblock {Sound and geographic visualization}. In {\em Visualization in Modern
  Cartography}; MacEachren, A., Taylor, F., Eds.; Academic Press, 1994. p. 149-166.
\newblock {\url{https://doi.org/10.1016/B978-0-08-042415-6.50015-6}}.

\bibitem[Madhyastha and Reed(1995)]{Madhyastha1995}
Madhyastha, T.; Reed, D.
\newblock {Data sonification: do you see what I hear?}
\newblock {\em IEEE Softw.} {\bf 1995}, {\em 12},~45--56.
\newblock {\url{https://doi.org/10.1109/52.368264}}.


\bibitem[Palom{\"{a}}ki(2006)]{palomaki06meaningsconveyed}
Palom{\"{a}}ki, H.
\newblock {Meanings conveyed by simple auditory rhythms}.
\newblock In Proceedings of the 12th International
  Conference on Auditory Display (ICAD2006),  June 19-24, 2006, London, UK.

\bibitem[Bernard et~al.(2022)Bernard, Monnoyer, Wiertlewski, and
  Ystad]{Bernard2022}
Bernard, C.; Monnoyer, J.; Wiertlewski, M.; Ystad, S.
\newblock {Rhythm perception is shared between audio and haptics}.
\newblock {\em Sci. Rep.} {\bf 2022}, {\em 12},~4188.
\newblock {\url{https://doi.org/10.1038/s41598-022-08152-w}}.

\bibitem[Visi et~al.(2014)Visi, Dothel, Williams, and
  Miranda]{visi2014unfolding}
Visi, F.; Dothel, G.; Williams, D.; Miranda, E.
\newblock Unfolding | Clusters: A Music and Visual Media Model of ALS
  Pathophysiology.
\newblock In Proceedings of the SoniHED Conference: Sonification
  of Health and Environmental Data, York, UK,  12th September, 2014. 

\bibitem[Carpendale(2003)]{carpendale2003considering}
Carpendale, S.
\newblock {\emph{Considering Visual Variables as a Basis for Information
  Visualisation}}; Technical Report; University of Calgary, 2500 University Drive NW, Calgary Alberta T2N 1N4, CANADA, 2003;
\newblock 
  {\url{https://doi.org/10.11575/PRISM/30495}}.

\bibitem[K{\"{o}}bben and Yaman(1995)]{kobben1995evaluating}
K{\"{o}}bben, B.; Yaman, M.
\newblock {Evaluating dynamic visual variables}.
\newblock In Proceedings of the Seminar on Teaching Animated
  Cartography, Madrid, Spain,  August 30-September 1, 1995; Ormeling, F., K{\"{o}}bben, B., P{\'{e}}rez-G{\'{o}}mez, R.,
  Eds., International Cartographic Association, 1996, pp. 45--51.

\bibitem[DiBiase et~al.(1992)DiBiase, MacEachren, Krygier, and
  Reeves]{DiBiase1992}
DiBiase, D.; MacEachren, A.M.; Krygier, J.B.; Reeves, C.
\newblock {Animation and the role of map design in scientific visualization}.
\newblock {\em Cartogr. Geogr. Inf. Syst.} {\bf 1992}, {\em
  19},~201--214.
\newblock {\url{https://doi.org/10.1559/152304092783721295}}.

\bibitem[Blok(1998)]{blok1998dynamic}
Blok, C.
\newblock {Dynamic visualization in a developing framework for the
  representation of geographic data}.
\newblock {\em Bull. Com. Fr. Cartogr.} {\bf 1998}, {\em
  156},~89--97.
\newblock {\url{https://doi.org/10.4000/cybergeo.509}}.


\bibitem[Baykal et~al.(2020)Baykal, Van~Mechelen, and
  Eriksson]{baykal2020collaborative}
Baykal, G.E.; Van~Mechelen, M.; Eriksson, E.
\newblock Collaborative technologies for children with special needs: A
  systematic literature review.
\newblock In Proceedings of the 2020 CHI Conference on Human
  Factors in Computing Systems,  	Honolulu HI USA April 25 - 30, 2020,     Association for Computing MachineryNew YorkNYUnited States; pp. 1--13.

\bibitem[Salminen et~al.(2020)Salminen, Guan, Jung, Chowdhury, and
  Jansen]{salminen2020literature}
Salminen, J.; Guan, K.; Jung, S.g.; Chowdhury, S.A.; Jansen, B.J.
\newblock A literature review of quantitative persona creation.
\newblock In Proceedings of the 2020 CHI Conference on Human
  Factors in Computing Systems,  	Honolulu HI USA April 25 - 30, 2020,     Association for Computing MachineryNew YorkNYUnited States; pp. 1--14.

\bibitem[Pettersson et~al.(2018)Pettersson, Lachner, Frison, Riener, and
  Butz]{pettersson2018bermuda}
Pettersson, I.; Lachner, F.; Frison, A.K.; Riener, A.; Butz, A.
\newblock A Bermuda triangle? A Review of method application and triangulation
  in user experience evaluation.
\newblock In Proceedings of the 2018 CHI Conference on Human
  Factors in Computing Systems,  Montreal QC Canada April 21 - 26, 2018; pp. 1--16.

\bibitem[Bargas-Avila and Hornb{\ae}k(2011)]{bargas2011old}
Bargas-Avila, J.A.; Hornb{\ae}k, K.
\newblock Old wine in new bottles or novel challenges: A critical analysis of
  empirical studies of user experience.
\newblock In Proceedings of the SIGCHI Conference on Human
  Factors in Computing Systems,  Vancouver BC Canada May 7 - 12, 2011; pp. 2689--2698.

\bibitem[Koelle et~al.(2020)Koelle, Ananthanarayan, and Boll]{koelle2020social}
Koelle, M.; Ananthanarayan, S.; Boll, S.
\newblock Social acceptability in HCI: A survey of methods, measures, and
  design strategies.
\newblock In Proceedings of the 2020 CHI Conference on Human
  Factors in Computing Systems,  Honolulu HI USA April 25 - 30, 2020; pp. 1--19.





\bibitem[Herman et~al.(2021)Herman, Omdal, Zeller, Richter, Samsel, Abram, and
  Keefe]{herman2021multi}
Herman, B.; Omdal, M.; Zeller, S.; Richter, C.A.; Samsel, F.; Abram, G.; Keefe,
  D.F.
\newblock Multi-Touch Querying on Data Physicalizations in Immersive AR.
\newblock {\em Proc.  ACM -Hum.-Comput. Interact.} {\bf
  2021}, {\em 5},~1--20.

\bibitem[Houben et~al.(2016)Houben, Golsteijn, Gallacher, Johnson, Bakker,
  Marquardt, Capra, and Rogers]{houben2016physikit_p25}
Houben, S.; Golsteijn, C.; Gallacher, S.; Johnson, R.; Bakker, S.; Marquardt,
  N.; Capra, L.; Rogers, Y.
\newblock {Physikit: data engagement through physical ambient visualizations in
  the home}.
\newblock In Proceedings of the 2016 CHI Conference on Human
  Factors in Computing Systems---CHI '16, San Jose, CA, USA,  May 7 - 12, 2016; Kaye, J., Druin, A., Lampe, C.,
  Morris, D., Hourcade, J.P., Eds.;     Association for Computing MachineryNew YorkNYUnited States, 2016; pp.
  1608--1619.
\newblock {\url{https://doi.org/10.1145/2858036.2858059}}.

\bibitem[Stevens(1946)]{stevens1946theory}
Stevens, S.S.
\newblock {On the theory of scales of measurement}.
\newblock {\em Science} {\bf 1946}, {\em 103},~677--680.

\bibitem[Djavaherpour et~al.(2021)Djavaherpour, Samavati, Mahdavi‐Amiri,
  Yazdanbakhsh, Huron, Levy, Jansen, and Oehlberg]{djavaherpour2021data}
Djavaherpour, H.; Samavati, F.; Mahdavi‐Amiri, A.; Yazdanbakhsh, F.; Huron,
  S.; Levy, R.; Jansen, Y.; Oehlberg, L.
\newblock Data to physicalization: a survey of the physical rendering process.
\newblock {\em Comput. Graph. Forum} {\bf 2021}, {\em 40},~569--598.
\newblock {\url{https://doi.org/10.1111/cgf.14330}}.

\bibitem[Nielsen(2012)]{Nielsen2012}
Nielsen, J.
\newblock {Usability 101: Introduction to Usability.}
\newblock{NN/g Nielsen Norman Group, 3rd January 2012}
\newblock {\url{https://www.nngroup.com/articles/usability-101-introduction-to-usability/}}.
\newblock last accessed 31st August 2022.

\bibitem[Saket et~al.(2016)Saket, Endert, and Stasko]{Saket2016}
Saket, B.; Endert, A.; Stasko, J.
\newblock {Beyond usability and performance: A review of user
  experience-focused evaluations in visualization}.
\newblock In Proceedings of the Beyond Time and Errors on
  Novel Evaluation Methods for Visualization---BELIV '16, Baltimore, Maryland,
  USA,  24 October 2016; Sedlmair, M.,
  Isenberg, P., Isenberg, T., Mahyar, N., Lam, H., Eds.;     Association for Computing MachineryNew YorkNYUnited States, 24th October, 2016; pp. 133--142.
\newblock {\url{https://doi.org/10.1145/2993901.2993903}}.

\bibitem[Hurtienne et~al.(2020)Hurtienne, Maas, Carolus, Reinhardt, Baur, and
  Wienrich]{hurtienne2020_p47}
Hurtienne, J.; Maas, F.; Carolus, A.; Reinhardt, D.; Baur, C.; Wienrich, C.
\newblock {Move\&Find: The value of kinaesthetic experience in a casual data
  representation}.
\newblock {\em IEEE Comput. Graph. Appl.} {\bf 2020}, {\em
  40},~61--75.
\newblock {\url{https://doi.org/10.1109/MCG.2020.3025385}}.

















\bibitem[Perovich et~al.(2021)Perovich, Wylie, and
  Bongiovanni]{perovich2021chemicals_p9}
Perovich, L.J.; Wylie, S.A.; Bongiovanni, R.
\newblock {Chemicals in the creek: designing a situated data physicalization of
  open government data with the community}.
\newblock {\em IEEE Trans. Vis. Comput. Graph.} {\bf
  2021}, {\em 27},~913--923.
\newblock {\url{https://doi.org/10.1109/TVCG.2020.3030472}}.

\bibitem[Ren and Hornecker(2021)]{ren2021comparing_p11}
Ren, H.; Hornecker, E.
\newblock {Comparing understanding and memorization in physicalization and VR
  visualization}.
\newblock In Proceedings of the Fifteenth International
  Conference on Tangible, Embedded, and Embodied Interaction,  Salzburg, Austria,
  February 14 - 17, 2021; Wimmer, R.,
  Kaltenbrunner, M., Murer, M., Wolf, K., Oakley, I., Eds.;     Association for Computing MachineryNew YorkNYUnited States, February 2021; pp. 1--7.
\newblock {\url{https://doi.org/10.1145/3430524.3442446}}.



\bibitem[Lee et~al.(2021)Lee, Kim, Kim, Hong, and Park]{lee2021adio_p56}
Lee, K.R.; Kim, B.; Kim, J.; Hong, H.; Park, Y.W.
\newblock {ADIO: An interactive artifact physically representing the intangible
  digital audiobook listening experience in everyday living spaces}.
\newblock In Proceedings of the CHI'21: CHI Conference on Human Factors in
  Computing Systems,  Virtual Event,  Yokohama Japan May 8 - 13, 2021; Kitamura, Y., Quigley, A., Isbister, K., Igarashi, T.,
  Bj{\o}rn, P., Drucker, S.M., Eds.;     Association for Computing MachineryNew YorkNYUnited States, May 2021; pp. 164:1--164:12.
\newblock {\url{https://doi.org/10.1145/3411764.3445440}}.

\bibitem[Sauv{\'{e}} et~al.(2020)Sauv{\'{e}}, Bakker, Marquardt, and
  Houben]{sauve2020_p43}
Sauv{\'{e}}, K.; Bakker, S.; Marquardt, N.; Houben, S.
\newblock {LOOP: Exploring physicalization of activity tracking data}.
\newblock In Proceedings of the NordiCHI '20: Shaping Experiences, Shaping
  Society, Proceedings of the 11th Nordic Conference on Human-Computer
  Interaction, Tallinn, Estonia,  October 25 - 29, 2020; Lamas, D., Sarapuu, H., L{\'{a}}rusd{\'{o}}ttir, M., Stage, J.,
  Ardito, C., Eds.;     Association for Computing MachineryNew YorkNYUnited States, 26th October 2020; pp. 52:1--52:12.
\newblock {\url{https://doi.org/10.1145/3419249.3420109}}.

\bibitem[Boem and Iwata(2018)]{boem2018vitalmorph_p42}
Boem, A.; Iwata, H.
\newblock {“It's like holding a human heart”: The design of Vital + Morph,
  a shape-changing interface for remote monitoring}.
\newblock {\em AI Soc.} {\bf 2018}, {\em 33},~599--619.
\newblock {\url{https://doi.org/10.1007/s00146-017-0752-1}}.

\bibitem[Daniel et~al.(2019)Daniel, Rivi{\`{e}}re, and
  Couture]{daniel2019cairnform_p8}
Daniel, M.; Rivi{\`{e}}re, G.; Couture, N.
\newblock {CairnFORM: A shape-changing ring chart notifying renewable energy
  availability in peripheral locations}.
\newblock In Proceedings of the Thirteenth International
  Conference on Tangible, Embedded, and Embodied Interaction, Tempe, Arizona,
  USA,  March 17 - 20, 2019; pp. 275--286.
\newblock {\url{https://doi.org/10.1145/3294109.3295634}}.

\bibitem[Veldhuis et~al.(2020)Veldhuis, Liang, and
  Bekker]{veldhuis2020coda_p10}
Veldhuis, A.; Liang, R.H.; Bekker, T.
\newblock {CoDa: collaborative data interpretation through an interactive
  tangible scatterplot}.
\newblock In Proceedings of the Fourteenth International
  Conference on Tangible, Embedded, and Embodied Interaction, Sydney, Australia,
  February February 9 - 12, 2020; van~den Hoven,
  E., Loke, L., Shaer, O., van Dijk, J., Kun, A.L., Eds.;     Association for Computing MachineryNew YorkNYUnited States, 9th Feb. 2020; pp. 323--336.
\newblock {\url{https://doi.org/10.1145/3374920.3374934}}.

\bibitem[Ang et~al.(2019)Ang, Samavati, Sabokrohiyeh, Garcia, and
  Elbaz]{ang2019physicalizing_p24}
Ang, K.D.; Samavati, F.F.; Sabokrohiyeh, S.; Garcia, J.; Elbaz, M.S.
\newblock Physicalizing cardiac blood flow data via 3D printing.
\newblock {\em Comput. Graph.} {\bf 2019}, {\em 85},~42--54.

\bibitem[{L{\'{o}}pez Garc{\'{i}}a} and Hornecker(2021)]{lopez-garcia2021_p28}
{L{\'{o}}pez Garc{\'{i}}a}, I.; Hornecker, E.
\newblock {Scaling data physicalization---How does size influence experience?}
\newblock In Proceedings of the TEI '21: Fifteenth International Conference on
  Tangible, Embedded, and Embodied Interaction, Online Event,  February 14 - 17, 2021; Wimmer, R., Kaltenbrunner, M.,
  Murer, M., Wolf, K., Oakley, I., Eds.;     Association for Computing MachineryNew YorkNYUnited States, 14th Feb. 2021; pp. 8:1--8:14.
\newblock {\url{https://doi.org/10.1145/3430524.3440627}}.

\bibitem[Keefe et~al.(2018)Keefe, Johnson, Altheimer, Hong, Hunter, Johnson,
  Rockcastle, Swackhamer, and Wittkamper]{keefe2018weather_p52}
Keefe, D.F.; Johnson, S.; Altheimer, R.; Hong, D.G.; Hunter, R.; Johnson, A.J.;
  Rockcastle, M.; Swackhamer, M.; Wittkamper, A.
\newblock {Weather Report: A site-specific artwork interweaving human
  experiences and scientific data physicalization}.
\newblock {\em IEEE Comput. Graph. Appl.} {\bf 2018}, {\em
  38},~10--16.
\newblock {\url{https://doi.org/10.1109/MCG.2018.042731653}}.

\bibitem[Cuya et~al.(2021)Cuya, Guarese, Johansson, Giambastiani, Iquiapaza,
  {de Jesus Oliveira}, Nedel, and Maciel]{cuya2021_p61}
Cuya, F.G.B.; Guarese, R.L.M.; Johansson, C.G.C.; Giambastiani, M.; Iquiapaza,
  Y.; {de Jesus Oliveira}, V.A.; Nedel, L.P.; Maciel, A.
\newblock {Vibrotactile data physicalization: exploratory insights for
  haptization of low-resolution images}.
\newblock In Proceedings of the SVR'21: 23rd Symposium on Virtual and Augmented
  Reality, Virtual Event, October 18 - 21, 2021; pp. 84--91.
\newblock {\url{https://doi.org/10.1145/3488162.3488171}}.

\bibitem[Stusak et~al.(2014)Stusak, Tabard, Sauka, Khot, and
  Butz]{stusak2014activity_p3}
Stusak, S.; Tabard, A.; Sauka, F.; Khot, R.A.; Butz, A.
\newblock {Activity sculptures: exploring the impact of physical visualizations
  on running activity}.
\newblock {\em IEEE Trans. Vis. Comput. Graph.} {\bf
  2014}, {\em 20},~2201--2210.
\newblock {\url{https://doi.org/10.1109/TVCG.2014.2352953}}.

\bibitem[Taher et~al.(2015)Taher, Hardy, Karnik, Weichel, Jansen, Hornb{\ae}k,
  and Alexander]{taher2015exploring_p37}
Taher, F.; Hardy, J.; Karnik, A.; Weichel, C.; Jansen, Y.; Hornb{\ae}k, K.;
  Alexander, J.
\newblock {Exploring interactions with physically dynamic bar charts}.
\newblock In Proceedings of the 33rd Annual ACM Conference
  on Human Factors in Computing Systems (CHI 2015), Seoul, Republic of Korea,  April 18 - 23, 2015; Begole, B., Kim, J.,
  Inkpen, K., Woo, W., Eds.;     Association for Computing MachineryNew YorkNYUnited States, 18th April 2015; pp. 3237--3246.
\newblock {\url{https://doi.org/10.1145/2702123.2702604}}.

\bibitem[Suzuki et~al.(2017)Suzuki, Stangl, Gross, and Yeh]{suzuki2017_p38}
Suzuki, R.; Stangl, A.; Gross, M.D.; Yeh, T.
\newblock {FluxMarker: Enhancing tactile graphics with dynamic tactile
  markers}.
\newblock In Proceedings of the 19th International ACM
  SIGACCESS Conference on Computers and Accessibility (ASSETS 2017), Baltimore, MD, USA,  20 October 2017- 1 November 2017 ; Hurst, A.,
  Findlater, L., Morris, M.R., Eds.;     Association for Computing MachineryNew YorkNYUnited States, 19, 2017; pp.
  190--199.
\newblock {\url{https://doi.org/10.1145/3132525.3132548}}.

\bibitem[Jansen et~al.(2013)Jansen, Dragicevic, and Fekete]{jansen2013_p36}
Jansen, Y.; Dragicevic, P.; Fekete, J.D.
\newblock {Evaluating the efficiency of physical visualizations}.
\newblock In Proceedings of the 2013 ACM SIGCHI Conference on Human Factors in
  Computing Systems (CHI'13), Paris, France,  27 April 2013- 2 May 2013; Mackay, W.E., Brewster, S.A., B{\o}dker, S.,
  Eds.;     Association for Computing MachineryNew YorkNYUnited States, 27 April 2013; pp. 2593--2602.
\newblock {\url{https://doi.org/10.1145/2470654.2481359}}.

\bibitem[Sauv{\'{e}} et~al.(2020)Sauv{\'{e}}, Potts, Alexander, and
  Houben]{sauve2020change_p1}
Sauv{\'{e}}, K.; Potts, D.; Alexander, J.; Houben, S.
\newblock {A change of perspective: how user orientation influences the
  perception of physicalizations}.
\newblock In Proceedings of the 2020 CHI Conference on Human
  Factors in Computing Systems, Honolulu, HI, USA,  April 25 - 30, 2020; Bernhaupt, R., Mueller, F.F., Verweij, D.,
  Andres, J., McGrenere, J., Cockburn, A., Avellino, I., Goguey, A., Bj{\o}n,
  P., Zhao, S.,  et~al., Eds.;     Association for Computing MachineryNew YorkNYUnited States, 23 April 2020; pp. 1--12.
\newblock {\url{https://doi.org/10.1145/3313831.3376312}}.

\bibitem[Hassenzahl(2004)]{hassenzahl2004interplay}
Hassenzahl, M.
\newblock {The interplay of beauty, goodness, and usability in interactive
  products}.
\newblock {\em Hum.-Comput. Interact.} {\bf 2004}, {\em 19},~319--349.
\newblock {\url{https://doi.org/10.1207/s15327051hci1904_2}}.

\bibitem[Law et~al.(2009)Law, Roto, Hassenzahl, Vermeeren, and Kort]{law2009}
Law, E.; Roto, V.; Hassenzahl, M.; Vermeeren, A.; Kort, J.
\newblock {Understanding, Scoping and Defining User eXperience: A Survey
  Approach}.
\newblock { In Proceedings of the CHI'09, ACM SIGCHI Conference on Human Factors in
  Computing Systems}, Boston MA USA April 4 - 9, 2009; pp. 719--728.
\newblock {\url{https://doi.org/10.1145/1518701.1518813}}.

\bibitem[Roth et~al.(2015)Roth, Ross, and MacEachren]{Roth2015}
Roth, R.E.; Ross, K.; MacEachren, A.
\newblock {User-centered design for interactive maps: A case study in crime
  analysis}.
\newblock {\em ISPRS Int. J. -Geo-Inf.} {\bf 2015}, {\em
  4},~262--301.
\newblock {\url{https://doi.org/10.3390/ijgi4010262}}.

\bibitem[ANSI(2001)]{ANSI2001}
ANSI.
\newblock {\emph{Common Industry Format for Usability Test Reports}};
\newblock Technical Report; 
\newblock {National Institute of Standards and Technology, US Department of Commerce, 100 Bureau Drive, Gaithersburg, MD 20899}
 2001. 
.

\bibitem[Jansen and Dragicevic(2013)]{jansen2013interaction}
Jansen, Y.; Dragicevic, P.
\newblock An interaction model for visualizations beyond the desktop.
\newblock {\em IEEE Trans. Vis. Comput. Graph.} {\bf
  2013}, {\em 19},~2396--2405.

\bibitem[Whetten(1989)]{Whetten1989}
Whetten, D.A.
\newblock What constitutes a theoretical contribution?
\newblock {\em Acad. Manag. Rev.} {\bf 1989}, {\em 14},~490--495.
\newblock {\url{https://doi.org/10.5465/amr.1989.4308371}}.

\bibitem[Weisstein(2022)]{weisstein2022fisher}
Weisstein, E.W.
\newblock {Fisher's exact test (From MathWorld - A Wolfram Web Resource)}.
\newblock{Wolfram Research, Inc.}, 2022.
\newblock {\url{https://mathworld.wolfram.com/FishersExactTest.html}}

\newblock last accessed September 10, 2022.

\bibitem[Chamberlin(1890)]{chamberlin1890method}
Chamberlin, T.C.
\newblock {The method of multiple working hypotheses}.
\newblock {\em Science} {\bf 1890}, {\em 15},~92--96.

\bibitem[Star(1989)]{star1989structure}
Star, S.L.
\newblock The structure of ill-structured solutions: {Boundary} objects and
  heterogeneous distributed problem solving. In {\em Distributed {Artificial}
  {Intelligence} ({Vol}. 2)}; Morgan Kaufmann Publishers Inc., San Francisco, CA, USA, 
  1989; pp. 37--54.

\bibitem[Star(2010)]{starThisNotBoundary2010}
Star, S.L.
\newblock This is not a boundary object: Reflections on the origin of a
  concept.
\newblock {\em Sci. Technol. Hum. Values} {\bf 2010}, {\em
  35},~601--617.
\newblock {\url{https://doi.org/10.1177/0162243910377624}}.

\bibitem[Vuillemot et~al.(2021)Vuillemot, Rivière, Beignon, and
  Tabard]{vuillemotBoundaryObjectsDesign2021}
Vuillemot, R.; Rivière, P.; Beignon, A.; Tabard, A.
\newblock Boundary objects in design studies: Reflections on the collaborative
  creation of isochrone maps.
\newblock {\em Comput. Graph. Forum} {\bf 2021}, {\em 40},~349--360.
\newblock {\url{https://doi.org/10.1111/cgf.14312}}.

\bibitem[Stolterman(2008)]{stolterman2008nature}
Stolterman, E.
\newblock The nature of design practice and implications for interaction design
  research.
\newblock {{\em Int. J. Des.} {\bf 2008}, {\em 2 (1)}, 55-65}. 


\bibitem[Perin(2021)]{perin2021whatstudentslearn}
Perin, C.
\newblock What students learn with personal data physicalization.
\newblock {\em IEEE Comput. Graph. Appl.} {\bf 2021}, {\em
  41},~48--58.
\newblock {\url{https://doi.org/10.1109/MCG.2021.3115417}}.

\bibitem[Thudt et~al.(2018)Thudt, Hinrichs, Huron, and
  Carpendale]{thudt2018selfreflection}
Thudt, A.; Hinrichs, U.; Huron, S.; Carpendale, S.
\newblock Self-reflection and personal physicalization construction.
\newblock In Proceedings of the 2018 {CHI} conference on
  human factors in computing systems ({CHI} 2018), Montreal, QC, Canada, April 21 - 26, 2018; Mandryk, R.L., Hancock, M.,
  Perry, M., Cox, A.L., Eds.;      Association for Computing Machinery, New York, NY, United States, 21 April 2018; p. 154.
\newblock {\url{https://doi.org/10.1145/3173574.3173728}}.

\bibitem[Ballatore et~al.(2019)Ballatore, Gordon, and
  Boone]{ballatore2019sonifying}
Ballatore, A.; Gordon, D.; Boone, A.P.
\newblock {Sonifying data uncertainty with sound dimensions}.
\newblock {\em Cartogr. Geogr. Inf. Sci.} {\bf 2019}, {\em
  46},~385--400.
\newblock {\url{https://doi.org/10.1080/15230406.2018.1495103}}.

\bibitem[Moorman et~al.(2020)Moorman, Djavaherpour, Etemad, and
  Samavati]{moorman2020geospatial}
Moorman, L.; Djavaherpour, H.; Etemad, K.; Samavati, F.F.
\newblock {Geospatial physicalization in geography education}.
\newblock {\em J. Geogr.} {\bf 2020}, {\em 120},~23--35.
\newblock {\url{https://doi.org/10.1080/00221341.2020.1832138}}.

\bibitem[Friske et~al.(2020)Friske, Wirfs-Brock, and
  Devendorf]{friske2020entangling}
Friske, M.; Wirfs-Brock, J.; Devendorf, L.
\newblock {Entangling the roles of maker and interpreter in interpersonal data
  narratives: explorations in yarn and sound}.
\newblock In Proceedings of the DIS'20: Designing Interactive Systems
  Conference 2020, Eindhoven, The Netherlands,  uly 6 - 10, 2020; Wakkary, R., Andersen, K., Odom, W., Desjardins, A.,
  Petersen, M.G., Eds.;     Association for Computing Machinery, New York, NY, United States, 3 July 2020; pp. 297--310.
\newblock {\url{https://doi.org/10.1145/3357236.3395442}}.

\bibitem[Allahverdi et~al.(2018)Allahverdi, Djavaherpour, Mahdavi-Amiri, and
  Samavati]{allahverdi2018landscaper}
Allahverdi, K.; Djavaherpour, H.; Mahdavi-Amiri, A.; Samavati, F.F.
\newblock Landscaper: {A} modeling system for {3D} printing scale models of
  landscapes.
\newblock {\em Comput. Graph. Forum} {\bf 2018}, {\em 37},~439--451.
\newblock {\url{https://doi.org/10.1111/cgf.13432}}.

\bibitem[Gorte and Degbelo(2022)]{gorte2022choriented}
Gorte, V.; Degbelo, A.
\newblock Choriented maps: Visualizing {SDG} data on mobile devices.
\newblock {\em  Cartogr. J.} {\bf 2022}, {\em 59},~35--54.
\newblock {\url{https://doi.org/10.1080/00087041.2021.1986616}}.

\bibitem[Dobson(1983)]{dobson1983visual}
Dobson, M.W.
\newblock {Visual information processing and cartographic communication: The
  utility of redundant stimulus dimensions}. In {\em Graphic Communication and
  Design in Contemporary Cartography}; Taylor, D.R.F., Ed.; John Wiley \& Sons:  Chichester, UK, 
1983; pp. 149--175.

\bibitem[Leis and Reinerman-Jones(2015)]{Leis2015}
Leis, R.; Reinerman-Jones, L.
\newblock {Methodological implications of confederate use for experimentation
  in safety-critical domains}.
\newblock {\em Procedia Manuf.} {\bf 2015}, {\em 3},~1233--1240.
\newblock {\url{https://doi.org/10.1016/j.promfg.2015.07.258}}.

\bibitem[Brown et~al.(1977)Brown, Lewis, and Monk]{Brown1977}
Brown, J.; Lewis, V.J.; Monk, A.F.
\newblock {Memorability, word frequency and negative recognition}.
\newblock {\em Q. J. Exp. Psychol.} {\bf 1977}, {\em
  29},~461--473.
\newblock {\url{https://doi.org/10.1080/14640747708400622}}.

\bibitem[Camina and G{\"{u}}ell(2017)]{Camina2017}
Camina, E.; G{\"{u}}ell, F.
\newblock {The neuroanatomical, neurophysiological and psychological basis of
  memory: current models and their origins}.
\newblock {\em Front. Pharmacol.} {\bf 2017}, {\em 8}.
\newblock {\url{https://doi.org/10.3389/fphar.2017.00438}}.

\bibitem[Cleveland and McGill(1986)]{Cleveland1986}
Cleveland, W.S.; McGill, R.
\newblock {An experiment in graphical perception}.
\newblock {\em Int. J. -Man-Mach. Stud.} {\bf 1986}, {\em
  25},~491--500.
\newblock {\url{https://doi.org/10.1016/S0020-7373(86)80019-0}}.

\bibitem[Comrey(1950)]{Comrey1950}
Comrey, A.L.
\newblock {A proposed method for absolute ratio scaling}.
\newblock {\em Psychometrika} {\bf 1950}, {\em 15},~317--325.
\newblock {\url{https://doi.org/10.1007/BF02289045}}.

\bibitem[Spence(1990)]{Spence1990}
Spence, I.
\newblock {Visual psychophysics of simple graphical elements}.
\newblock {\em J. Exp. Psychol. Hum. Percept. Perform.} {\bf 1990}, {\em 16},~683--692.
\newblock {\url{https://doi.org/10.1037/0096-1523.16.4.683}}.

\end{thebibliography}
\end{document}